\documentclass[twocolumn]{aastex631}
\usepackage{graphicx}
\usepackage{epsfig}
\usepackage{natbib}
\usepackage{multirow}
\usepackage{amsmath}
\newcommand{\eg}{{\rm e.g.}}

\newcommand{\g}{{\rm\thinspace g}}

\newcommand{\s}{{\rm\thinspace s}}

\newcommand{\fesc}{$f_{\rm esc}$}
\newcommand{\fesclya}{$f_{{\rm esc,Ly}\alpha}$}
\newcommand{\fabs}{$f_{\rm abs}$}
\newcommand{\logten}{$\log_{\rm 10}$}

\newcommand{\lya}{Ly$\alpha$}

\newcommand{\G}{{\rm G}}
\newcommand{\K}{{\rm K}}
\newcommand{\hi}{H\thinspace{\sc i}}

\newcommand{\oiii}{[O\thinspace{\sc iii}]}

\newcommand{\oii}{[O\thinspace{\sc ii}]}
\newcommand{\oi}{[O\thinspace{\sc i}]}
\newcommand{\oipermit}{O\thinspace{\sc i}}

\newcommand{\neiii}{[Ne\thinspace{\sc iii}]}

\newcommand{\sii}{[S\thinspace{\sc ii}]}

\newcommand{\sitwo}{Si\thinspace{\sc ii}}
\newcommand{\sithree}{Si\thinspace{\sc iii}}
\newcommand{\cii}{C\thinspace{\sc ii}}

\newcommand{\nv}{N\thinspace{\sc v}}
\newcommand{\sigsfr}{$\Sigma_{\rm SFR}$}

\newcommand{\fratio}{$F_{\lambda {\rm LyC}}/F_{\lambda {\rm 1100}}$}

\newcommand{\Msol}{\hbox{\thinspace M$_{\sun}$}}

\shorttitle{Multivariate Prediction of Escape Fraction}

\begin{document}

\title{Multivariate Predictors of LyC Escape I: A Survival Analysis of the Low-redshift Lyman Continuum Survey\footnote{Based on observations made with the NASA/ESA Hubble Space Telescope, obtained at the Space Telescope Science Institute, which is operated by the Association of Universities for Research in Astronomy, Inc., under NASA contract NAS 5-26555. These observations are associated with programs GO-15626, GO-13744, GO-14635, GO-15341, and GO-15639.}}

\author[0000-0002-6790-5125]{Anne E. Jaskot}
\affiliation{Department of Astronomy, Williams College, Williamstown, MA 01267, USA}
\author{Anneliese C. Silveyra}
\affiliation{Department of Astronomy, Williams College, Williamstown, MA 01267, USA}
\affiliation{Department of Physics, University of Nevada, Reno, NV 89557, USA}
\author[0000-0001-6281-7727]{Anna Plantinga}
\affiliation{Department of Mathematics \&\ Statistics, Williams College, Williamstown, MA 01267, USA}
\author[0000-0002-0159-2613]{Sophia R. Flury}
\affiliation{Department of Astronomy, University of Massachusetts Amherst, Amherst, MA 01002, USA}
\author[0000-0001-8587-218X]{Matthew Hayes}
\affiliation{Department of Astronomy, Oskar Klein Centre, Stockholm University, AlbaNova, SE-10691 Stockholm, Sweden}
\author[0000-0002-0302-2577]{John Chisholm}
\affiliation{Department of Astronomy, University of Texas at Austin, Austin, TX 78712, USA}
\author[0000-0001-6670-6370]{Timothy Heckman}
\affiliation{Department of Physics and Astronomy, Johns Hopkins University, Baltimore, MD 21218, USA}
\author[0000-0001-8940-6768]{Laura Pentericci}
\affiliation{INAF - Osservatorio Astronomico di Roma, via Frascati 33, 00078, Monteporzio Catone, Italy}
\author[0000-0001-7144-7182]{Daniel Schaerer}
\affiliation{Observatoire de Gen{\`e}ve, Universit{\'e} de Gen{\`e}ve, Chemin Pegasi 51, 1290 Versoix, Switzerland}
\author[0000-0002-6849-5375]{Maxime Trebitsch}
\affiliation{Astronomy, Kapteyn Astronomical Institute, Landleven 12, 9747 AD Groningen, The Netherlands}
\author[0000-0002-2201-1865]{Anne Verhamme}
\affiliation{Observatoire de Gen{\`e}ve, Universit{\'e} de Gen{\`e}ve, Chemin Pegasi 51, 1290 Versoix, Switzerland}
\affiliation{Univ. Lyon, Univ. Lyon 1, ENS de Lyon, CNRS, Centre de Recherche Astrophysique de Lyon UMR5574, 69230 Saint-Genis-Laval, France}
\author[0000-0003-4166-2855]{Cody Carr}
\affiliation{Center for Cosmology and Computational Astrophysics, Institute for Advanced Study in Physics, Zhejiang University, Hangzhou 310058, China}
\affiliation{Institute of Astronomy, School of Physics, Zhejiang University, Hangzhou 310058, China}
\author[0000-0001-7113-2738]{Henry C. Ferguson}
\affiliation{Space Telescope Science Institute, 3700 San Martin Dr., Baltimore, MD 21218, USA}
\author[0000-0001-7673-2257]{Zhiyuan Ji}
\affiliation{Steward Observatory, University of Arizona, Tucson, AZ 85721, USA}
\author[0000-0002-7831-8751]{Mauro Giavalisco}
\affiliation{Department of Astronomy, University of Massachusetts Amherst, Amherst, MA 01002, USA}
\author[0000-0002-6586-4446]{Alaina Henry}
\affiliation{Space Telescope Science Institute, 3700 San Martin Dr., Baltimore, MD 21218, USA}
\author[0000-0001-8442-1846]{Rui Marques-Chaves}
\affiliation{Observatoire de Gen{\`e}ve, Universit{\'e} de Gen{\`e}ve, Chemin Pegasi 51, 1290 Versoix, Switzerland}
\author[0000-0002-3005-1349]{G{\"o}ran {\"O}stlin}
\affiliation{Department of Astronomy, Oskar Klein Centre, Stockholm University, AlbaNova, SE-10691 Stockholm, Sweden}
\author[0000-0001-8419-3062]{Alberto Saldana-Lopez}
\affiliation{Department of Astronomy, Oskar Klein Centre, Stockholm University, AlbaNova, SE-10691 Stockholm, Sweden}
\author[0000-0002-9136-8876]{Claudia Scarlata}
\affiliation{Minnesota Institute for Astrophysics, School of Physics and Astronomy, University of Minnesota, 316 Church St. SE, Minneapolis, MN 55455, USA}
\author[0000-0003-0960-3580]{G{\'a}bor Worseck} 
\affiliation{VDI/VDE Innovation+Technik, Berlin, Germany}
\author[0000-0002-9217-7051]{Xinfeng Xu}
\affiliation{Center for Interdisciplinary Exploration and Research in Astrophysics, Northwestern University, Evanston, IL 60201, USA}

\begin{abstract}
To understand how galaxies reionized the universe, we must determine how the escape fraction of Lyman Continuum (LyC) photons (\fesc) depends on galaxy properties. Using the $z\sim0.3$ Low-redshift Lyman Continuum Survey (LzLCS), we develop and analyze new multivariate predictors of \fesc. These predictions use the Cox proportional hazards model, a survival analysis technique that incorporates both detections and upper limits. Our best model predicts the LzLCS \fesc\ detections with a root-mean-square (RMS) scatter of 0.31 dex, better than single-variable correlations. According to ranking techniques, the most important predictors of \fesc\ are the equivalent width (EW) of Lyman-series absorption lines and the UV dust attenuation, which track line-of-sight absorption due to \hi\ and dust. The \hi\ absorption EW is uniquely crucial for predicting \fesc\ for the strongest LyC emitters, which show properties similar to weaker LyC emitters and whose high \fesc\ may therefore result from favorable orientation. In the absence of \hi\ information, star formation rate surface density (\sigsfr) and \oiii/\oii\ ratio are the most predictive variables and highlight the connection between feedback and \fesc. We generate a model suitable for $z>6$, which uses only the UV slope, \sigsfr, and \oiii/\oii. We find that \sigsfr\ is more important in predicting \fesc\ at higher stellar masses, whereas \oiii/\oii\ plays a greater role at lower masses. We also analyze predictions for other parameters, such as the ionizing-to-non ionizing flux ratio and \lya\ escape fraction. These multivariate models represent a promising tool for predicting \fesc\ at high redshift.

\end{abstract}
%\keywords{Galaxies: evolution --- Galaxies: starburst --- intergalactic medium --- ISM: general --- Radiative transfer --- Stars: massive}

%--------------------------------------------
\section{Introduction}
\label{sec:intro}

Star-forming galaxies likely caused one of the most significant transformations in cosmic history: the reionization of  hydrogen in the intergalactic medium (IGM) at $z \gtrsim 6$. Current constraints on the galaxy and quasar luminosity functions at $z>6$ and on galaxies' ionizing photon production efficiencies favor stars as the dominant source of ionizing, Lyman continuum (LyC) photons \citep[\eg,][]{bouwens15, bouwens16, finkelstein15, finkelstein19, robertson15, ricci17, shen20, fauchergiguere20, debarros19, endsley21}. Exactly which star-forming galaxies contribute these LyC photons remains unclear, however. Studies disagree as to whether \mbox{low-,} \mbox{intermediate-,} or high-luminosity galaxies dominate reionization \citep[\eg,][]{razoumov10, wise14, paardekooper15, finkelstein19, cain21, begley22, rosdahl22, saldana23, wyithe13, naidu20, izotov21, ma20, matthee22} or whether other properties such as concentrated star formation, nebular ionization, or starburst age demarcate the LyC-emitting galaxy population \citep[\eg,][]{heckman01, clarke02, alexandroff15, sharma16, marchi18, jaskot13, nakajima14, izotov18b, zastrow13, ma15, trebitsch17, naidu22}. 

Observations with the {\it James Webb Space Telescope (JWST)} are revealing that proposed LyC-emitting galaxy populations exist at high redshift \citep[\eg,][]{schaerer22b, williams23, endsley23, fujimoto23, tang23, mascia24, atek24}, although it remains to be seen whether they are sufficiently numerous at $z>6$. However, identifying the galaxies responsible for reionization also requires knowing \fesc, the fraction of LyC photons that escape into the IGM. The value of \fesc\ for $z>6$ galaxies is not known, and its physical connection with galaxy properties may be complex. LyC escape can depend on a galaxy's interstellar gas geometry, dust content, stellar feedback, and gravitational potential and may vary with time \citep[\eg,][]{heckman01, heckman11, wise14, ma15, sharma16, trebitsch17, chisholm18, barrow20, gazagnes20, mauerhofer21, saldana22}. LyC escape is an inherently multi-parameter problem.

Because of attenuation in the IGM, detecting the LyC flux from galaxies becomes unlikely above $z\sim4$ \citep[\eg,][]{inoue14}. As a result, the astronomy community has relied on lower-redshift samples in order to investigate \fesc\ and its dependence on galaxy properties. Studies of $z\sim2-4$ galaxies find that \fesc\ may be enhanced in galaxies with lower UV luminosities, lower dust attenuation, higher \lya\ equivalent widths (EWs), strong \oiii~$\lambda$5007 emission, and/or compact sizes \citep[\eg,][]{marchi18, steidel18, bassett19, fletcher19, nakajima20, begley22, saxena22}. More local samples, at $z\sim0.3$, likewise identify strong \lya\ emission, elevated \oiii/\oii\ ratios, high star formation rate surface densities (\sigsfr), and low dust attenuation as characteristics of LyC emitters (LCEs) \citep[\eg,][]{borthakur14, izotov16b, izotov18b, verhamme17, chisholm18}. Nevertheless, \fesc\ can have a range of values, even for galaxies with similar properties \citep[\eg,][]{izotov18b, vanzella16, rutkowski17, fletcher19, bian20, marqueschaves21}. 

The recently completed Low-redshift Lyman Continuum Survey (LzLCS; \citealt{flury22a}) offers a new opportunity to investigate all these physical properties simultaneously in a large statistical sample of LCEs and non-emitters. Through the LzLCS and archival programs \citep{izotov16a, izotov16b, izotov18a, izotov18b, izotov21, wang19}, 89 galaxies at $z\sim0.3$ have LyC observations from the {\it Hubble Space Telescope (HST)} Cosmic Origins Spectrograph (COS), and 50 of these galaxies are detected in the LyC. This combined sample, hereafter the LzLCS+, covers a wide range of luminosities, metallicities, H$\beta$ and Ly$\alpha$ EWs, \oiii/\oii, and \sigsfr\ \citep{flury22a}, which enables it to more clearly reveal the trends and scatter between \fesc\ and galaxy properties. By quantifying the relationship between \fesc\ and observables, the LzLCS+ can generate predictions for \fesc\ in $z>6$ galaxies, where direct LyC detections are inaccessible. 

The results from the LzLCS+ confirm that \fesc\ correlates with a variety of galaxy properties, from line-of-sight measurements such as \hi\ covering fraction, dust attenuation, and \lya\ escape fraction to global properties such as \oiii/\oii\ and \sigsfr\ \citep[\eg,][]{saldana22, flury22b, chisholm22, xu23}. The latter properties hint at the possible role of mechanical or radiative feedback in creating low optical depth sightlines that allow ionizing photons to escape. Nevertheless, all correlations between \fesc\ and observable properties show significant scatter \citep{wang21, saldana22, flury22b, chisholm22, xu23}. 

A combination of properties might more accurately predict \fesc\ than a single variable alone. For instance, several studies have generated multivariate predictions of \lya\ properties for low- and high-redshift galaxies \citep[\eg,][]{yang17, trainor19, runnholm20}. \citet{runnholm20} present such an analysis for \lya\ luminosity using galaxies in the Lyman Alpha Reference Sample (LARS; \citealt{hayes13,ostlin14}). By applying multivariate regression using physical and observable galaxy properties, they predict the \lya\ luminosities of the LARS galaxies with a root-mean-square residual scatter of 0.2-0.3 dex. 

Similarly, \citet{maji22} and \citet{choustikov24} perform multivariate linear fits for \fesc\ using simulated galaxies from the SPHINX cosmological simulations \citep{rosdahl18}. \citet{maji22} find that a combination of four variables (escaping \lya\ luminosity, gas mass, gas metallicity, and recent star formation rate) can explain 85\% of the variance in escaping LyC luminosity, while three variables (escaping \lya\ luminosity, recent star formation rate, and gas mass) explain 66\% of the variance in \fesc. However, some of the relevant variables (e.g., \lya\ luminosity and gas mass) will be difficult or impossible to obtain for many galaxies at $z>6$. In contrast, \citet{choustikov24} focus on multivariate \fesc\ predictions using observable properties, such as \oiii/\oii, UV magnitude, and UV slope. While promising, the \citet{maji22} and \citet{choustikov24} simulation results require further testing and confirmation using observational samples. 

Other recent studies have taken a more empirical approach by using the $z\sim0.3$ LyC observations to derive relationships between \fesc\ and observable properties. \citet{lin24} fit for the probability of a galaxy having detectable vs.\ undetected \fesc\ as a function of UV magnitude, UV slope, and \oiii/\oii\ and use these results to assess the likelihood of LyC escape from galaxies at $z>6$. \citet{mascia23} develop a model that predicts \fesc\ based on some of the strongest correlating variables in the LzLCS: $\beta_{\rm 1550}$, UV half-light radius, and \oiii/\oii.

In this paper, we investigate a variety of empirical multivariate predictions for \fesc\ using the LzLCS+, the largest observational sample of LyC measurements at low redshift, which enables both reliable individual \fesc\ measurements and comprehensive rest-frame UV and optical ancillary data. Whereas the \lya\ data used by \citet{runnholm20} includes only \lya-emitters and the simulated LyC data used by \citet{maji22} and \citet{choustikov24} include measurements down to \fesc$=0$, the LzLCS+ sample includes 39 galaxies with non-detected LyC corresponding to \fesc\ upper limits of 0.03-5.9\%. Standard multivariate regression models do not account for upper limits, but a valid statistical analysis of the full LzLCS+ dataset should include information from both detections and non-detections \citep[\eg,][]{isobe86}. Hence, we turn to the statistical approach of survival analysis, which can handle censored data like that of the LzLCS. We apply the Cox proportional hazards model \citep{cox72}, a semi-parametric survival analysis method, to the full LzLCS+ dataset to derive multivariate predictions of \fesc. By testing different sets of variables and assessing their predictive ability, we explore which physical properties combine to set a galaxy's \fesc. We adopt a cosmology of  $H_0 = 70$ km s$^{-1}$ Mpc$^{-1}$, $\Omega_m = 0.3$, and $\Omega_\Lambda = 0.7$.

\section{Methods}
\label{sec:methods}
\subsection{Sample: The Low-redshift Lyman Continuum Survey}
\label{sec:lzlcs}

To derive multivariate predictors of \fesc, we use the Low-redshift Lyman Continuum Survey (LzLCS; \citealt{flury22a}), the largest sample of LCEs at low redshift. \citet{flury22a} describe the survey sample, observations, data processing, and measurements in full, but we review some of the key details here. The LzLCS is a 134-orbit Cycle 26 {\it HST} program (GO-15626; PI Jaskot) that obtained far-ultraviolet spectra, including the rest-frame LyC, with the COS G140L grating for 66 star-forming galaxies at $z\sim0.3$. The targeted galaxies each fulfill one or more hypothesized selection criteria for LCEs: high nebular ionization (O32 = \oiii~$\lambda$5007/\oii~$\lambda$3727 $\geq 3$), blue UV slopes (power law index $\beta < -2$), and/or concentrated star formation (\sigsfr\ $> 0.1$ \Msol\ yr$^{-1}$ kpc$^{-2}$). We combine the LzLCS with archival COS observations from \citet{izotov16a, izotov16b, izotov18a, izotov18b, izotov21} and \citet{wang19}. We exclude one galaxy, J1333+6246 \citep{izotov16b}, whose \oiii$\lambda\lambda$5007,4959, H$\alpha$, and H$\beta$ line fluxes may be inaccurate; these lines appear truncated in the SDSS spectrum and the galaxy's H$\alpha$/H$\beta$ and H$\beta$/H$\gamma$ line ratios are unphysical. Excluding this problematic galaxy, the combined LzLCS and archival galaxies (hereafter the LzLCS+ sample) consists of 88 low-redshift galaxies with LyC observations. The sample spans a range of $10^8-10^{10}$ \Msol\ in stellar mass, $12+\log_{10}(\rm{O/H})=7.5-8.5$, and -21.5 to -18.3 in observed (not corrected for internal reddening) 1500\AA\ absolute magnitude. 

We process the raw COS spectra using the \textsc{calcos} pipeline (v3.3.9) and the \textsc{FaintCOS} software routines \citep{worseck16, makan21} to reduce and calibrate the spectra and account for the backgrounds due to dark current and scattered geocoronal \lya. We correct all spectra for Milky Way attenuation using the \citet{green18} dust maps and \citet{fitzpatrick99} attenuation law. The LyC flux measurements represent the flux in a 20 \AA-wide region near rest-frame 900 \AA; we exclude wavelengths above an observed wavelength of 1180~\AA\ because of telluric contamination. For consistency, we re-process and re-measure the archival observations using this same methodology. Following the criteria in \citet{flury22a}, we define LyC detections as observations that have a probability $< 0.02275$ of originating from background counts. By this definition, 49 of the 88 total galaxies have detected LyC. As in \citet{flury22a}, we adopt the 84th percentile of the background count distribution as the upper limit for non-detections.

Using these LyC fluxes, we then calculate the absolute \fesc\ from different estimates of the intrinsic LyC, as described in \citet{flury22a}; for this paper, we adopt the \fesc\ values from \citet{saldana22}, derived from \textsc{Starburst99} \citep{leitherer11a, leitherer14} spectral energy distribution (SED) fits to the UV continuum, following the methods outlined in \citet{chisholm19}. As discussed in \citet{flury22a}, the UV SED-fitting method is more reliable than H$\beta$-derived \fesc\ estimates for the diverse LzLCS sample, because the presence of underlying, older ($> 10$ Myr old) stellar populations that emit optical emission can bias the H$\beta$ \fesc\ results. The H$\beta$ estimates of intrinsic LyC also do not account for LyC photons absorbed by dust and assume isotropic escape. With our preferred UV-based \fesc\ estimate, the LzLCS+ LyC detections cover \fesc$=0.4-88.9\%$ and the non-detections have upper limits ranging from 0.03-5.9\%. We also consider an alternative, purely empirical measure of LyC escape, the $F_{\lambda {\rm LyC}}/F_{\lambda {\rm 1100}}$ flux ratio in Section \S\ref{sec:results:altlyc}. 

In addition to LyC measurements, the LzLCS+ dataset contains a wealth of ancillary measurements, described in full in  \citet{flury22a} and \citet{saldana22}. Each galaxy in the survey has rest-frame optical photometry and spectroscopy through the Sloan Digital Sky Survey (SDSS; \citealt{blanton17}) and UV photometry via the {\it Galaxy Evolution Explorer} ({\it GALEX}; \citealt{martin03}). We derive stellar masses ($M_*$) using \textsc{Prospector} \citep{leja17, johnson19} fits to the SDSS and {\it GALEX} photometry (\citealt{flury22a}, Ji et al.\ in prep.). We measure nebular line fluxes by fitting the SDSS spectra with multiple Gaussian profiles and then iteratively derive the electron temperature ($T_e$), electron density, stellar absorption, and nebular $E(B-V)$ from the \oiii~$\lambda\lambda$5007,4959,4363, \sii~$\lambda\lambda$6716,6731, and Balmer lines. Where the \sii\ doublet or \oiii~$\lambda$4363 are undetected, we adopt $n_e=100$ cm$^{-3}$ and the estimated \oiii~$\lambda$4363 flux from the \citet{pilyugin06} ``ff-relation", respectively. We then derive oxygen abundances via the direct method as implemented in \textsc{pyneb} \citep{luridiana15}. To estimate the star formation rate (SFR), we use the dust-corrected H$\beta$ luminosities, Case B H$\alpha$/H$\beta$ ratio \citep{storey95}, and \citet{kennicutt12} SFR calibration.

The COS data provide estimates of additional physical parameters. We use the COS near-UV acquisition images to calculate the UV half-light radius. The sources all appear to be very compact within the central 1\arcsec\ diameter of the COS aperture, so that vignetting will not strongly affect our radius measurements. From the measured radius, we also derive \sigsfr\ as
\begin{equation}
\Sigma_{\rm SFR} = \frac{\rm SFR}{2 \pi r^2_{\rm 50,NUV}}.
\end{equation}
From the \textsc{Starburst99} SED fits to the UV spectra, we fit for the dust excess E(B-V). We denote the E(B-V) derived from the UV spectral fits as E(B-V)$_{\rm UV}$ to distinguish it from E(B-V)$_{\rm neb}$ derived from the Balmer line ratios. We also extrapolate the UV SED fits to infer the ``observed" (non-extinction corrected) absolute magnitude at 1500 \AA\ ($M_{1500}$) and the power law index slope at 1550 \AA\ ($\beta_{1550}$; \citealt{saldana22}). 

The G140L spectra also cover Ly$\alpha$. For our \lya\ measurements, we use spectra extracted using a slightly wider spatial aperture (30 pixel vs.\ 25 pixel), because the scattered \lya\ emission may be more spatially extended than the continuum light \citep{flury22a}. After masking the \sithree~$\lambda$1206 and \nv~$\lambda$1240 regions, we linearly fit the continuum within 100 \AA\ of \lya\ using an iterative sigma clipping algorithm and integrate the \lya\ fluxes relative to this continuum level. This linear continuum fit typically agrees within $<5$\%\ with the continuum estimated from SED fits to the 25-pixel aperture extracted spectra \citep{saldana22}. We do not correct for the small contribution of stellar \lya\ absorption, but we conservatively adopt a 25\%\ uncertainty on the \lya\ continuum estimate \citep{flury22a}. From these fluxes, we derive Ly$\alpha$ luminosities, equivalent widths (EWs), and escape fractions (\fesclya); we use the dust-corrected H$\beta$ flux and the galaxies' measured electron temperatures and densities to infer the intrinsic Ly$\alpha$ flux. The reported Ly$\alpha$ measurements represent the sum of both absorption and emission along the line of sight. Even galaxies with net Ly$\alpha$ emission may have underlying Ly$\alpha$ absorption troughs, which are commonly detected in samples at $z<0.1$ \citep[\eg][]{mckinney19, hu23}. At the higher redshifts of our sample, z$\sim0.3$, the COS aperture covers a larger physical area, and these troughs may experience more infilling due to scattered \lya\ emission. Because the radii of our \lya\ spectral apertures are $\sim$2.6 times larger than the UV continuum half-light radius \citep{flury22a}, we expect our \lya\ measurements to capture most of the scattered emission \citep[\eg][]{hayes13}; future \lya\ imaging of LzLCS+ galaxies will test this assumption (HST GO-17069, P.I. Hayes). Nine galaxies in our sample have significant \lya\ absorption that overlaps with the \sithree~$\lambda$1206 absorption feature. To account for the uncertain strength of \sithree~$\lambda$1206 contamination in these galaxies, we re-measure the \lya\ fluxes and EWs, excluding wavelengths within 500 km s$^{-1}$ of \sithree~$\lambda$1206, and we increase our \lya\ uncertainties to account for this possible flux difference. We note that this effect is minor, changing EWs by $<3$\AA\ and \fesclya\ by $\leq0.017$, with a median change of only 0.001 in \fesclya\ for the nine affected galaxies.

Lastly, we use the G140L spectra to measure the EWs and residual fluxes near line center ($R_{l}$) for a variety of low-ionization absorption lines, as described in \citet{saldana22}. The UV absorption line measurements use the same (25 pixel aperture) spectral extraction as the LyC measurements, and the SED fits are used to estimate the continuum level. For optically thick gas, well-resolved absorption lines, and assuming a uniform dust screen the residual flux is related to the gas covering fraction ($C_f$) by
\begin{equation}
R(\lambda_i) = 1-C_f(\lambda_i)
\end{equation}
(see e.g., \citealt{gazagnes18} for a discussion of geometry effects on $R$). This equation assumes that all lines are saturated; while this assumption appears to hold for most of the \hi\ lines, the observational uncertainties are too high to draw conclusions about saturation for the metal lines (see \citealt{saldana22} for more information). In this work, we consider some of the strongest low-ionization metal absorption lines (\sitwo~$\lambda$1260, \cii~$\lambda$1334, \hi~Ly$\beta+$\oipermit). To improve the $S/N$ of the measurements, we also consider the inverse-variance weighted average of the EWs and $R_l$ values for a given galaxy, calculated for the available \hi\ Lyman-series lines (Ly$\beta$-Ly6) or for the available low-ionization metal (LIS) lines (\sitwo\ $\lambda\lambda1190,1193$, \sitwo\ $\lambda$1260, \oi\ $\lambda$1302, \cii\ $\lambda$1334; see \citealt{saldana22} for details). We refer to these average absorption line measurements as EW(\hi,abs), $R_l$(\hi,abs), EW(LIS), and $R_l$(LIS). Because the measured residual intensity depends on resolution, we caution that the measured $R_l$ will not perfectly correspond to the true $R_l$ of the system; for the LzLCS+, this systematic error is of order 10-20\%\ and is within the reported uncertainty \citep{saldana22}. While trends between $R_l$ and \fesc\ will still be apparent \citep[\eg,][]{saldana22}, future investigations should use care when applying models derived from the LzLCS+ $R_l$ data to spectroscopic data with a different resolution.

\subsection{Survival Analysis}
\label{sec:survive}
As described above, one of the strengths of the LzLCS+ dataset is its large and diverse sample, spanning a wide range of measured \fesc. The full sample of 88 galaxies includes 39 non-detections with \fesc\ upper limits of only a few percent. These non-detections can convey essential information about which properties do and do not distinguish LCEs from non-LCEs, and omitting non-detections can bias fitted relations \citep[c.f., ][]{isobe86}. Standard multivariate linear regression does not account for censored data, that is, data with upper limits. Therefore, we instead apply survival analysis techniques to the LzLCS+ sample, as these techniques properly treat censored data.

As implied in the name, survival analysis originated in the field of medicine (see reviews by \citealt{clark03} and \citealt{bradburn03}, and see \citealt{feigelson85} and \citealt{isobe86} for applications to astronomy). In a medical context, the censored data often consist of known and unknown lifetimes for individuals participating in a medical study. People who are alive at the end of the study have an unknown lifetime; all we know is that they will live longer than their current age. Survival analysis techniques ultimately seek to describe the probability of a particular lifetime for a population or to compare how the survival probability changes between different populations, based on data that include some known lifetimes and some limits. 

To set up an analogous scenario for predicting the probability of a particular \fesc, we do the following. Instead of a ``death" or known lifetime, we have a detection. Most survival analyses involve lower limits (``right-censored" data). For simplicity in following common techniques, we therefore adopt the absorbed fraction of LyC (\fabs\ $=1-$\fesc) and its corresponding lower limits rather than \fesc\ and its upper limits. With this setup, increasing \fabs\ (decreasing \fesc) is equivalent to increasing the ``age" of the study participant. A non-detection, i.e., a lower limit to \fabs\ (and upper limit on \fesc), is analogous to a lifetime greater than some threshold. Instead of starting at time $= 0$ and proceeding until we record a death or the study ends, we can imagine trying to observe a galaxy with \fesc\ $=1$ (\fabs\ $=0$) and proceeding with deeper and deeper observations until we have a detection or cease our efforts. Just as the threshold between recording a measured lifetime vs.\ a limit will depend on factors such as the length of the study and age of the participant, the threshold between detected \fesc\ and an upper limit will depend on aspects of our study such as the observation depth and galaxy brightness.

\subsubsection{The Cox Proportional Hazards Model}
\label{sec:cox}
The Cox proportional hazards regression model \citep{cox72} describes the probability of an event in an infinitesimally small window of time, given a combination of independent variables and assuming the event did not already occur. In the Cox model, this event probability (or ``hazard function") at time $t$ for a set of variables $x$ is modeled as
\begin{equation}
h(t|x) = h_0(t)\exp[\sum_{i=1}^{n} b_i(x_i - {\bar x_i})],
\label{eqn:coxhazard}
\end{equation}
where $b_i$ are fitted coefficients for each variable $x_i$ and where $h_0(t)$ represents the baseline hazard, the probability of an event given average values of all variables ($\bar x_i$). An increase in variable $x_i$ relative to that variable's average value in the sample results in an exponential increase or decrease in the detection probability, depending on the sign of the coefficient $b_i$. In our case, instead of modeling the probability of an event at time $t$, the hazard function represents the probability of a LyC detection in an infinitesimally small window of \fesc\ for a set of independent variables, assuming the galaxy was not already detected at a higher value of \fesc.

The Cox model is semi-parametric. One of its strengths is that it does not assume a particular functional form for the baseline hazard function but rather estimates it non-parametrically. In other words, the event times or detection values do not have to obey a known statistical distribution \citep{bradburn03}. Because the Cox model predicts the detection {\it probability} at each possible \fesc\ value, rather than predicting the value of \fesc\ itself, the model results are identical for \fesc\ and for \logten(\fesc). In other words, the non-parametric estimate of $h_0$(\fabs) for an input array of \fesc\ values would be identical to $h_0(\log_{10}$(\fabs)) for the equivalent input array of \logten(\fesc) values.

Yet, the parametric nature of the hazard function equation (Equation~\ref{eqn:coxhazard}) with its reported fitted coefficients makes it straightforward to analyze the relative importance of variables and generate predictions for future datasets. On the other hand, this fixed functional form implicitly assumes that the effect of the independent variables is multiplicative and does not depend on $t$, in our case, \fabs. This functional form would not be appropriate, for example, if one variable tended to introduce a step function in \fesc, while the others had more of a smooth correlation. However, the adopted functional form has some flexibility in that independent variables are allowed to take any functional form (e.g., $x_i$ can be replaced with $\log_{\rm 10}(x_i)$, $x_i^2$, etc). Another limitation of the Cox model is that, while it handles censored dependent variable values, it cannot handle censored data in the independent variables. Consequently, we have to limit our analysis to input variables, $x_i$, that are available for all (or nearly all) of the sample galaxies, as we explain below. In summary, the Cox proportional hazards model serves as a multivariate regression model for describing and predicting dependent variables with upper limits. Many other common statistical methods either do not treat upper limits (e.g., multivariate linear regression, principal component analysis), are univariate analyses (e.g., Kaplan-Meier analysis), or are fully parametric, less-flexible models (e.g., the Weibull model).

In fitting the Cox model to the LzLCS+, we experiment with different combinations of independent variables from the parent list given in Table~\ref{table:all_variables}. Where possible, we take the base 10 logarithm of the relevant variable. We cannot take the logarithm for variables that contain both positive and negative values, such as the UV LIS and \lya\ lines; these lines range from absorption to emission within the sample. Putting variables on a logarithmic scale serves two purposes. First, it puts variables on a consistent scale of similar order of magnitude. In addition, the coefficients in Equation~\ref{eqn:coxhazard} have a simple interpretation; increasing a variable by an order of magnitude leads to a factor of $e^{b_i}$ increase in the probability $h(f_{\rm abs}|x)$. Exploring alternative functional forms for the input variables is outside the scope of this paper, but we test the effects of switching between logarithmic and linear scalings in the following subsection (\S\ref{sec:interpret}). While some statistical methods require putting all variables on an identical scale, such as scaling from 0 to 1, this further rescaling is not necessary in the Cox model. Equation~\ref{eqn:coxhazard} effectively rescales variables; the sample mean is subtracted from each input variable measurement, and the $b_i$ parameter translates a linear change in variable $x_i$ to a change in the probability of measuring a given \fesc. Any normalization would change the derived values of $b_i$ but not the statistical significance of input variables nor their ultimate effect on \fesc.

In selecting a subset of variables to use in our fits, we include multiple variables from each category in Table~\ref{table:all_variables}, but we avoid any variables that are closely related to each other and which may therefore be highly collinear. Collinearity can result in multiple best-fit solutions and can cause the Cox model to fail to converge. Consequently, we typically consider either \sigsfr\ alone or SFR and $r_{\rm 50,NUV}$, but not all three variables at once. Similarly, we choose only one variable that traces nebular ionization, one variable to trace UV color, and limited subsets of the UV absorption line variables. We discuss the effect of selecting different variables in Section \S\ref{sec:modified_fiducial}. As mentioned above, the Cox proportional hazard model cannot account for missing or censored independent variable data. Hence, we exclude a galaxy from our fits if it is missing any of the required variables. Numbers in brackets in Table~\ref{table:all_variables} indicate the number of galaxies missing a given measurement. Most of the models discussed in this paper exclude only one galaxy, J1046+5827, a non-leaker that lacks a reported \sigsfr\ and $r_{\rm 50, NUV}$. We apply the Cox model to \fesc\ as our primary dependent variable, but we also perform and discuss model fits to the \fratio\ ratio, (\S\ref{sec:results:altlyc}), LyC luminosity (\S\ref{sec:results:altlyc}), \fesclya, and \lya\ luminosity (\S\ref{sec:results:fesclya}). %

\begin{deluxetable*}{ll}
\tablecaption{Complete List of Independent Variables}
\label{table:all_variables}
\tablehead{
\colhead{Category} & \colhead{Variables}}
\startdata
Mass and Luminosity & \logten($M_*$); \logten(SFR)-H$\beta$; $M_{\rm 1500}$ \\
Nebular Properties & \logten(EW(H$\beta$)); E(B-V)$_{\rm neb}$; 12+\logten(O/H) \\
Nebular Ionization & \logten(\neiii~$\lambda$3869/\oii~$\lambda$3727); \logten(O32=\oiii~$\lambda$5007/\oii~$\lambda$3727); \logten(\oii~$\lambda$3727/H$\beta$) \\
Morphology & \logten(\sigsfr) [1 missing]; \logten($r_{\rm 50, NUV}$) [1 missing] \\
UV Color and attenuation & $\beta_{\rm 1200}$; $\beta_{\rm 1550}$; E(B-V)$_{\rm UV}$ \\
\lya & EW(\lya); $L$(\lya); \fesclya \\
UV Absorption Lines & EW(Ly$\beta$) [9 missing]; EW(\sitwo~$\lambda$1260) [2 missing]; EW(\cii~$\lambda$1334) [11 missing]; EW(\hi,abs); EW(LIS) \\
& $R_l$(Ly$\beta$) [3 missing]; $R_l$(\sitwo~$\lambda$1260); $R_l$(\cii~$\lambda$1334) [4 missing]; $R_l$(\hi,abs) [5 missing]; $R_l$(LIS) [2 missing] 
\enddata
\tablecomments{Numbers within brackets denote the number of galaxies within the LzLCS+ sample missing these measurements. Positive values of EW(H$\beta$) and EW(\lya) denote net emission; positive values of the UV absorption lines denote net absorption.}
\end{deluxetable*} %
We apply the Cox proportional hazards model to our data using the python package {\tt lifelines} \citep{davidson19}. The {\tt lifelines} CoxPHFitter routine returns the best-fit coefficients for our selected independent variable set, their p-values, and various goodness-of-fit statistics. As shown by Equation~\ref{eqn:coxhazard}, the Cox proportional hazards model gives the {\it probability} of observing a particular \fesc\ value given a set of physical or observable properties. Instead, our goal is to predict the expected {\it value} of \fesc\ given a set of properties. As described below, to find the expected \fesc, we find the median of the probability distribution, where the true \fesc\ has an equal probability of lying above or below this adopted value \citep[\eg,][]{bradburn03, davidson19};  \fesc\ is predicted to be above the median 50\% of the time and below it the other 50\% of the time. 

The median \fesc\ = 1 - \fabs\ represents the \fesc\ value where the survival function $S$($f_{\rm abs}$), the probability that there is no detection at $f_{\rm abs,detect} < f_{\rm abs}$, reaches 0.5. This probability is equivalent to the probability that $f_{\rm esc,detect}$ is detected at $f_{\rm esc,detect} < f_{\rm esc}$. The survival function is calculated as \citep[\eg,][]{cox72, bradburn03, davidson19, mclernon23}
\begin{equation}
S(f_{\rm abs}) = \exp[-{\rm HF}_0(f_{\rm abs})\cdot{\rm ph}(x)],
\label{eqn:survive}
\end{equation}
where HF$_0$ is the baseline cumulative hazard function:
\begin{equation}
{\rm HF}_0(f_{\rm abs}) = \int_0^{f_{\rm abs}} h_0(f) df
\end{equation}
and ph($x$) is the partial hazard function for a set of variables x:
\begin{equation}
{\rm ph}(x) = \exp[\sum_{i=1}^{n} b_i(x_i - {\bar x_i})].
\end{equation}
Occasionally, $S$(\fabs) for a set of parameters will never reach 0.5 and the predicted median \fesc\ is indeterminate. This situation corresponds to an arbitrarily small predicted \fesc\ $\sim0$. 

The best fit coefficients for the Cox proportional hazards model are found by maximizing the partial likelihood, which compares ph($x$) for each detection with the sum of ph($x$) for all galaxies with higher \fabs\ (lower \fesc), including both detections and non-detections. The resulting coefficients $b_i$ are those that best sort \fabs\ in order. Once the coefficients are determined, Breslow's estimator \citep{breslow72} determines the baseline cumulative hazard function, HF$_0$(\fabs), using the number of detections with values lower than each \fabs\ (i.e., detections with higher values of \fesc) plus the ph($x$) values for all detections and limits higher than \fabs\ (i.e., with lower \fesc). The model fits HF$_0$(\fabs) non-parametrically, reporting a value of HF$_0$ for each of the input \fabs\ values of the LzLCS+ sample.

One can also use the survival function to evaluate the values at which $S$ reaches 0.159 and 0.841, the bounds corresponding to the Normal-theory 1$\sigma$ uncertainty of the predicted \fesc. These probabilities account for the scatter in the correlations between \fesc\ and the independent variables. The scatter may arise from observational uncertainty in the measurements as well as inherent variation among the population. In addition to this method, we have also used a Monte Carlo (MC) method to explore how the \fesc\ predictions change if we vary each variable within its uncertainties. We randomly resample each independent variable measurement and each dependent variable detection using their uncertainties, re-run the Cox fit, and obtain new estimates of the median predicted \fesc\ for each galaxy. Using the distribution of predicted \fesc, we then calculate the 15.9 and 84.1 percentiles. For nearly all galaxies, the uncertainties estimated using the survival function are greater than or equal to the uncertainties determined from this MC method, which indicates that the scatter in the correlations is the dominant effect. We conclude that the survival function sufficiently represents the uncertainty in predicted \fesc\ in most cases. 

\subsubsection{Interpreting Cox Model \fesc\ Predictions}
\label{sec:interpret}

As shown by Equation~\ref{eqn:survive}, the Cox model does not represent an equation that shows how \fesc\ itself depends on particular variables. Rather, it describes how the probability of observing a given \fesc\ changes for different sets of galaxy properties. The baseline cumulative hazard function HF$_0$(\fabs) describes the expected probability distribution for galaxies that have the average properties of the LzLCS+ dataset. In this case, $x_i = {\bar x_i}$ for all variables, and the reference probability that the observed $f_{\rm esc,detect}$ is less than a particular value of $f_{\rm esc}$ is simply $S_{\rm ref}(f_{\rm esc}) = \exp[-{\rm HF}_0(f_{\rm esc})]$. If we increase one variable by an increment of 1, $x_i-{\bar x_i} = 1$, ${\rm ph}(x) = \exp(b_i)$, and the new probability $S(f_{\rm esc}) = S_{\rm ref}(f_{\rm esc})^{\exp(b_i)}$. Because the probability $S_{\rm ref}(f_{\rm esc}) < 1$, raising it to the power of $\exp(b_i)$ decreases the probability, which means that low values of $f_{\rm esc,detect} <$ \fesc\ are less likely. The probability distribution therefore shifts such that the probability corresponding to $S$=0.5 occurs at higher \fesc, resulting in a larger predicted median \fesc. Changing a second variable, $x_j$, by an increment of 1 changes the new probability again by a power of $\exp(b_j)$, such that $S(f_{\rm esc}) = [S_{\rm ref}(f_{\rm esc})^{\exp(b_i)}]^{\exp(b_j)}]$.

\begin{figure*}
\gridline{\fig{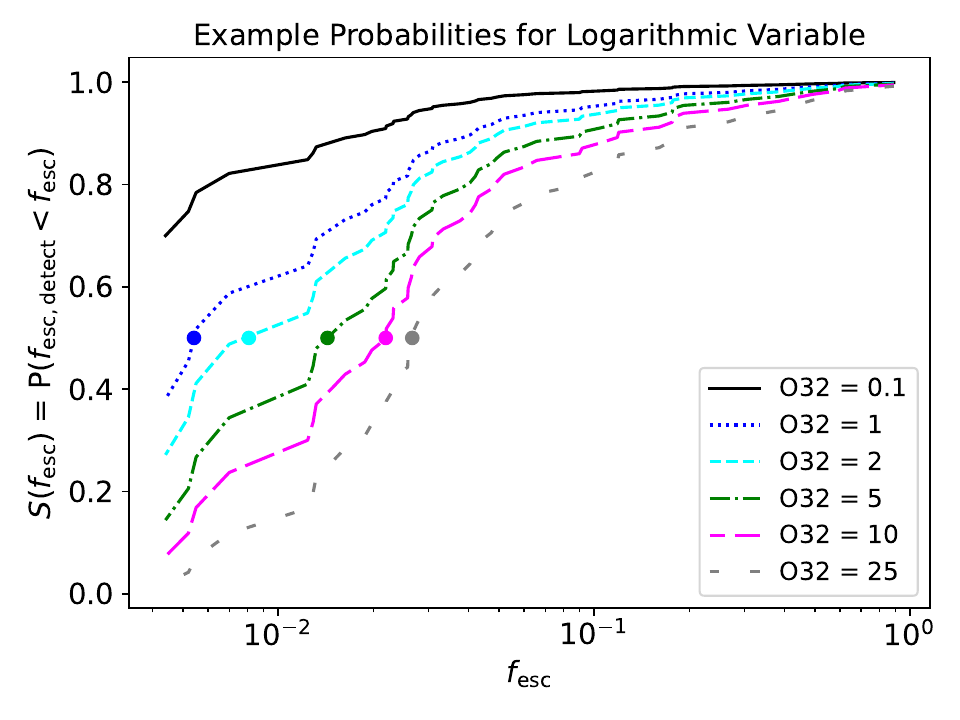}{0.5\textwidth}{(a)}
	\fig{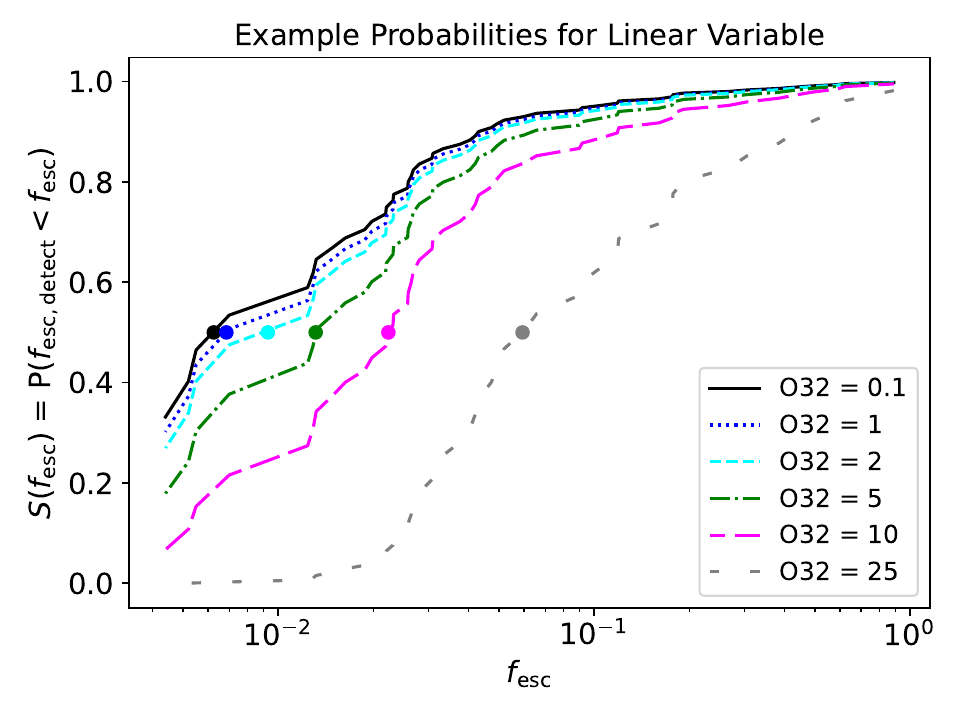}{0.5\textwidth}{(b)}
}
\caption{Examples of the survival function probabilities predicted from the Cox model. The left panel uses $\beta_{\rm 1550}$, \logten(\sigsfr), and \logten(O32) as input variables, and the right panel uses $\beta_{\rm 1550}$, \logten(\sigsfr), and O32 instead of \logten(O32). Lines show the probability that the detected $f_{\rm esc,detect}$ is less than a particular value of \fesc\ or equivalently, the probability that $f_{\rm abs,detect}$ is {\it not} less than a given $f_{\rm abs}$. The lines are plotted at \fesc\ values corresponding to the measured \fesc\ for the LzLCS+ dataset, from which the baseline cumulative hazard function is estimated non-parametrically.  Data points show the median predicted \fesc, where $S$(\fesc)$=0.5$. The different lines show predictions for different input values of O32. As O32 increases, the probability distribution shifts to higher values of \fesc, and the extent of the shift depends on whether the variable is linear or logarithmic. The black solid line in the left panel (\logten(O32)$=-1$) shows an example where $S$ never reaches 0.5, indicating an inferred median \fesc$\sim 0$. 
\label{fig:probplots}}
\end{figure*}

We illustrate an example of the survival function probabilities, the associated median \fesc, and the dependence of these quantities on input variables in Figure~\ref{fig:probplots}. In the left panel, we show a simple Cox model with three input variables: $\beta_{\rm 1550}$, \logten(\sigsfr), and \logten(O32). Dots represent the predicted median \fesc\ where $S$(\fesc)$ = 0.5$ for each value of O32. For \logten(O32)$ = 1$, $S$(\fesc)$ = 0.5$ at \fesc$ = 0.005$. Changing only the \logten(O32) value by 1, from ${\rm O32} = 1$, the blue dotted line, to ${\rm O32}=10$, the magenta long-dashed line, shifts $S$(\fesc$ = 0.005$) from 0.5 down to $0.5^{\exp(0.996)} = 0.153$, where 0.996 is the best-fit coefficient $b_i$ for the \logten(O32) variable. Thus, according to this model, high \fesc\ values are more common among galaxies with high O32. In this example, we changed only one parameter, but changing the other variables at same time could either further shift the probability distribution to higher \fesc\ or counteract the change caused by increasing O32. We note that, unlike an equation for \fesc\ as a function of these input variables, which could conceivably reach unphysical values of \fesc\ $> 1$ for extreme parameters, the probability distribution can shift toward higher values of \fesc, but it never extends beyond \fesc\ $ = 1$.  

Figure~\ref{fig:probplots} also shows how the predicted probability responds to linear vs.\ logarithmic variables. As outlined above, the probabilities shift by a power of $\exp(b_i)$ for a step size of 1 in $x_i$, which corresponds to an order of magnitude increase if $x_i$ represents the base 10 logarithm of a variable. In the left panel, the model uses \logten(O32) as the input variable, and in the right panel, we show the effect of changing the form of the input variable from \logten(O32) to linear O32. The differences are most apparent where a large linear spacing (e.g., changing O32 from 10 to 25) is not equivalent to a large logarithmic spacing and vice versa (e.g., O32 from 0.1 to 1). However, the change in the predicted median \fesc\ values is largely minor. In changing to a linear scale, \fesc\ changes from 0 to 0.006 for ${\rm O32} = 0.1$ and from 0.027 to 0.059 for ${\rm O32}=25$. For the other plotted values, the change in \fesc\ is $<0.002$. The best-fit coefficients $b_i$ and HF$_0$(\fesc) values also differ between the two models and are optimized to match the observed distribution of \fesc\ in the LzLCS+. Consequently, although the choice of variable form does affect predictions for galaxies at the extremes of the input parameter space, our main results, including the overall quality of our model predictions, are not highly sensitive to the functional form of the input variables. 

We do not thoroughly explore functional forms here, but we experiment with changing the ISM absorption measurements (EW(\hi,abs) and $R_l$(\hi,abs)) and \lya\ variables (EW(\lya), $L$(\lya), \fesclya) to logarithmic forms. For the logarithmic \lya\ variables, we exclude all galaxies with net \lya\ absorption from the dataset. We find the same variables are statistically significant in our models, although their exact coefficients and p-values necessarily change. The goodness-of-fit metrics (\S\ref{sec:metrics}) are also similar, typically changing by $\leq0.02$. The change in variable form also does not affect our main conclusions regarding the most important ranked variables (\S\ref{sec:selection}), although it can result in minor changes in ranked variable order. 

Observationally, for most variables, \logten(\fesc) does seem to change significantly with the logarithm of a variable \citep{flury22b}. These trends could indicate that \fesc\ has a power law dependence on many variables, although, as noted above, such relationships could only apply over the range where the predicted \fesc\ $\leq1$. One exception to this logarithmic dependence is the UV absorption line measurements, which do show a dependence of \logten(\fesc) on the linear form of the variables \citep{saldana22}. This functional form may indicate a more complex dependence of \fesc\ on the gas geometry. We find that our predictions using the linear EW(\hi,abs) better match the observed \fesc\ by every metric compared to models using \logten(EW(\hi,abs)), and we therefore choose to keep EW(\hi,abs) in the linear form. Models using the linear or logarithmic form of $R_l$(\hi,abs) perform comparably well, but with the linear form better matching the observed \fesc\ for the strongest LCEs. Given their comparable or improved performance and for consistency with the LIS line measurements, which can reach negative values, we therefore keep all absorption line variables in their linear form. We also choose to keep the \lya\ measurements in a linear form so as to not bias our dataset by excluding non-\lya\ emitters. We emphasize that our fit quality and main results are not sensitive to these choices regarding variable form. 

In summary, the Cox model predicts the probability of observing particular \fesc\ values for different galaxy populations and can identify which variables most affect that probability. Statistically significant variables signify that the \fesc\ probability is highly responsive to an incremental change in that variable: either a linear or logarithmic increase, depending on the variable form. While this work identifies these significant variables, future theoretical or observational programs could endeavor to derive the exact functional dependence of \fesc\ on these parameters.

\subsection{Model Assessment}
\label{sec:metrics}
We assess the goodness-of-fit for our models in a few different ways. Our primary method is the concordance index \citep{harrell82}, which applies to censored data. The concordance index considers all possible pairs of data points and how the observed rank order of \fesc\ compares to the order predicted by the model. The model classifies each pair as concordant if the data point with a higher observed \fesc\ also has a higher predicted \fesc\ and discordant if it does not. Pairs can also be tied, if their predicted \fesc\ values are identical. For some pairs with upper limits, their rank is ambiguous. The concordance index is
\begin{equation}
C = \frac{n_c+0.5n_t}{n_c+n_d+n_t},
\end{equation}
where $n_c$ is the number of concordant pairs, $n_t$ is the number of tied pairs, and $n_d$ is the number of discordant pairs. The concordance index ranges from 0 (perfect disagreement) to 0.5 (perfectly random) to 1.0 (perfect concordance; e.g., \citealt{davidson19}).\footnote{For the concordance index calculation in the {\tt lifelines} package, pairs are ranked by their partial hazards function ph($x$), rather than by median \fesc\ or \fabs. The ph($x$) will scale with \fesc, except for the cases where $S$(\fesc) is indeterminate and median \fesc$\sim$0. Two galaxies that both have \fesc$\sim$0 could still have slightly different ph($x$), corresponding to different probability distributions for \fesc$>0$.} We also calculate the $R^2$ statistic for the galaxies with \fesc\ detections as
\begin{equation}
R^2=1-\frac{\sum_i(y_i-f_i)^2}{\sum_i(y_i-{\bar y})^2},
\label{eqn:r2}
\end{equation}
where $y_i$ are the observed values of \logten(\fesc), ${\bar y}$ is their mean value, and $f_i$ are the model-predicted \logten(\fesc) values. Following \citet{maji22}, we also calculate the adjusted $R^2$:
\begin{equation}
R^2_{\rm adj} = 1-(1-R^2)\frac{n-1}{n-p-1},
\end{equation}
where $n$ is the number of data points and $p$ is the number of independent variables in the model. The $R^2_{\rm adj}$ parameter measures whether additional variables improve the model fit more than would be expected by random chance. Lastly, we also calculate the root-mean-square (RMS) dispersion about the model predictions:
\begin{equation}
{\rm RMS}=\sqrt{\frac{\sum_i(y_i-f_i)^2}{n}}.
\end{equation}
We can only calculate these three quantities, $R^2$, $R^2_{\rm adj}$, and RMS, for the detections in our sample. Although they cannot provide a complete picture of model performance, they can tell us how well our predictions work for the galaxies with known \fesc\ and can quantify our model performance at the high \fesc\ end. Each of these metrics correlates strongly with the others and with the concordance index, and our main results are insensitive to the exact metric used. 

We choose to evaluate both $R^2$ and RMS using \logten(\fesc) rather than the linear \fesc, because the scatter in the predicted \fesc\ values is approximately constant in logarithmic space over the full observed \fesc\ range. A single RMS value is therefore representative of both low and high \fesc\ galaxies. In contrast, in linear space, the RMS changes systematically from low to high observed \fesc. For instance, for our fiducial model (\S\ref{sec:fiducial}), the RMS is 0.02 for galaxies with \fesc\ in the lowest third of the sample, 0.16 for the next third, and 0.22 for the final third, whereas the logarithmic RMS is $\sim$0.3-0.4 dex across the full sample range. Because $R^2$ and RMS use \logten(\fesc), these statistics only include galaxies with both detected \fesc\ and non-zero predicted \fesc. In contrast, the concordance index $C$ incorporates the full dataset, including galaxies with observed upper limits and those with predicted \fesc=0.

We also assess the predictive ability of our models using cross validation, applying the model to data that was not used in generating the model itself. We perform a $k$-fold cross validation, in which we randomly split the dataset into $k$ groups. We combine $k-1$ of the groups to become our training set, which we use to fit the model. We then test the model on the remaining group using the $C$, $R^2$, and RMS metrics described above. ($R^2_{\rm adj}$ is often undefined, due to runs where $n$ for the small test set is equal to $p$). We then repeat this process until each group has been used as the test group with the remaining groups used as the training set. Following \citet{runnholm20}, we adopt $k=3$, repeat the group selection 100 times, and average the final results. A value of $k=3$ gives us sufficient galaxies in the test set to calculate statistics, and the repetition of the process ensures our results are not sensitive to the exact group selected \citep{runnholm20}. We provide the results of these tests in our evaluation of the models and distinguish the metrics derived from this cross-validation analysis by the subscript `CV'. 

\subsection{Variable Selection}
\label{sec:selection}
We explore different subsets of independent variables and discuss the resulting model fits in Section \S\ref{sec:modified_fiducial}. With these different selections, we test the effect of switching the absorption line probed, of including or excluding \lya, and of limiting the variable set to observables accessible to {\it JWST} (Section \S\ref{sec:jwstmodels}). However, we also wish to analyze the relative importance of each independent variable and identify the variable subsets that most accurately predict \fesc. To evaluate the importance of each variable, we turn to the tools of forward and backward selection \citep[\eg,][]{runnholm20, maji22}. 

In forward selection, we run the Cox model with each independent variable individually in turn and determine which variable provides a fit with the highest concordance. We select this variable and then combine it with each remaining variable in turn to determine which combination of two variables provides the highest concordance. We proceed in this manner until we have a ranked order of the most significant variables. Conversely, with backward selection, we start with the full set of independent variables, and remove each one in turn. The combination of variables that gives the highest concordance identifies the least important excluded variable. We exclude it and proceed with the remaining variables, identifying and removing the next-least important variable each time to obtain a rank-ordered list. Importantly, each round of forward or backward selection only ranks variables with respect to the current best-performing model. Hence, some variables may be poorly ranked if they provide similar information as an already-selected variable. Variable rankings can reveal the most important predictors of \fesc, but a poor ranking does not necessarily imply that a variable is uncorrelated with \fesc.

When we perform our forward and backward selections, we do not use the full list of variables in Table~\ref{table:all_variables} but limit it in the following ways. Firstly, we include only the EW(\hi,abs), EW(LIS), $R_l$(\hi,abs), and $R_l$(LIS) measurements, which are averages of multiple lines, rather than the measurements of individual absorption lines. This choice avoids overly limiting our sample sizes, since 22 galaxies are missing one or more individual absorption line measurements. Secondly, we avoid highly collinear variables and variables that measure very similar properties. We include only \sigsfr\ rather than SFR and $r_{\rm 50, NUV}$, we use only one ionization-sensitive line ratio (O32 = \oiii/\oii), and we use only one measure of UV dust attenuation ($\beta_{\rm 1550}$ or E(B-V)$_{\rm UV}$). When included separately, alternative measures of these properties end up with the same or nearly the same ranks, which suggests that they are indeed largely interchangeable. We explore three different sets of variables in our rankings: one set includes $\beta_{\rm 1550}$ and all remaining variables, one set corresponds to the variables in our best-performing Cox model, and one set is limited to variables accessible at high redshift. We also test our final variable rankings using an MC method. We re-run the forward and backward selection processes after randomly sampling the independent and dependent variables 100 times using their uncertainties. We present the results of the forward and backward selection methods in Section \S\ref{sec:results:variablerank}. 

\section{Predicting \fesc}
\label{sec:results:fesc}
\subsection{Fiducial Model}
\label{sec:fiducial}

\begin{figure*}[htb]
\plotone{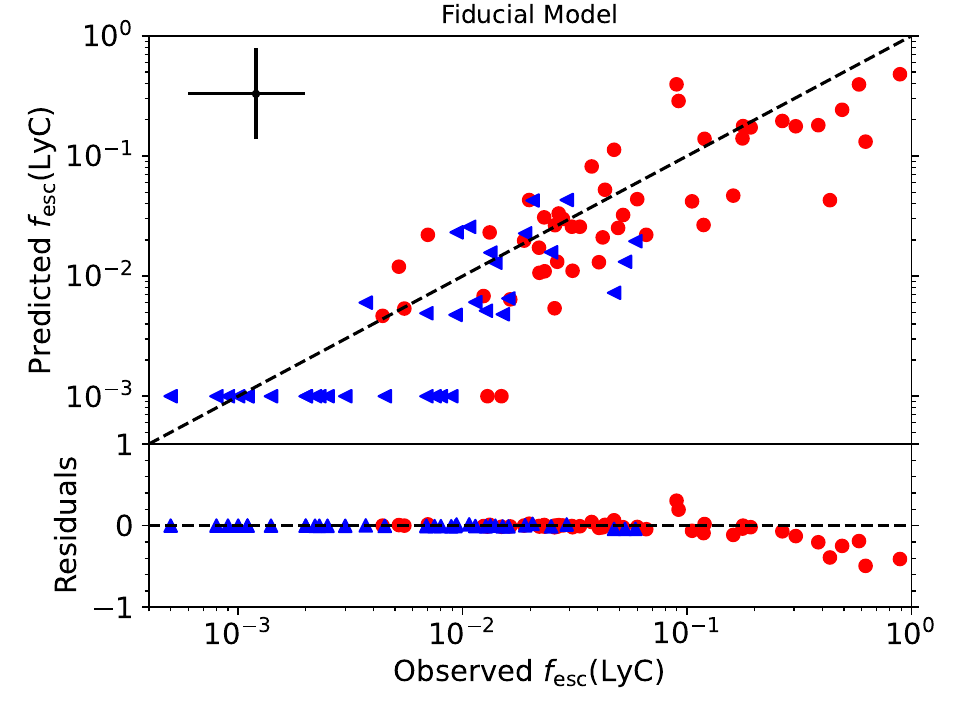}
\caption{The \fesc\ predictions from the fiducial Cox model compared with the observed \fesc\ values for the LzLCS+ sample. Red circles indicate detections, and blue triangles represent the 1$\sigma$ upper limit on non-detections. The dashed line shows a one-to-one correspondence. The black cross in the upper left indicates the median size of the uncertainties in the observed and predicted \fesc. The uncertainties in the predicted \fesc\ represent the 15.9 and 84.1 percentiles of the \fesc\ probability distribution. We plot galaxies with predicted \fesc$\sim$0 at the bottom of the figure with their predicted \fesc\ set to $10^{-3}$. The bottom panel shows a plot of the residuals on a linear scale. This model predicts \fesc\ for the LyC detections with an RMS scatter of 0.36 dex.
\label{fig:fiducialfit}}
\end{figure*}

For our fiducial model, we choose the following variables: \logten($M_*$), $M_{\rm 1500}$, \logten(EW(H$\beta$)), E(B-V)$_{\rm neb}$, 12+\logten(O/H), \logten(O32), \logten(\sigsfr), E(B-V)$_{\rm UV}$, \fesclya, EW(LIS). We present the fitted coefficients and significance for these variables in Table~\ref{tab:fiducial} and compare the model-predicted median \fesc\ with the observations in Figure~\ref{fig:fiducialfit}. We list the goodness-of-fit metrics for this model in Table~\ref{table:metrics}. Statistically significant variables, with $p$-values $\leq0.05$ are \fesclya, E(B-V)$_{\rm UV}$, \logten(\sigsfr), E(B-V)$_{\rm neb}$, and \logten(O32). Not surprisingly, these same variables individually correlate or anti-correlate with \fesc, as discussed in \citet{flury22b} and \citet{saldana22}. We choose to include both O32 and EW(H$\beta$) in the fiducial model, even though they correlate with each other \citep[\eg,][]{flury22b}, because they may have subtly different relationships with \fesc; a young starburst may have both high EW(H$\beta$) and high O32, but high global \fesc\ will tend to increase O32, while decreasing EW($H\beta$) \citep[\eg,][]{zackrisson13, nakajima14, jaskot16}. However, we find that EW(H$\beta$) is not significant and its coefficient is near zero, which indicates that EW(H$\beta$) does not contribute meaningful information to the fit beyond what O32 already provides. For the fiducial model, we choose EW(LIS) instead of EW(\hi), because EW(LIS) is less affected by IGM and circumgalactic medium opacity and is therefore a viable indirect indicator across a wider redshift range. We consider the effect of substituting EW(\hi) or other measurements of absorption line strength in \S\ref{sec:modified_fiducial}.

Although most variables show the behavior expected from their individual relationship with \fesc, one exception is the nebular attenuation, E(B-V)$_{\rm neb}$, which correlates with \fesc\ in the Cox model instead of anti-correlating as might be expected. However, the model already contains the UV E(B-V) and \fesclya, two parameters that show strong trends with E(B-V)$_{\rm neb}$. The nebular E(B-V) therefore essentially operates as a second-order effect, where the nebular attenuation would increase \fesc\ at a fixed value of line-of-sight UV attenuation or \fesclya. This extra nebular E(B-V) correlation could indicate a trend with some other physical property not included in the model, such as starburst age, global gas content, or gas clumpiness. In the case of \fesclya, E(B-V)$_{\rm neb}$ may help quantify how much of the \lya\ escape is due to a low \hi\ optical depth. At fixed \fesclya, a higher E(B-V)$_{\rm neb}$ implies a greater contribution of dust and weaker contribution of \hi\ to the \lya\ absorption. As expected, if we remove both \lya\ and E(B-V)$_{\rm UV}$ from the model, the correlation with E(B-V)$_{\rm neb}$ disappears.

As discussed in \S\ref{sec:interpret}, the best-fit coefficients in Table~\ref{tab:fiducial} quantify how the probability of observing a given \fesc\ responds to an incremental change in each variable. The probability $S$ of observing $f_{\rm esc,detect} < $\fesc\ changes from the original probability $S_{\rm ref}$ by $S = S_{\rm ref}^{\exp(b_i \Delta x_i)}$ for a change in variable $x_i$. The coefficients in Table~\ref{tab:fiducial} indicate that changing \fesclya\ by 0.1 results in raising the probability to the power of $\exp(0.7)=2.0$, leading to a lower probability of low \fesc\ values (see \S\ref{sec:interpret}. Conversely, increasing E(B-V)$_{\rm UV}$ by 0.1 increases the probability of low \fesc\ values by an even greater factor; the probability is raised to $\exp(-1.49)\sim1/5$. For the other statistically significant variables, a 0.1 change in E(B-V)$_{\rm neb}$ also raises the probability to the power of $\sim2$, while 0.1 changes in either \logten(\sigsfr) or \logten(O32) raise the probability to powers of 1.2 and 1.3, respectively. 

\begin{deluxetable*}{lll}[htb]
\tablecaption{Fiducial Model Coefficients}
\label{tab:fiducial}
\tablehead{
\colhead{Variable} & \colhead{Coefficient $b$} & \colhead{p-value}}
\startdata
\fesclya & 7.00 & 6.0E-9 \\
E(B-V)$_{\rm UV}$ & -14.90 & 6.3E-5 \\
\logten(\sigsfr) & 1.83 & 1.1E-3 \\
E(B-V)$_{\rm neb}$ & 6.82 & 2.8E-3 \\
\logten(O32) & 2.65 & 1.6E-2 \\
\logten($M_*$) & 0.63 & 0.18 \\
12+\logten(O/H) & 1.46 & 0.22 \\
$M_{\rm 1500}$ & -0.46 & 0.21 \\
EW(LIS) & -0.26 & 0.32 \\
\logten(EW(H$\beta$)) & -0.04 & 0.98 \\
\enddata
\tablecomments{Variables are listed in order of their $p$-value significance. Positive values of EW(LIS) represent net absorption; positive values of EW(H$\beta$) represent net emission.}
\end{deluxetable*} %

As illustrated in Figure~\ref{fig:fiducialfit} and Table~\ref{table:metrics}, the fiducial model predicts the observed \fesc\ values for detections with an RMS scatter of 0.36 dex for both the full sample and when the dataset is split into test and training sets (RMS$_{\rm CV}$ = 0.36; see \S\ref{sec:metrics}). The model's high concordance, $C=0.89$ ($C_{\rm CV}=0.85$), shows that it successfully ranks galaxies on the basis of \fesc. The $R^2$, $R^2_{\rm adj}$, and $R^2_{\rm CV}$ metrics are 0.60, 0.49, and 0.55 respectively. This successful fit shows that including multiple variables results in a substantial improvement over predictions using a single variable. Even for some of the best single-variable correlations, \fesc\ spans 2 dex at a given galaxy property \citep[\eg,][]{flury22b, chisholm22}. 

The improved accuracy of the fiducial multivariate model is driven by only a handful of variables. Table~\ref{tab:fiducial} shows that only four variables are statistically significant in the fiducial model, and limiting the model to only these four variables achieves comparable accuracy, with the $R^2$, RMS, and $C$ metrics changing by 0.01-0.02. As can be seen in Table 2, highly insignificant variables tend to have best-fit coefficients near 0, which means the variable plays almost no role in the prediction. Excluding these variables therefore has little effect on the model, and because these extra parameters have negligible effects, overfitting is not a major concern.

Despite some overall improvement, the scatter in the fiducial Cox model is still significant, which demonstrates the difficulty in predicting the line-of-sight \fesc\ even given a large amount of information. The residuals in the bottom panel of Figure~\ref{fig:fiducialfit} also show that the fiducial model systematically underpredicts the value of \fesc\ for the strongest LCEs, even though it does typically identify them as having \fesc$>0.1$. \citet{mascia23} find a similar result and suggest that it may arise from the limited number of strong LCEs in the LzLCS+. We suggest that this under-prediction arises from a different effect, where the strongest LCEs represent a subset of galaxies with favorably oriented low-column density channels but whose global \fesc\ may actually be substantially lower. We elaborate on this possibility in \S\S\ref{sec:modified_fiducial} and \ref{sec:outliers}. Because LyC may escape from narrow channels, the chance orientation of a galaxy can introduce scatter in relationships between global galaxy properties and the measured line-of-sight \fesc\ \citep[\eg,][]{cen15,flury22b,seive22}. Although the fiducial model explicitly includes variables that trace line-of-sight properties, such as EW(LIS) and E(B-V)$_{\rm UV}$, the LIS metals may imperfectly trace the LyC-absorbing \hi\ gas. 

Other effects may also contribute to the observed scatter. Simulations show that \fesc\ may fluctuate in time, which may introduce scatter between \fesc\ and observables that are sensitive to a different timescale \citep[\eg,][]{trebitsch17, barrow20}. Other possible causes of the model scatter and underprediction of \fesc\ are uncertainties in the observed \fesc\ values or a disconnect between the global properties we measure and the local properties of the LyC-emitting region \citep[\eg,][]{martin15, kim23}. \citet{riverathorsen22} point out that a small number of massive stars can potentially dominate the escaping LyC emission; the properties of this LyC source region may differ from both the global galaxy properties and the properties inferred from the non-ionizing UV spectrum. Finally, because different properties may regulate LyC escape in different types of galaxies \citep[\eg,][]{flury22b}, one relation may not suffice to predict \fesc\ in all types of galaxies. We explore this possibility further in Section \S\ref{sec:subsamples}. 

\subsection{Modifications to the Fiducial Model}
\label{sec:modified_fiducial}
Modifications to the fiducial model demonstrate that properties sensitive to dust and \hi\ column density are the most useful predictors of \fesc. We summarize the performance of the fiducial model and compare it with different modifications in Table~\ref{table:metrics}. We do not test all possible combinations of the variables in Table~\ref{table:all_variables}, but we experiment with modifying the fiducial model by swapping alternative measures of some properties and by dropping each statistically significant variable.

We first substitute different absorption line tracers (Table~\ref{table:all_variables}) for EW(LIS). Of these different tracers, only EW(Ly$\beta$), EW(\hi,abs), and $R_l$(\hi,abs) give statistically significant coefficients with $p<0.05$. Because the sample size changes slightly for the other low-ionization metal absorption line selections (see Table~\ref{table:all_variables}), we cannot compare the metrics for the other models in detail. However, the fit quality appears comparable to the fiducial model (RMS $=0.33-0.39$ and $C=0.88-0.89$). Only EW(\hi,abs) clearly improves the model fit, with RMS $=0.31$ and $C=0.91$.

\begin{figure*}
\plotone{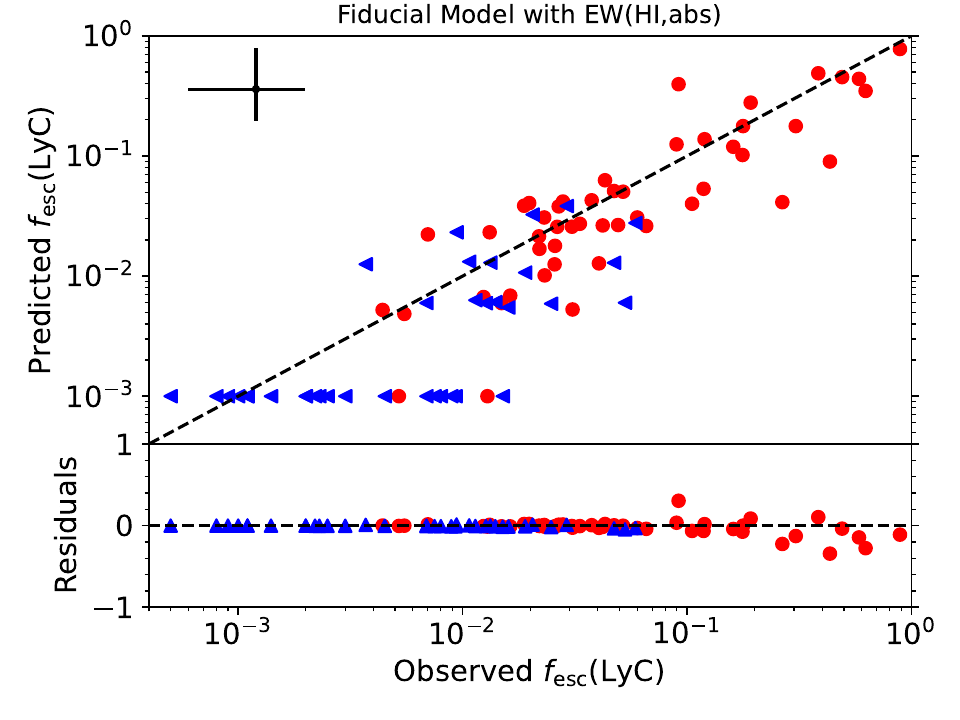}
\caption{The \fesc\ predictions from the Cox model improve when we include EW(\hi,abs) instead of EW(LIS). Symbols are the same as in Figure~\ref{fig:fiducialfit}.
\label{fig:graphT}}
\end{figure*}

The fit with EW(\hi,abs) (Figure~\ref{fig:graphT}) is also our best-performing model, and it scores better than the fiducial model on all metrics (Table~\ref{table:metrics}). Moreover, this model more accurately predicts \fesc\ for strong LCEs (Figure~\ref{fig:graphT}). Hence, the \fesc\ underprediction in the fiducial model arises, at least in part, from the fact that EW(LIS) and the other variables imperfectly trace the line-of-sight \hi\ gas. The Cox model is predicting the median \fesc\ for a given set of parameters, and the strongest LCEs may not be noticeably distinct from more moderate LCEs in the fiducial model's set of parameters. In contrast, the low EW(\hi,abs) values of the strongest leakers do distinguish them from almost all of the moderate LCEs \citep[\eg,][]{saldana22}. The exact conversion from EW(LIS) to \hi\ absorption and \fesc\ depends on metallicity and gas geometry, since the trace metals will only show up in absorption for a sufficiently large \hi\ column. While a range of physical conditions could cause weak LIS absorption, weak \hi\ absorption more directly indicates LyC escape. Lastly, the observational uncertainty in EW(LIS) is also higher than in EW(\hi,abs), and these uncertainties will be the most problematic for the strongest LCEs with the weakest absorption lines. In conclusion, while LIS lines can identify LCEs, \hi\ absorption lines appear necessary to accurately determine their line-of-sight \fesc.

\begin{deluxetable*}{lllllllll}
\tablecaption{Goodness-of-Fit for Cox \fesc\ Models}
\label{table:metrics}
\tablehead{
\colhead{Model} & \colhead{Sample Size} & \colhead{$R^2$} & \colhead{$R^2_{\rm adj}$} & \colhead{RMS} & \colhead{$C$} & \colhead{$R^2_{\rm CV}$} & \colhead{RMS$_{\rm CV}$} & \colhead{$C_{\rm CV}$}}
\startdata
Fiducial & 87 & 0.60 & 0.49 & 0.36 & 0.89 & 0.55 & 0.36 & 0.85 \\ 
\hline
\multicolumn{8}{c}{Fiducial Model with Modifications\tablenotemark{a}} \\
\hline
EW(\hi,abs) instead of EW(LIS) & 87 & 0.69 & 0.60 & 0.31 & 0.91 & 0.64 & 0.31 & 0.88 \\ 
with both EW(\lya) \&\ \fesclya & 87 & 0.61 & 0.48 & 0.36 & 0.89 & 0.56 & 0.35 & 0.86 \\ 
EW(Ly$\beta$) instead of EW(LIS) & 78\tablenotemark{b} & 0.64 & 0.52 & 0.34 & 0.89 & 0.58 & 0.34 & 0.85 \\ 
EW(\cii) instead of EW(LIS) & 77\tablenotemark{b} & 0.59 & 0.46 & 0.37 & 0.89 & 0.54 & 0.36 & 0.85 \\ 
EW(\sitwo) instead of EW(LIS) & 85\tablenotemark{b} & 0.63 & 0.52 & 0.33 & 0.89 & 0.56 & 0.33 & 0.85 \\
\oii/H$\beta$ instead of O32 & 87 & 0.62 & 0.51 & 0.35 & 0.88 & 0.56 & 0.34 & 0.85 \\
$R_l$(\hi,abs) instead of EW(LIS) & 82\tablenotemark{b} & 0.61 & 0.50 & 0.34 & 0.88 & 0.56 & 0.33 & 0.85 \\ 
$\beta_{\rm 1550}$ instead of E(B-V)$_{\rm UV}$ & 87 & 0.58 & 0.46 & 0.37 & 0.88 & 0.53 & 0.37 & 0.85 \\ 
$R_l$(\sitwo) instead of EW(LIS) & 87 & 0.58 & 0.47 & 0.37 & 0.88 & 0.53 & 0.37 & 0.85 \\ 
$R_l$(\cii) instead of EW(LIS) & 84\tablenotemark{b} & 0.55 & 0.41 & 0.38 & 0.88 & 0.49 & 0.37 & 0.85 \\ 
\neiii/\oii\ instead of O32 & 87 & 0.57 & 0.44 & 0.38 & 0.88 & 0.51 & 0.37 & 0.85 \\ 
$R_l$(Ly$\beta$) instead of EW(LIS) & 84\tablenotemark{b} & 0.52 & 0.37 & 0.39 & 0.88 & 0.45 & 0.39 & 0.84 \\ 
$R_l$(LIS) instead of EW(LIS) & 85\tablenotemark{b} & 0.60 & 0.49 & 0.36 & 0.88 & 0.54 & 0.36 & 0.84 \\ 
without O32 & 87 & 0.57 & 0.46 & 0.38 & 0.88 & 0.51 & 0.37 & 0.85 \\ 
without E(B-V)$_{\rm neb}$ & 87 & 0.54 & 0.42 & 0.38 & 0.88 & 0.48 & 0.38 & 0.84 \\ 
EW(\lya) instead of \fesclya\ & 87 & 0.54 & 0.41 & 0.39 & 0.88 & 0.49 & 0.39 & 0.83 \\ 
$\beta_{\rm 1200}$ instead of E(B-V)$_{\rm UV}$ & 87 & 0.51 & 0.36 & 0.40 & 0.88 & 0.43 & 0.39 & 0.85 \\ 
without \sigsfr & 88\tablenotemark{b} & 0.49 & 0.37 & 0.41 & 0.87 & 0.41 & 0.40 & 0.84 \\ 
without E(B-V)$_{\rm UV}$ & 87 & 0.30 & 0.13 & 0.48 & 0.86 & 0.23 & 0.47 & 0.83 \\ 
$L$(\lya) instead of \fesclya\ & 87 & 0.41 & 0.24 & 0.44 & 0.85 & 0.34 & 0.44 & 0.82 \\ 
without \fesclya & 87 & 0.35 & 0.19 & 0.46 & 0.84 & 0.27 & 0.46 & 0.79 \\ 
\hline
\multicolumn{8}{c}{JWST Models} \\
\hline
Full JWST Model & 87 & 0.29 & 0.14 & 0.47 & 0.83 & 0.21 & 0.47 & 0.79 \\
$\beta_{\rm 1550}$+\logten(\sigsfr)+\logten(O32) & 87 & 0.34 & 0.29 & 0.46 & 0.83 & 0.27 & 0.45 & 0.81 \\
$\beta_{\rm 1550}$+\logten(\sigsfr)+\logten(\neiii/\oii) & 87 & 0.40 & 0.35 & 0.44 & 0.83 & 0.32 & 0.43 & 0.82 \\
\enddata
\tablenotetext{a}{Models are listed in order of decreasing concordance, $C$. We substitute different measures of dust, UV absorption lines, \lya, and nebular ionization as indicated. We also experiment with removing each of the statistically significant variables in the fiducial model.}
\tablenotetext{b}{These models apply to fewer or more galaxies than the fiducial model, and some differences in goodness-of-fit metrics relative to the fiducial model may result from the changed sample.}
\end{deluxetable*}

Like EW(\hi,abs), \lya\ measurements can contribute key information about \hi\ column density. Indeed, \citet{maji22} find that \lya\ luminosity, $L$(\lya), is the most important variable in predicting the LyC \fesc\ from simulations. In our fiducial model, \fesclya\ was the most statistically significant coefficient, consistent with observational and theoretical studies that find a connection between LyC and \lya\ \citep[\eg,][]{dijkstra16, verhamme17, steidel18, izotov20}. In the fiducial model, \fesclya\ serves as the best estimate of the \hi\ optical depth. When EW(\hi,abs) is substituted in the model, \fesclya\ remains statistically significant, but its p-value drops from 6E-9 to 3E-3 and its best-fit coefficient declines from 7 to 4. Because it is less sensitive to scattered emission, EW(\hi,abs) more directly traces the line-of-sight \hi\ optical depth, which reduces the model's reliance on \fesclya. Without EW(\hi,abs), however, \fesclya\ provides crucial information in the fiducial model. If we exclude it, the fit quality worsens by all metrics; the RMS rises from 0.36 to 0.46, $C$ drops from 0.89 to 0.84, and $R^2$ drops from 0.60 to 0.35. Substituting $L$(\lya) for \fesclya\ also decreases the quality of the fit (RMS = 0.44, $C$ = 0.85, $R^2$ = 0.41), whereas EW(\lya) performs almost as well as \fesclya\ (RMS = 0.39, $C$ = 0.88, $R^2$ = 0.54). Interestingly, when included together, both EW(\lya) and \fesclya\ show up as significant coefficients. The two variables may provide slightly different information relevant to LyC escape. While \fesclya\ may be more directly linked to the fraction of escaping LyC, the EW provides additional information about starburst properties such as the intrinsic \lya\ production and underlying continuum. However, including both EW(\lya) and \fesclya\ results in marginal, if any, improvement in the fit. All metrics change by only 0-0.01. 

Of all the variables in the fiducial model, excluding E(B-V)$_{\rm UV}$ or \fesclya\ has the most detrimental effect on the model fit. Without information on the line-of-sight dust attenuation from E(B-V)$_{\rm UV}$, $R^2$ drops to 0.30 and the RMS scatter rises by $\sim$0.1 dex to 0.48. The concordance also drops slightly to $C=0.86$. Excluding \fesclya\ has the strongest negative impact on the concordance ($C=0.84$) and the second largest impact on $R^2$ and RMS. The concordance metric includes non-detections, whereas the $R^2$ and RMS parameters do not. Hence, the larger effect of \fesclya\ on concordance may illustrate that even without information on dust, a lack of \lya\ emission is generally sufficient to identify non-leakers in the LzLCS+, even if the presence of \lya\ does not necessarily imply LyC escape. Conversely, predicting accurate \fesc\ for LCEs requires measurements of the line-of-sight dust attenuation.

Substituting alternate measures of dust attenuation or ionization has little effect on the fiducial model. Using $\beta_{\rm 1550}$ instead of E(B-V)$_{\rm UV}$ changes the goodness-of-fit metrics by only 0-0.03, consistent with the observed tight correlation between $\beta_{\rm 1550}$ and E(B-V)$_{\rm UV}$ in the LzLCS+ \citep{chisholm22}. However, the $\beta_{\rm 1200}$ parameter is not as successful at tracing dust content \citep[\eg,][]{chisholm22} and worsens the fit quality by all metrics. Likewise, alternate measures of ionization such as \neiii/\oii\ or \oii/H$\beta$ work as well as O32, with no change in $C$ and minor ($\leq$0.03) changes in $R^2$ and RMS. Of the three measures of ionization, \oii/H$\beta$ performs best in all metrics, but this improvement is marginal.

\subsection{Most Important Variables}
\label{sec:results:variablerank}
\begin{figure*}
\gridline{\fig{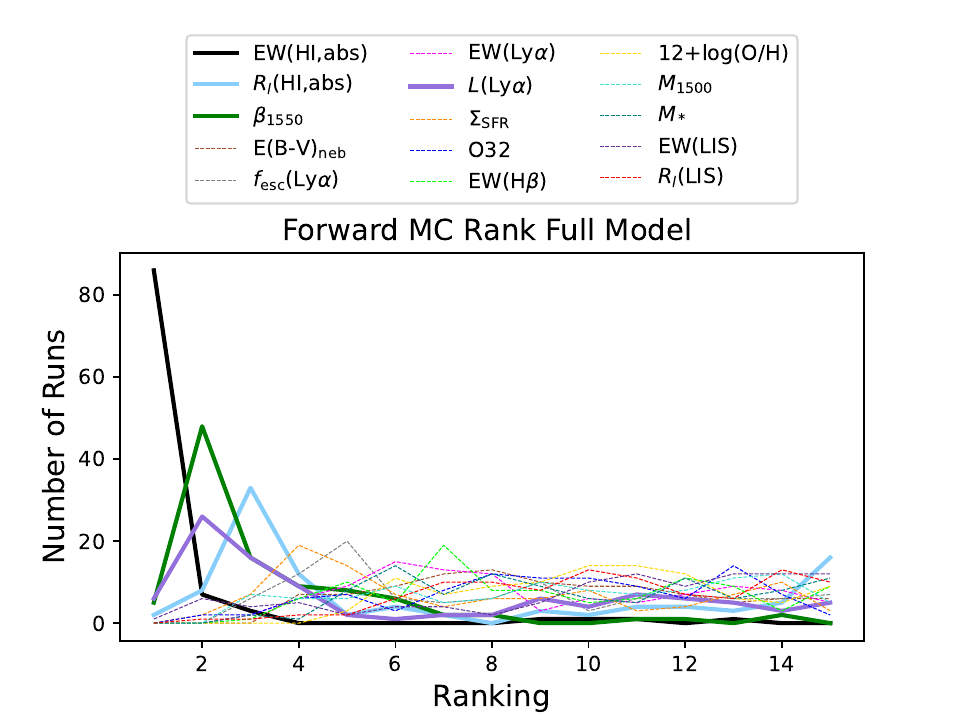}{0.5\textwidth}{(a)}
	\fig{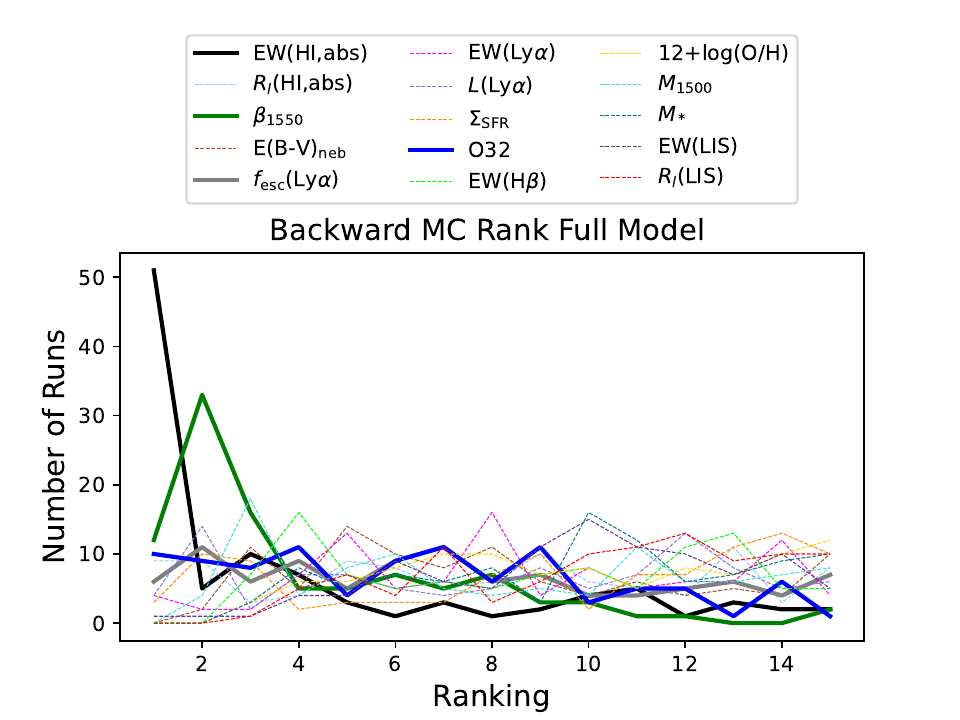}{0.5\textwidth}{(b)}
}
\gridline{\fig{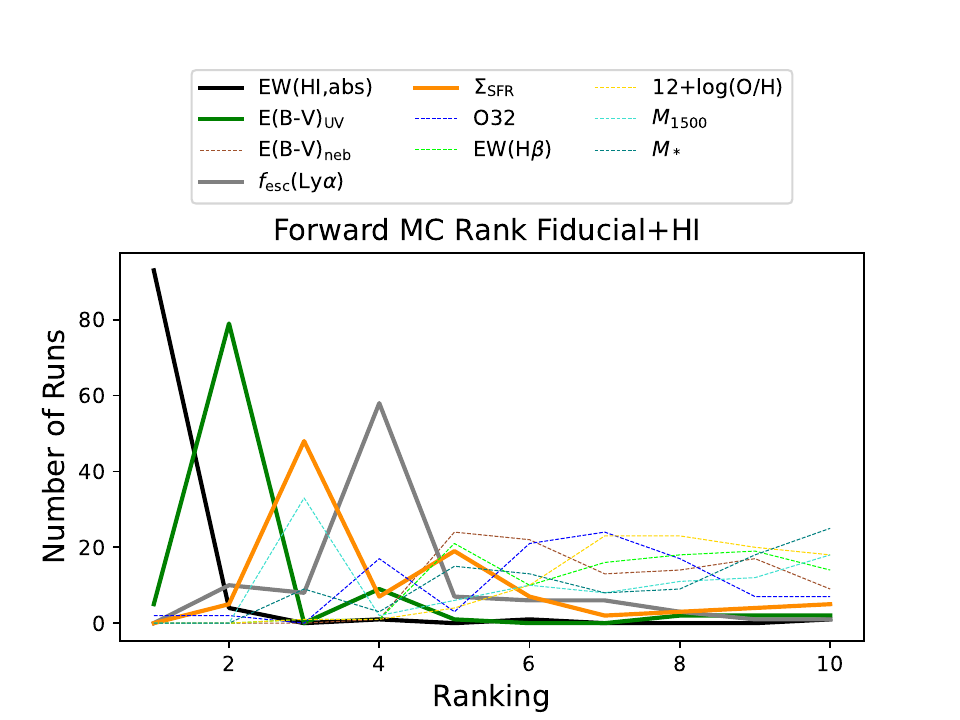}{0.5\textwidth}{(c)}
	\fig{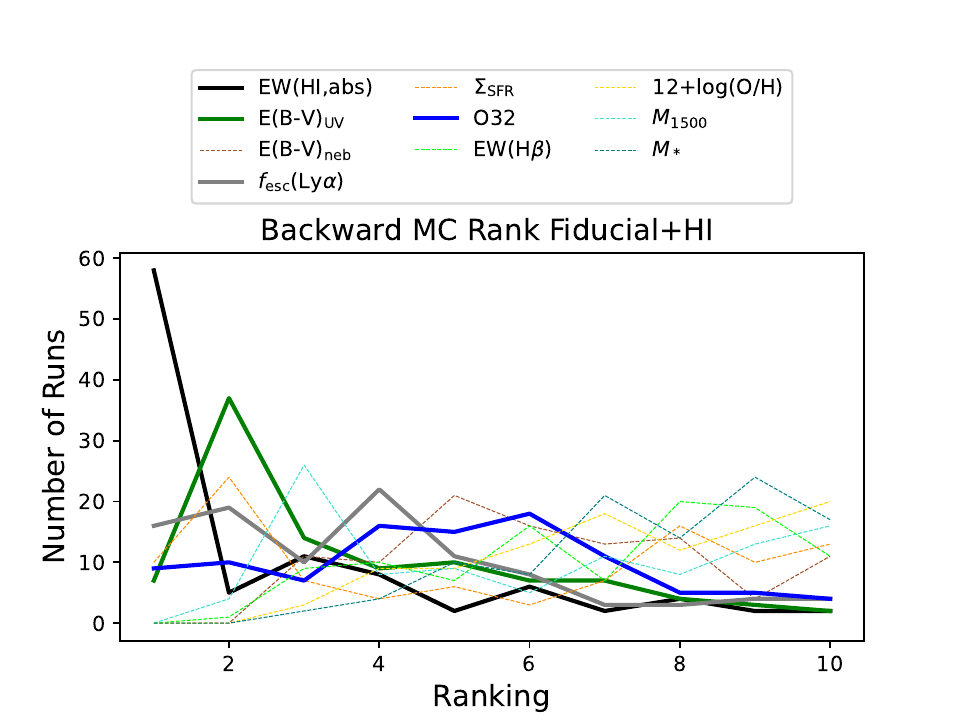}{0.5\textwidth}{(d)}
}
\gridline{\fig{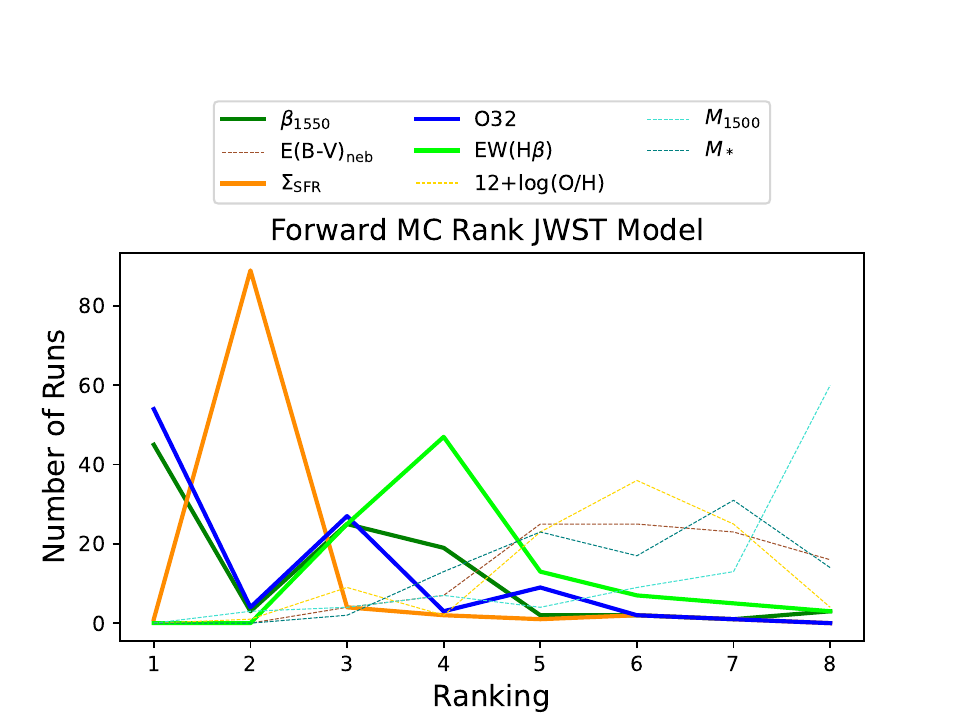}{0.5\textwidth}{(e)}
	\fig{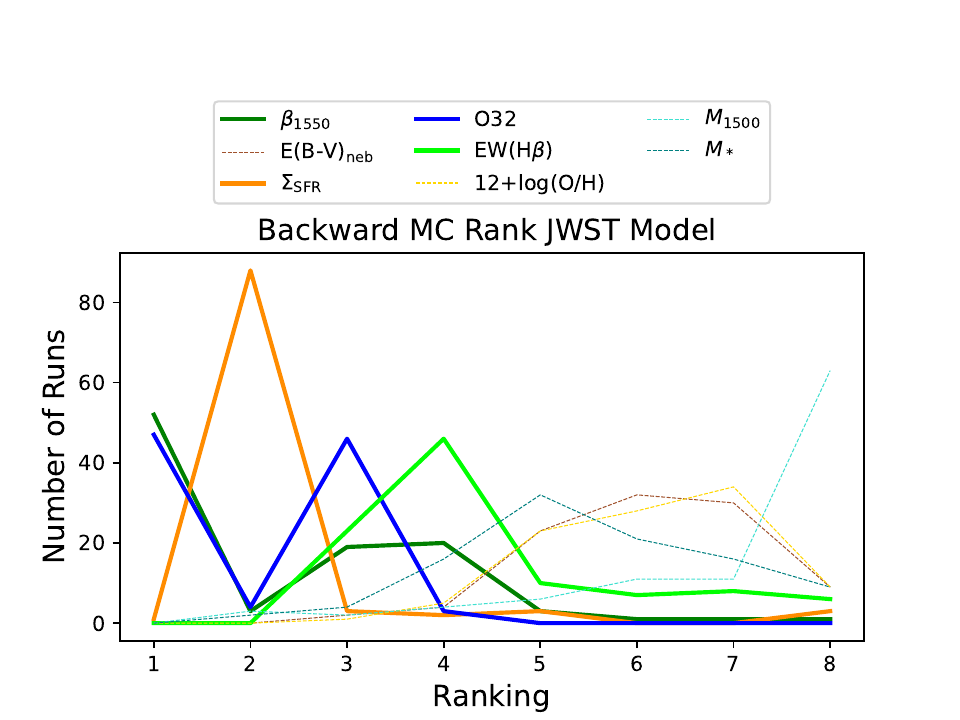}{0.5\textwidth}{(f)}
}
\caption{The distribution of variable rankings from 100 forward selection (left) and backward selection (right) MC runs, after resampling each variable using its observational uncertainties. Variable combinations are the same as in Table~\ref{table:rankorder}. Thick solid lines indicate the four variables with the lowest (best) mean MC ranks.
\label{fig:mcrank}}
\end{figure*}

As discussed in Section \S\ref{sec:selection}, we perform forward and backward selection to determine which variables have the greatest effect on the model fit quality. In Table~\ref{table:rankorder}, we show a ranked ordering of variables based on forward and backward selection for three representative models. The ``Full Model" includes $\beta_{\rm 1550}$ instead of E(B-V)$_{\rm UV}$ and includes some variables that measure similar but not highly collinear properties, in order to test which ones perform best. For instance, we include \fesclya, EW(\lya), and $L$(\lya), which are related but may have different relationships with \fesc. Our second set of variables is the more limited list from the Fiducial+HI model, our best-performing Cox model. This model also differs from the Full Model by using E(B-V)$_{\rm UV}$. The third model, the ``JWST Model" excludes absorption line measurements, which are difficult or impossible to measure for most galaxies at $z>6$, and \lya\ measurements, which are heavily affected by the IGM at $z>6$; we discuss this model in Section \S\ref{sec:jwstmodels}. We also list the mean MC ranks for each variable obtained by sampling the observational uncertainties and rerunning the ranking process 100 times. We plot the distribution of these ranks in Figure~\ref{fig:mcrank}.

\movetabledown=2.25in
\begin{rotatetable*}
\begin{deluxetable*}{lllllllll}
\tablecaption{Variables by Ranked Order}
\label{table:rankorder}
\tablehead{
\colhead{Rank} & \multicolumn{2}{c}{Full Model} & \multicolumn{2}{c}{Fiducial+HI Model} & \multicolumn{2}{c}{JWST Model} \\
\colhead{} & \colhead{Forward} & \colhead{Backward} & \colhead{Forward} & \colhead{Backward} & \colhead{Forward} & \colhead{Backward}}
\startdata
1 & EW(\hi,abs) [1.52] &  EW(\hi,abs) [3.91] & EW(\hi,abs) [1.21] & \fesclya\ [3.96] & $\beta_{\rm 1550}$ [2.55] & $\beta_{\rm 1550}$  [2.31] \\
2 & $\beta_{\rm 1550}$ [3.42] & $\beta_{\rm 1550}$  [4.14] & E(B-V)$_{\rm UV}$ [2.58] &  \logten(\sigsfr) [5.40] & \logten(\sigsfr)\tablenotemark{a} [2.23] & \logten(\sigsfr)\tablenotemark{a} [2.33] \\
3 & $M_{\rm 1500}$ [9.41] & $M_{\rm 1500}$ [8.19] & \logten($M_*$)\tablenotemark{a} [7.32] & E(B-V)$_{\rm UV}$ [3.79] & \logten(O32) [2.19] & \logten(O32) [2.05] \\
4 & $R_l$(\hi,abs) [7.01] & $R_l$(\hi,abs) [7.86] & \logten(EW(H$\beta$)) [7.39] & EW(\hi,abs) [2.63] & \logten(EW(H$\beta$)) [4.29] &  \logten(EW(H$\beta$)) [4.49]\\
5 & EW(\lya)\tablenotemark{a} [8.82]& EW(\lya)\tablenotemark{a} [8.56] & \fesclya\ [4.32] & \logten($M_*$)\tablenotemark{a} [7.65] & \logten($M_*$) [6.04] &   \logten($M_*$) [5.50] \\
6 & \logten(EW(H$\beta$)) [9.18] &  \logten($M_*$) [9.72] & \logten(\sigsfr) [4.43] &  \logten(EW(H$\beta$)) [6.90] & E(B-V)$_{\rm neb}$ [6.04] & E(B-V)$_{\rm neb}$ [6.11] \\
7 &  \logten(\sigsfr) [7.80]  & $R_l$(LIS) [10.21] & \logten(O32) [6.52] & \logten(O32) [4.99] & 12+\logten(O/H) [5.75] &  12+\logten(O/H) [6.16] \\
8 & $L$(\lya) [6.13] & EW(LIS) [9.72] & 12+\logten(O/H) [7.92] & 12+\logten(O/H) [7.34] & $M_{\rm 1500}$ [6.91] &$M_{\rm 1500}$ [7.05] \\ 
9 & $R_l$(LIS) [10.26]  &  \fesclya\ [7.32]  &  E(B-V)$_{\rm neb}$ [7.02] & E(B-V)$_{\rm neb}$ [6.23] & --- & --- \\
10 & \logten($M_*$) [9.62]  & 12+\logten(O/H) [9.66] &  $M_{\rm 1500}$ [6.29] & $M_{\rm 1500}$ [6.11] & ---  & --- \\
11 & EW(LIS) [10.12]  & E(B-V)$_{\rm neb}$ [8.10] & --- & --- & ---  & --- \\
12 & \fesclya\  [8.22]   & \logten(O32) [6.51] & --- & --- & ---  & --- \\
13 & 12+\logten(O/H) [10.13]  &  \logten(\sigsfr) [9.14]  & --- & --- & ---  & --- \\
14 & E(B-V)$_{\rm neb}$ [9.07] & $L$(\lya) [8.18] & --- & --- & ---  & --- \\
15 &  \logten(O32) [9.29]  &  \logten(EW(H$\beta$)) [8.76] & --- & --- & ---  & --- \\
\enddata
\tablenotetext{a}{After this variable, adding variables improves $C$ by $<0.01$.}
\tablecomments{Variables in rank order from 1 (most important) to 15 (least important) by forward or backward selection. Numbers in brackets indicate the mean of the variable's ranks in the 100 MC runs. We plot the full distribution of ranks in the MC runs in Figure~\ref{fig:mcrank}. The Full Model includes most variables in Table~\ref{table:all_variables} aside from those that are highly collinear. The Fiducial+HI Model includes the variables in the best-performing Cox Model. The JWST Model includes only variables accessible at $z>6$.}
\end{deluxetable*} %
\end{rotatetable*}

We find that the most important factors for predicting \fesc\ are almost always the strength of \hi\ absorption (EW(\hi,abs)) and dust attenuation (E(B-V)$_{\rm UV}$ or $\beta_{\rm 1550}$). These factors are naturally connected to the line-of-sight \fesc, since the two sources of LyC absorption are \hi\ atoms and dust. Previous studies have used the combination of \hi\ absorption lines and dust to derive predictions for \fesc\ \citep[\eg,][]{reddy16b, chisholm18, saldana22}, and the $\beta_{\rm 1550}$ parameter alone predicts much of the variation in \fesc\ \citep{chisholm22}. Importantly, unlike other tracers of \hi\ content and dust such as \lya\ and the nebular E(B-V), the \hi\ absorption lines and UV measures of dust attenuation are more direct tracers of the \hi\ and dust absorption {\it along the line of sight to the UV-emitting stars} and can thus substantially aid the prediction of the line-of-sight \fesc. The combination of EW(\hi,abs) and one of the UV dust measures alone achieves a concordance $C=0.84$.

The MC method also finds that these two parameters, \hi\ absorption and UV dust attenuation, have the top mean ranking. The MC mean ranks for the other variables are not necessarily consistent with their rank order in Table~\ref{table:rankorder}, which reflects the fact that the ranked order varies considerably among different MC runs (Figure~\ref{fig:mcrank}). Since $C$ barely changes after adding the first three to five variables, the order of most variables may be largely random, consistent with the broad ranking distribution in Figure~\ref{fig:mcrank}. As also noted in \S\ref{sec:fiducial}, only the most significant variables dominate the model predictions, and limiting the model to this smaller set of variables gives comparable results to a model generated using a larger variable set. Consequently, we recommend using only the most statistically significant or top-ranked variables when using Cox models to predict \fesc.

Table~\ref{table:rankorder} and Figure~\ref{fig:mcrank} show that \lya\ measurements are often some of the more important variables, although generally not as important as EW(\hi,abs). For all rankings that involve \lya, including the MC rankings, a \lya\ measurement, either EW(\lya), $L$(\lya), or \fesclya, achieves a rank between 1-5. In the MC runs, for the Fiducial+HI model, \fesclya\ is the third most important variable after EW(\hi,abs) and E(B-V)$_{\rm UV}$.  In Section~\ref{sec:modified_fiducial}, we found that excluding \fesclya\ had one of the most detrimental effects on the fiducial model. Crucially, the fiducial model lacks EW(\hi,abs), so \fesclya\ provided essential information on the \hi\ optical depth in its place. With EW(\hi,abs) included, \fesclya\ plays a lesser role but is still ranked among the top variables.

The rankings for the Full Model in Table~\ref{table:rankorder} show that EW(\hi,abs) is a better predictor of \fesc\ in the LzLCS+ sample than the residual intensity $R_l$(\hi,abs). This ranking may result from the low resolution of the FUV spectra. Low spectral resolution, which depends on the spatial size of the source, will artificially increase the observed residual intensity, $R_l$(\hi,abs). In contrast, the EW(\hi,abs) is less sensitive to spectral resolution and may serve as a more accurate measure of the line-of-sight gas. Indeed, \citet{saldana22} find that at a fixed $R_l$, stronger LCEs have lower EWs. At the same time, the observational uncertainties are also higher in $R_l$, such that \fesc\ trends may appear more readily with EW. To investigate the effect of observational uncertainty on the rankings, we compare each variable's median uncertainty with standard deviation of that variable in the LzLCS+ sample. If the measurement uncertainty is comparable to the standard deviation, we may not be able to discern trends with that variable across the LzLCS+. For most variables, the median uncertainty is much lower, $\lesssim20$\%\ of the standard deviation. These variables span the full range of possible ranks and MC ranks, which suggests that we can distinguish variables that do and do not affect the \fesc\ predictions. However, a few variables ($R_l$(\hi,abs), \logten($M_*$), EW(LIS), and $R_l$(LIS)) have higher ratios of median uncertainty to standard deviation (0.50-1.03). These variables all have high MC ranks in the Full Model (6.73-10.81) and any genuine trends with \fesc\ may be hidden by the uncertainties in their measurements. As a test, we insert \logten(\fesc) as a dummy variable, as it should correlate perfectly with itself. Its MC rank begins to deviate from 1.00 when we give it an uncertainty $\geq0.6$ times the standard deviation. To further test the effect of uncertainty, we re-run the MC rankings after doubling the uncertainty in the EW(\hi,abs) measurements, such that the EW(\hi,abs) and $R_l$(\hi,abs) variables have similar ratios of uncertainty to standard deviation. The EW(\hi,abs) MC ranks of 1.21-4.90 increase to 2.6-7.2 when the uncertainty is doubled. We conclude that the higher uncertainty in $R_l$(\hi,abs), \logten($M_*$), EW(LIS), and $R_l$(LIS) may prevent us from determining the importance of these variables. Higher resolution data are necessary to test whether EW(\hi,abs) or $R_l$(\hi,abs) better predicts \fesc. 

In their multivariate analysis of \fesc\ from cosmological simulations, \citet{maji22} also use forward and backward selection to rank the importance of variables. They find that the three most important predictors of \fesc\ are $L$(\lya), the SFR, and galaxy gas mass. Our rankings share some broad similarities with these simulation results. We also find that gas content (as measured by EW(\hi,abs)), SFR (in the form of the UV luminosity), and \lya\ emerge as important parameters. However, our sample and the \citet{maji22} simulated galaxies have some crucial differences. First, we measure the line-of-sight \fesc, whereas \citet{maji22} measure the total global \fesc\ through all sightlines; consequently, the relevant parameter for our analysis is the line-of-sight \hi\ rather than the global \hi. Secondly, the galaxies in the \citet{maji22} sample are much less massive than the LzLCS+ sample, with a median stellar mass of \logten($M_*/$\Msol)=6.41 vs.\ 8.8. Dust may play a greater role in determining \fesc\ in the higher mass, more enriched galaxies in our sample, which would account for its higher importance in our predictions.

\subsection{Predictions at $z>6$ with {\it JWST}}
\label{sec:jwstmodels}
Although the fiducial model and the models using the top-ranked variables predict \fesc\ reasonably well, we cannot apply these models to {\it JWST} observations of galaxies in the epoch of reionization. {The Gunn-Peterson trough and Lyman-series IGM absorption at lower redshifts will prevent \hi\ absorption line measurements,} and the partially neutral IGM at $z>6$ can suppress \lya\ emission \citep[\eg,][]{stark11,schenker14}. Measuring LIS absorption lines instead requires high signal-to-noise observations of the rest-UV continuum, which will be difficult for faint galaxies. Consequently, we explore alternative models that use parameters that can be derived from $z>6$ {\it JWST} observations. 

\begin{figure*}
\gridline{\fig{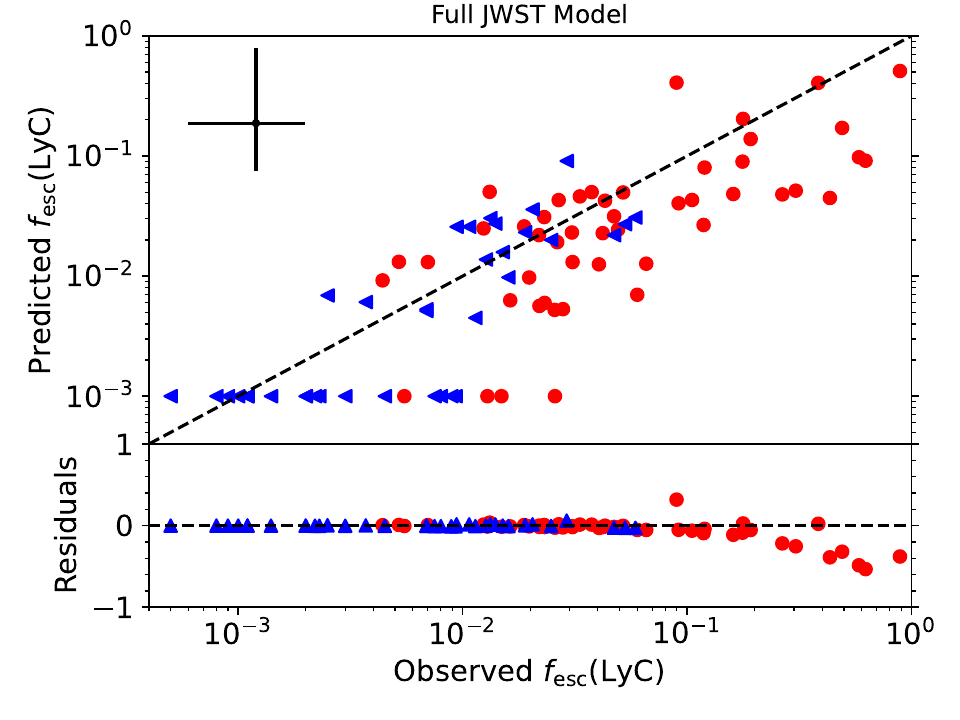}{0.5\textwidth}{(a)}
	\fig{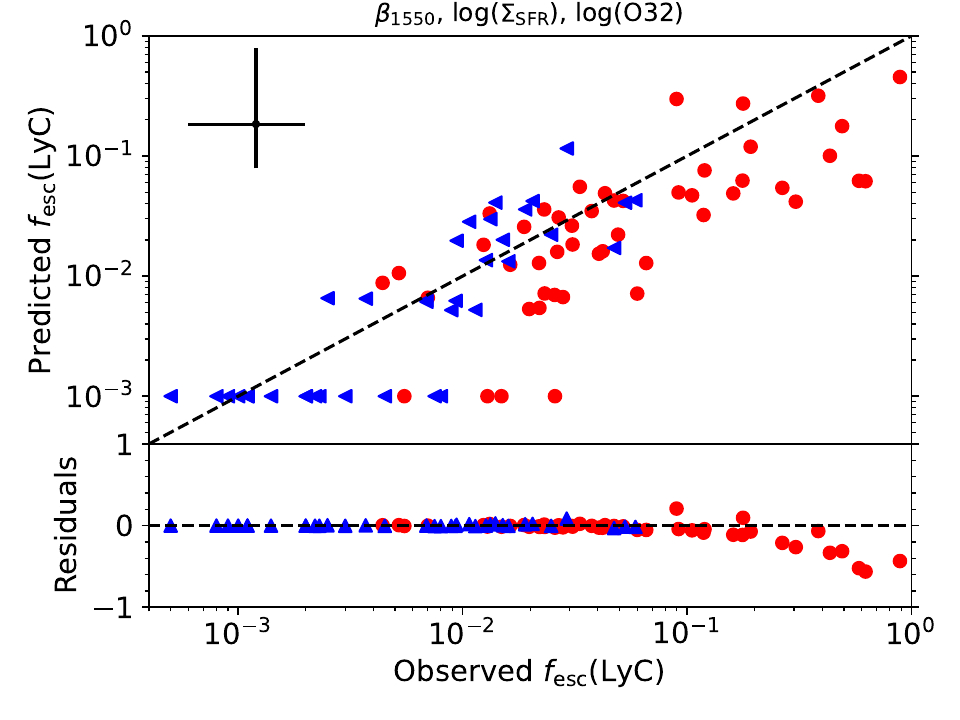}{0.5\textwidth}{(b)}}
\caption{The \fesc\ predictions from JWST models with the full set of variables (a) and top-ranked variables only (b). Symbols are the same as in Figure~\ref{fig:fiducialfit}. The scatter in these models is higher than in the fiducial model, but they still reproduce the observed \fesc\ with an RMS scatter of 0.46-0.47 dex.
\label{fig:jwstmodel}}
\end{figure*}

We create a JWST model by modifying the fiducial model to exclude EW(LIS) and \fesclya. We also choose $\beta_{\rm 1550}$ instead of E(B-V)$_{\rm UV}$ as it is more easily inferred from observations without any required stellar population modeling \citep[\eg,][]{chisholm22} or assumptions about the dust-attenuation law. The JWST model thus includes the following variables:  \logten($M_*$), $M_{\rm 1500}$, \logten(EW(H$\beta$)), E(B-V)$_{\rm neb}$, 12+\logten(O/H), \logten(O32), \logten(\sigsfr), $\beta_{\rm 1550}$. We show the resulting Cox model fit in Figure~\ref{fig:jwstmodel}a and list the goodness-of-fit metrics in Table~\ref{table:metrics}. The scatter is noticeably higher by 0.1 dex than in the fiducial model, with RMS=0.47 dex. Like the fiducial model, the JWST model also tends to under-predict \fesc\ in several of the strongest LCEs with \fesc$>0.2$. This reduced fit quality shows that information about the line-of-sight \hi\ content from absorption lines and \lya\ is essential to precisely predict \fesc. Nevertheless, other observable properties can still provide rough \fesc\ estimates and distinguish strong LCEs from non-leakers. Since the JWST model relies more on global properties rather than line-of-sight \hi\ to predict \fesc, the underprediction for the most extreme LCEs suggests that their global properties are not distinct from more moderate LCEs. The model-predicted \fesc\ may indicate the typical \fesc\ value for this combination of parameters. Moreover, if the most extreme \fesc\ occurs only along favorable, nearly transparent sightlines, the \fesc\ predicted from global parameters could be a better estimate of these galaxies' global average \fesc. The strongest LCEs in the LzLCS+ sample do tend to have high nebular EWs (EW(H$\beta \gtrsim 200$ \AA; \citealt{flury22b}), which shows that they cannot be devoid of absorbing gas in all directions.

Only three variables are statistically significant coefficients in the JWST model: $\beta_{\rm 1550}$, \logten(\sigsfr), and \logten(O32). These same three parameters are ranked as the most important in forward and backward selection and have the lowest mean ranks from forward and backward selection with MC sampling (Table~\ref{table:rankorder}, Figure~\ref{fig:mcrank}). When we limit the model to just these three variables, the fit quality is similar to or better than the JWST model with the full set of variables (Table~\ref{table:metrics}). We also find that the $C_{\rm CV}$ for the cross-validation analysis is closer to the $C$ derived from the full sample; with fewer variables in the fit, even a smaller training sample can generate a reliable model. We display the predicted \fesc\ from this model in Figure~\ref{fig:jwstmodel}b. Using \neiii/\oii\ instead of O32 results in a similar quality fit (Table~\ref{table:metrics}) and may be more suitable for observations covering a limited spectral range, observations at $z\gtrsim11$ where \oiii\ is redshifted out of the NIRSpec wavelength range, or observations with uncertain nebular dust attenuation \citep{levesque14}. These fits demonstrate that the multivariate Cox model can predict \fesc\ for high-redshift galaxies using {\it JWST} observables. We apply Cox models to high-redshift galaxy samples in a forthcoming paper (Jaskot et al., in prep.), where we also provide all parameters needed to apply these models to future samples.

Our JWST model includes similar variables as the multivariate \fesc\ predictions derived by \citet{choustikov24} from the SPHINX simulations, but we find a different dependence on some of these variables. \citet{choustikov24} use the following observables: UV slope $\beta$, E(B-V)$_{\rm neb}$, H$\beta$ luminosity, EW(H$\beta$), UV magnitude, R23$ = $(\oiii~$\lambda\lambda$5007,4959+\oii$\lambda$3727)/H$\beta$, O32, and half-light radius. Like the \citet{choustikov24} model, we find that $\beta_{\rm 1550}$ is statistically significant and that high \fesc\ is associated with blue UV slopes. We likewise find that O32 is statistically significant. However, in the LzLCS+ sample, O32 correlates with \fesc, whereas the SPHINX simulations find an anti-correlation. Strong LCEs with high O32 do not appear within the SPHINX galaxy population \citep{choustikov24}. This disagreement may result from the different properties of the observed vs.\ simulated galaxy populations, such as the stellar mass range. Alternatively, it may indicate the need for more efficient radiative feedback and/or resolving smaller-scale turbulent gas structure, to allow LyC escape from younger clusters \citep[\eg,][]{kimm19, kakiichi21, choustikov24}. 

Our model also disagrees with \citet{choustikov24} regarding the role of \sigsfr. Whereas we find that \sigsfr\ is one of the most important predictors of \fesc\ in the JWST model and is statistically significant, \citet{choustikov24} find that compactness is insignificant and shows little relationship with \fesc. One of main differences between the LzLCS+ and SPHINX galaxy datasets is the mass range probed; the LzLCS+ sample has a median mass of \logten($M_*$/\Msol)$=8.8$ and extends up to \logten($M_*$/\Msol)$=10.8$, whereas the majority of the SPHINX galaxies have masses \logten($M_*$/\Msol)$<9$. At even lower masses, \logten($M_*$/\Msol)$<7$, simulations suggest that LyC escape may occur in a more extended mode, driven by star clusters in low-column density regions in the outskirts of galaxies \citep{kostyuk23}. The LzLCS+ likewise hints that the link between \sigsfr\ and \fesc\ may be mass dependent. Several of the LCEs with high \sigsfr\ in the LzLCS+ sample have high masses (\logten($M_*$/\Msol)$=9.7-10.5$), and as we discuss in the following section, \sigsfr\ may be a particularly significant predictor of \fesc\ in high-mass galaxies.

\subsection{Variations with Galaxy Properties}
\label{sec:subsamples}
\citet{flury22b} discuss the possibility that the LyC escape process may vary among different types of galaxies. For example, the dominant feedback mechanism may shift from radiative feedback to mechanical supernova feedback as a starburst ages or in galaxies with different masses or metallicities \citep[\eg,][]{jaskot19, kimm19, jecmen23}. In addition, the strongest LCEs could have a nearly density-bounded gas geometry, whereas LyC photons may escape along narrow channels in weaker LCEs \citep[\eg,][]{gazagnes20, flury22b}. We therefore explore whether the Cox model predictions change if we limit our analysis to different subpopulations of the larger LzLCS+ sample. We first divide our galaxy sample into two bins, above and below the median value of stellar mass, \logten($M_*$/\Msol)$=8.8$. 

Different variables appear important for the low- vs.\ high-mass sample fits. For the fiducial model, \sigsfr\ is statistically significant in the high-mass fit only, with a coefficient value $\sim5\times$ higher than that found in the low-mass fit. Conversely, $M_{\rm 1500}$ and O32 are only statistically significant in the low-mass fit, with coefficients $\sim8$ and $\sim5\times$ stronger than in the high-mass fit. The same pattern occurs with the JWST models, with \sigsfr\ significant in the high-mass fits and  $M_{\rm 1500}$ and O32 significant in the low-mass fits. 

\begin{figure*}
\gridline{\fig{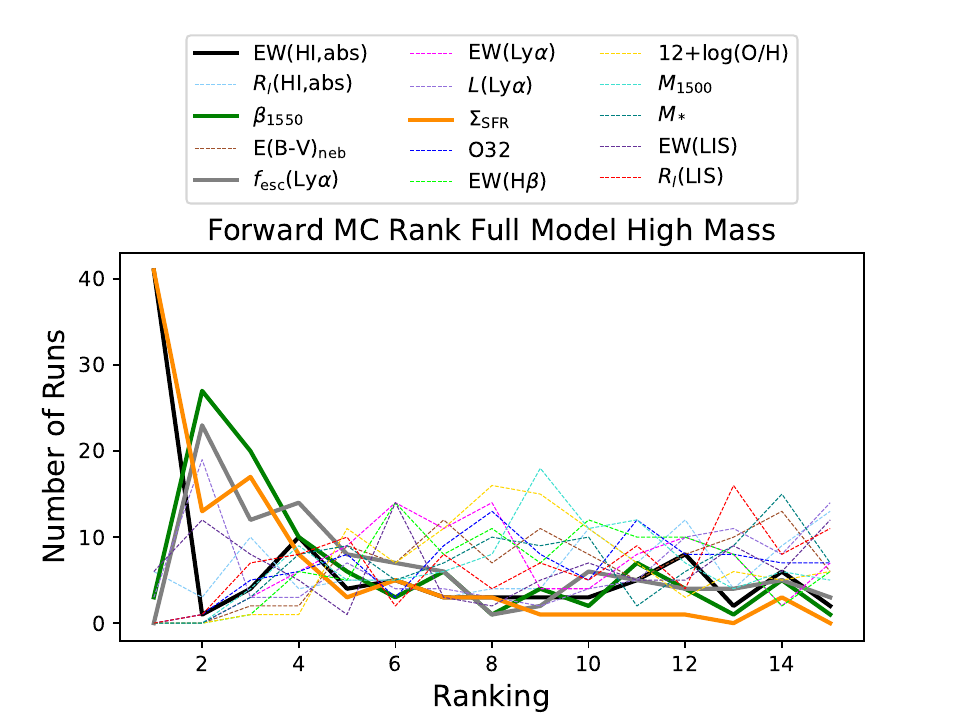}{0.5\textwidth}{(a)}
	\fig{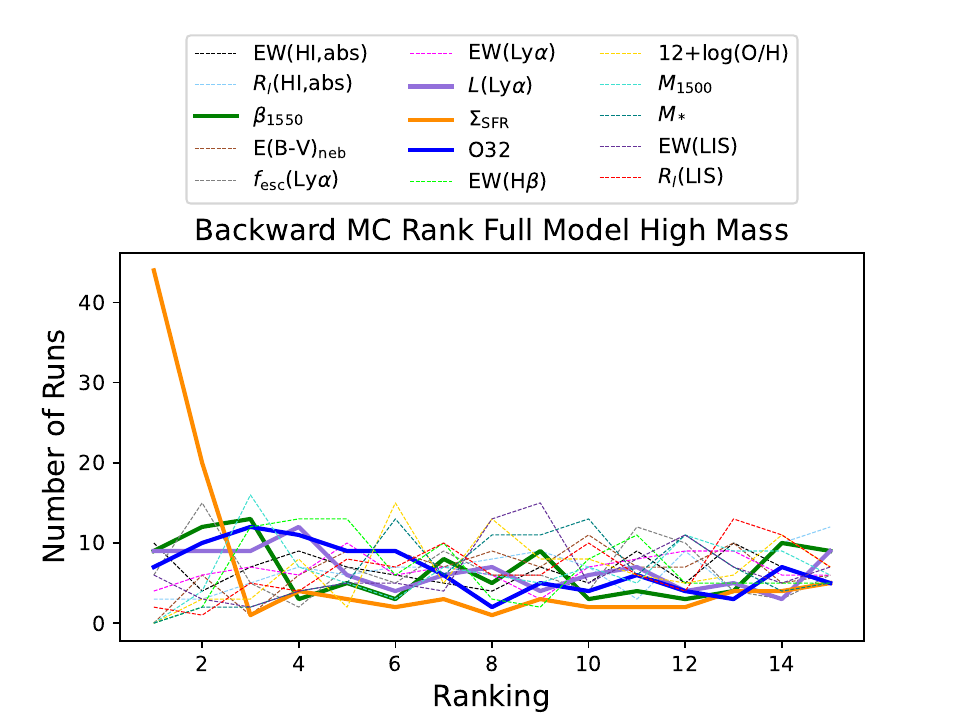}{0.5\textwidth}{(b)}
}
\gridline{\fig{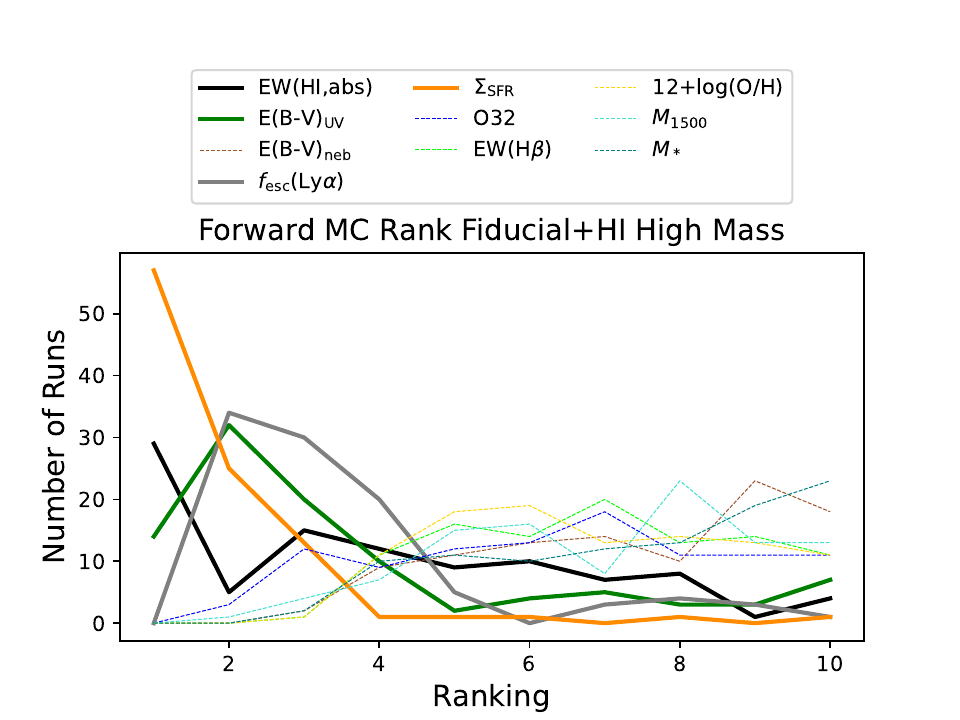}{0.5\textwidth}{(c)}
	\fig{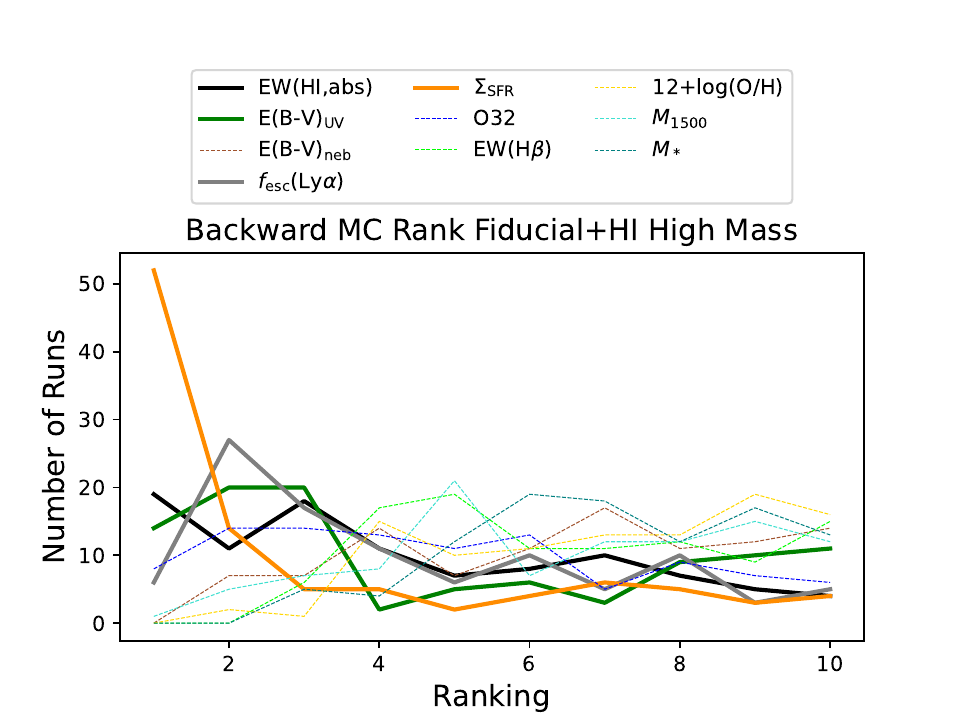}{0.5\textwidth}{(d)}
}
\gridline{\fig{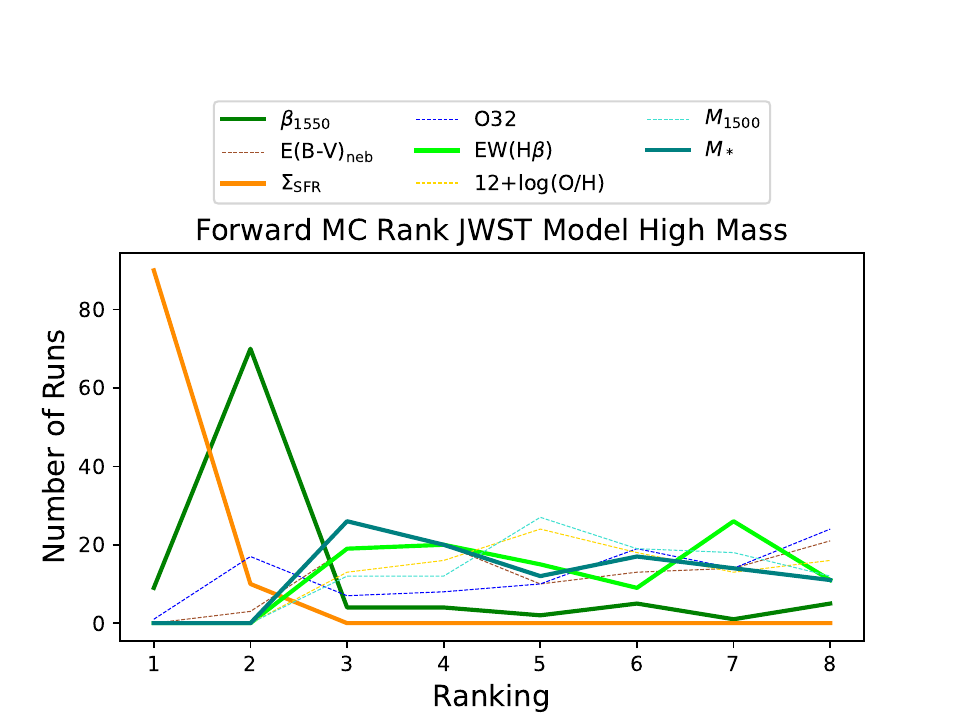}{0.5\textwidth}{(e)}
	\fig{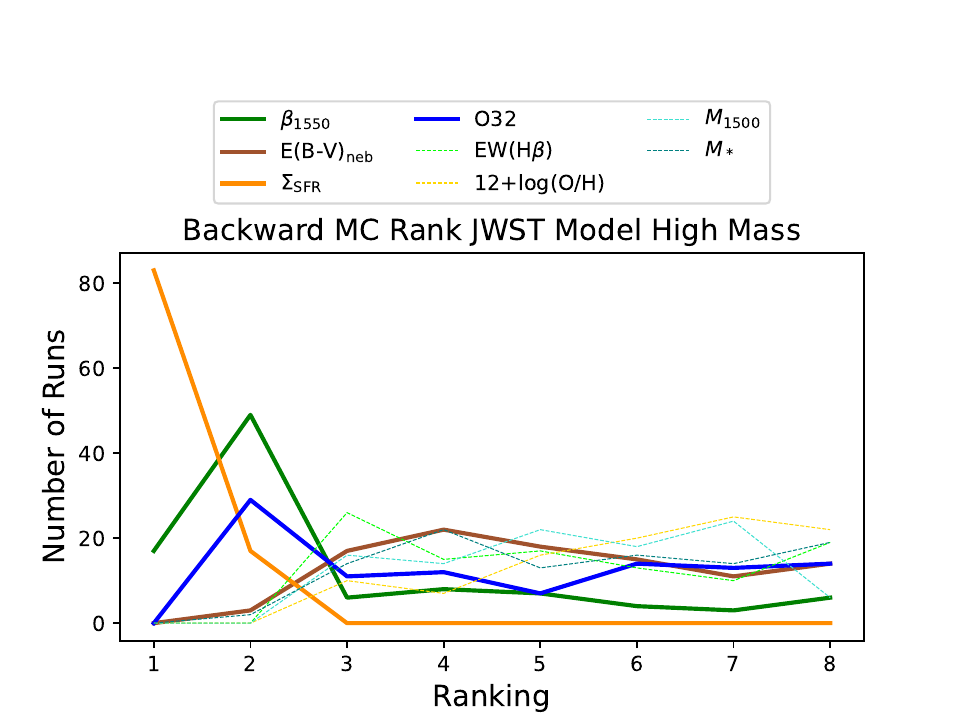}{0.5\textwidth}{(f)}
	}
\caption{The distribution of MC variable rankings from forward selection (left) and backward selection (right) for galaxies with \logten($M_*$/\Msol)$\geq8.8$. Variable combinations are the same as in Table~\ref{table:rankorder}. Thick solid lines indicate the four variables with the lowest (best) mean MC ranks.
\label{fig:mcrank_highm}}
\end{figure*}

\begin{figure*}
\gridline{\fig{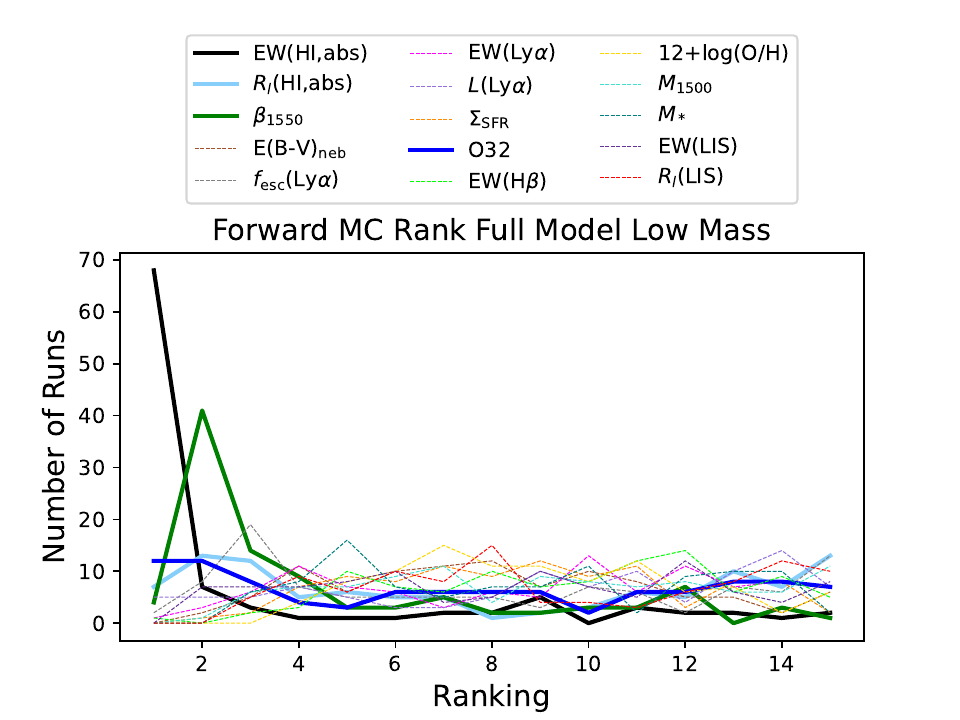}{0.5\textwidth}{(a)}
	\fig{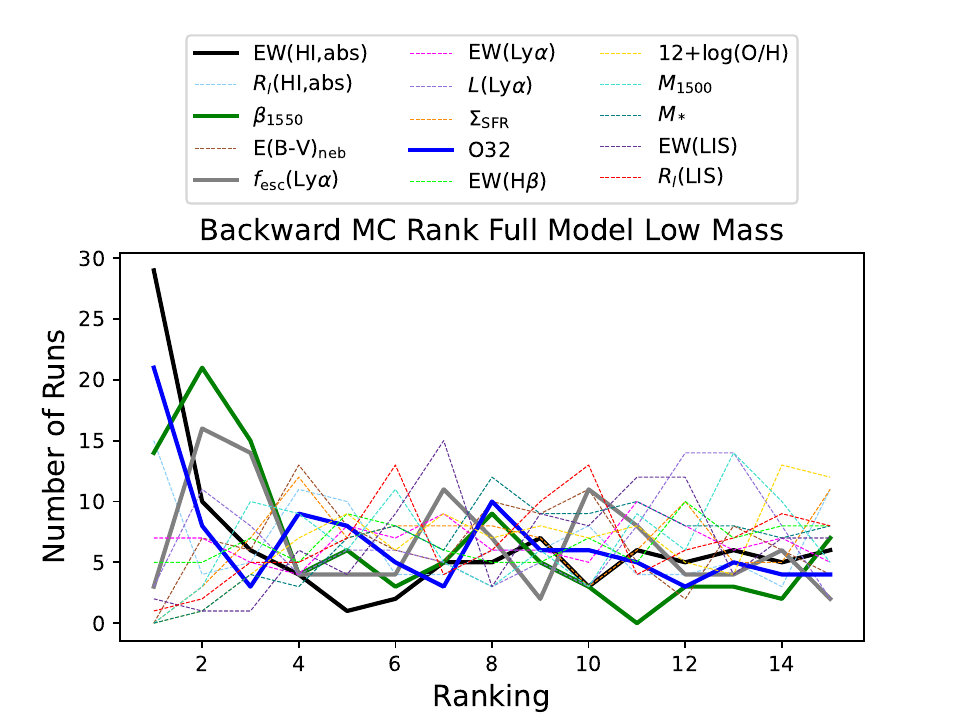}{0.5\textwidth}{(b)}
	}
\gridline{\fig{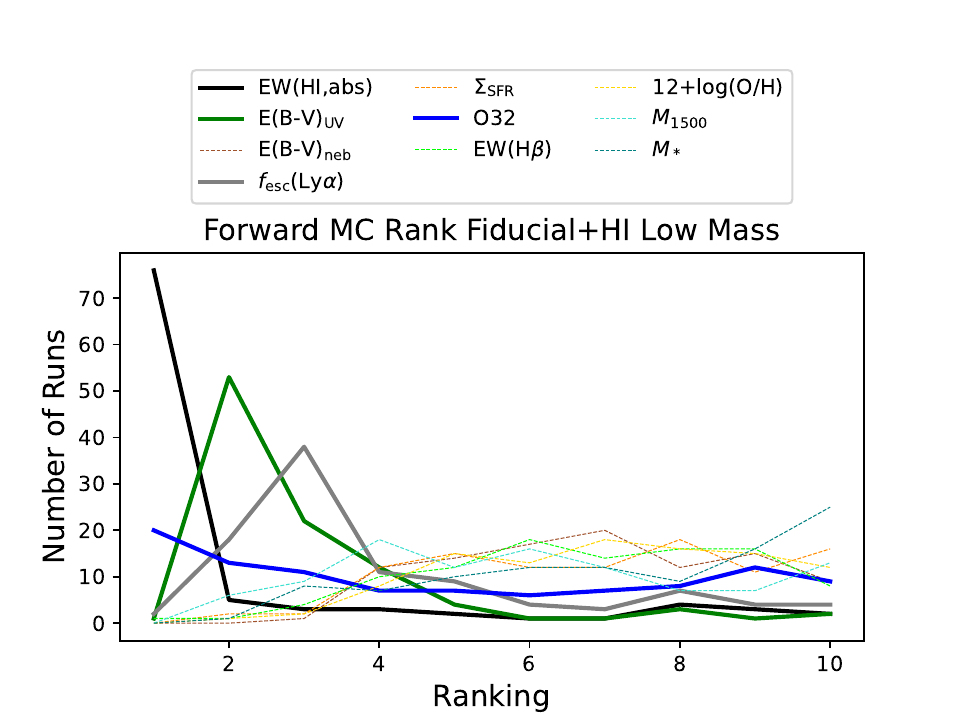}{0.5\textwidth}{(c)}
	\fig{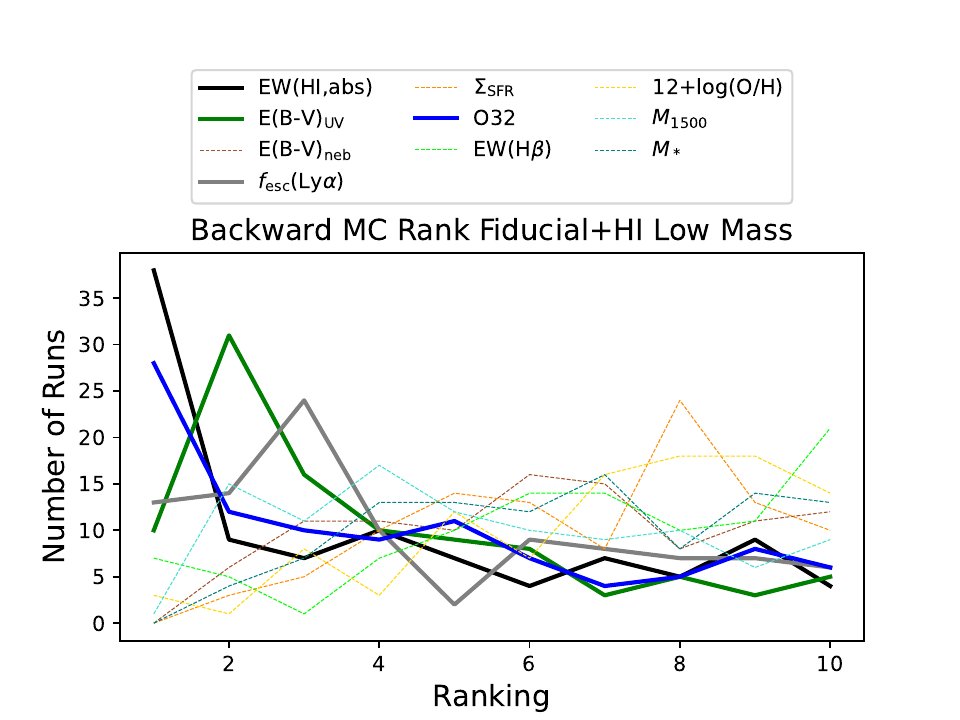}{0.5\textwidth}{(d)}
	}
\gridline{\fig{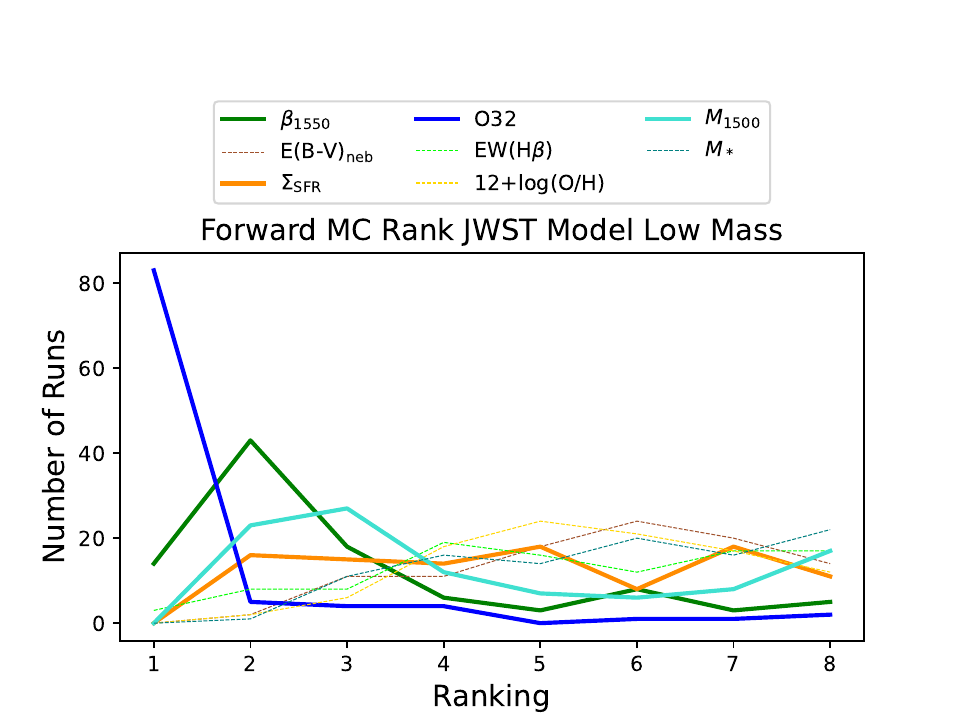}{0.5\textwidth}{(e)}
	\fig{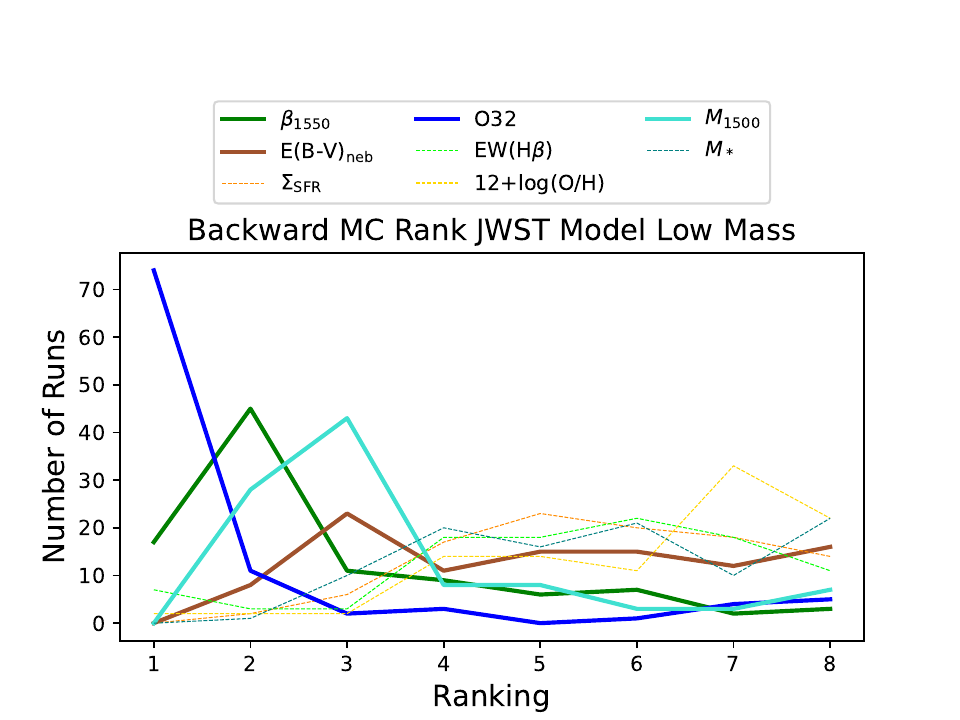}{0.5\textwidth}{(f)}
}
\caption{The distribution of MC variable rankings from forward selection (left) and backward selection (right) for galaxies with \logten($M_*$/\Msol)$<8.8$. Variable combinations are the same as in Table~\ref{table:rankorder}. Thick solid lines indicate the four variables with the lowest (best) mean MC ranks.
\label{fig:mcrank_lowm}}
\end{figure*}

The variable ranks by forward and backward selection (Tables~\ref{table:rankorder_highm}-\ref{table:rankorder_lowm}, Figures~\ref{fig:mcrank_highm}-\ref{fig:mcrank_lowm}) likewise suggest that different variables are important for the two subsamples. As with the full sample (Table~\ref{table:rankorder}), variables tracing dust attenuation, \hi\ absorption, and \lya\ emission are some of the most important variables for both the high- and low-mass subsamples. However, for the high-mass subsample (Table~\ref{table:rankorder_highm}), \logten(\sigsfr) consistently ranks as the most important variable in all models and in its mean MC rank values. In the low-mass subsample (Table~\ref{table:rankorder_lowm}), \logten(O32) actually appears at the bottom of the ranked list for the full set of variables (rank 15), but by mean MC rank, \logten(O32) is one of the four most important variables in all models and is the most important variable in the JWST model runs. This seemingly discrepant rank may suggest that O32 provides similar information as some of the other top-ranked variables. Figure~\ref{fig:mcrank_lowm} shows that O32 replaces EW(\hi,abs) as the top-ranked variable $\sim$10-25\%\ of the time, and high O32 may generally imply weak \hi\ absorption. In the JWST model, when EW(\hi,abs) is not included, O32 takes over as the most important variable in a majority of the MC runs. To test whether the smaller sample size of the high-mass and low-mass subsets affects these results, we re-run the MC variable ranking using bootstrap resampling to change the selected sample. The resulting ranking distributions are similar to those in Figures~\ref{fig:mcrank_highm}-\ref{fig:mcrank_lowm} with only minor changes to the top four ranked variables. Our main conclusions remain unchanged. \sigsfr\ consistently holds the top MC rank for every model for the high-mass subset. Dust, \hi\ absorption, \lya\ and O32 are important for the low-mass subset, and O32 is the top-ranked variable for the low-mass JWST model.

The separate high- and low-mass models do a slightly better job at reproducing the observed \fesc\ values in LCEs than the full fiducial model with all galaxies included. The separate models show a slightly higher $R^2=0.64-0.70$ and lower RMS$=0.28-0.32$. However, the $R^2_{\rm adj}$ values of 0.31-0.43 are lower than the full model $R^2_{\rm adj}=0.49$, which suggests the improvement mainly comes from fitting a smaller sample. One exception is the JWST Model, where the high-mass subsample predictions are substantially improved compared to the full model; $R^2$ rises to 0.54 from 0.29, $R^2_{\rm adj}$ is 0.23 instead of 0.14, the RMS drops from 0.47 to 0.34, and $C$ increases from 0.83 to 0.90. The \sigsfr\ variable takes on greater importance in the JWST model, since it lacks \hi\ and \lya\ information. The high-mass subsample \fesc\ depends strongly on \sigsfr, and the best-fit coefficient for \sigsfr\ increases by a factor of four compared to the full sample JWST model. Evidently \sigsfr\ is key for accurately predicting \fesc\ in the higher mass galaxies within the LzLCS+. 

Interestingly, the importance of \sigsfr\ vs.\ O32 for different subsamples only emerges when we split the sample by stellar mass, not by UV luminosity. Almost all the highly ionized galaxies, with O32$\gtrsim$5 are in the lower stellar mass bin. Some of these galaxies can be quite luminous for their mass, so they are not as distinguishable by luminosity. The galaxies with the highest O32 ratios also have high nebular emission line EWs, suggestive of young ages. Lower-mass galaxies may have more bursty star formation histories in general \cite[\eg,][]{lee07}. Among the low-mass galaxies in the LzLCS+, the high O32, high EW galaxies may represent currently bursting galaxies with extremely young average ages for their UV-emitting stellar populations \citep[\eg,][]{izotov11}.

The noticeably different roles of the \sigsfr\ and O32 variables in the high- vs. low-mass samples suggest that the causes of LyC escape may indeed differ between these galaxy populations. In the low-mass sample, radiative feedback in young starbursts may drive high LyC escape over wide opening angles \citep[\eg,][]{flury22b}. Galaxies with high luminosities and highly ionized gas would have higher \fesc. At the same time, if the youngest stellar population dominates the UV light output, the measured line-of-sight dust attenuation and \hi\ absorption would also be representative of the conditions near the LyC-emitting stars and would be sufficient to constrain \fesc. In contrast, escape in higher-mass galaxies, with potentially longer-duration starburst episodes may rely on the cumulative effect of supernova feedback punching holes in the interstellar medium (ISM; e.g., \citealt{flury22b}). The ability to drive these required outflows may be linked to the galaxy's \sigsfr \citep[\eg,][]{heckman01, heckman11, kim20}. If the size of these holes is sufficiently small, the measured dust attenuation and \hi\ EW across the full starburst may not reflect the conditions at the LyC escape site, especially if the youngest LyC-emitting populations do not dominate the total UV light. This disconnect could account for the reduced importance of $\beta_{\rm 1550}$ and EW(\hi,abs) and enhanced importance of \sigsfr in the high-mass models. 

\movetabledown=2.25in
\begin{rotatetable*}
\begin{deluxetable*}{lllllll}
\tablecaption{Most Important Variables for High-Mass Sample (\logten($M_*$) $\geq 8.8$)}
\label{table:rankorder_highm}
\tablehead{
\colhead{Rank} & \multicolumn{2}{c}{Full Model} & \multicolumn{2}{c}{Fiducial+HI Model} & \multicolumn{2}{c}{JWST Model} \\
\colhead{} & \colhead{Forward} & \colhead{Backward} & \colhead{Forward} & \colhead{Backward} & \colhead{Forward} & \colhead{Backward}}
\startdata
\multicolumn{7}{c}{} \\
\multicolumn{7}{c}{Top-Ranked Variables} \\
\hline
1 & \logten(\sigsfr)  & \logten(\sigsfr) & \logten(\sigsfr) & \logten(\sigsfr) & \logten(\sigsfr) & \logten(\sigsfr) \\
2 & \fesclya  &  \fesclya   & \fesclya  & \fesclya\ & $\beta_{\rm 1550}$ & $\beta_{\rm 1550}$ \\
3 & $\beta_{\rm 1550}$ & $\beta_{\rm 1550}$ & E(B-V)$_{\rm UV}$ & E(B-V)$_{\rm UV}$ & \logten($M_*$) & \logten($M_*$) \\
4 & $R_l$(\hi,abs) & $R_l$(\hi,abs) & EW(\hi,abs) & EW(\hi,abs) & $M_{\rm 1500}$ & $M_{\rm 1500}$ \\
\multicolumn{7}{c}{} \\
\multicolumn{7}{c}{Top Variables by MC Rank Order} \\
\hline
& \logten(\sigsfr) [3.24] & \logten(\sigsfr) [4.35] & \logten(\sigsfr) [1.79] & \logten(\sigsfr) [2.98] &  \logten(\sigsfr) [1.10] & \logten(\sigsfr) [1.17] \\
& $\beta_{\rm 1550}$ [5.34] & \logten(O32) [6.75] & \fesclya\ [3.53] & EW(\hi,abs) [4.33] & $\beta_{\rm 1550}$ [2.64] & $\beta_{\rm 1550}$ [2.93] \\
& EW(\hi,abs) [5.38] & $L$(\lya) [7.17] & E(B-V)$_{\rm UV}$ [3.68] & \fesclya\ [4.37]  & \logten($M_*$) [5.06] & \logten(O32) [4.61] \\
& \fesclya\ [6.18] & $\beta_{\rm 1550}$ [7.41]& EW(\hi,abs) [3.99] & E(B-V)$_{\rm UV}$ [4.76] & \logten(EW(H$\beta$)) [5.36]  & E(B-V)$_{\rm neb}$ [5.14]  \\
\enddata
\tablecomments{Numbers in brackets indicate the mean of the variable's ranks in the 100 MC runs.}
\end{deluxetable*} %
\end{rotatetable*}

\movetabledown=2.25in
\begin{rotatetable*}
\begin{deluxetable*}{lllllll}
\tablecaption{Most Important Variables for Low-Mass Sample (\logten($M_*$) $< 8.8$)}
\label{table:rankorder_lowm}
\tablehead{
\colhead{Rank} & \multicolumn{2}{c}{Full Model} & \multicolumn{2}{c}{Fiducial+HI Model} & \multicolumn{2}{c}{JWST Model} \\
\colhead{} & \colhead{Forward} & \colhead{Backward} & \colhead{Forward} & \colhead{Backward} & \colhead{Forward} & \colhead{Backward}}
\startdata
\multicolumn{7}{c}{Top-Ranked Variables} \\
\hline
1 & EW(\hi,abs) & EW(\hi,abs) & EW(\hi,abs)  & EW(\hi,abs) & $\beta_{\rm 1550}$ & $\beta_{\rm 1550}$ \\
2 & $\beta_{\rm 1550}$  & $\beta_{\rm 1550}$ & E(B-V)$_{\rm UV}$ & E(B-V)$_{\rm UV}$ & \logten(O32) & \logten(O32) \\
3 & \fesclya & $M_{\rm 1500}$ & \fesclya  & $M_{\rm 1500}$ & $M_{\rm 1500}$ & $M_{\rm 1500}$  \\
4 & \logten($M_*$) & EW(\lya) & \logten(\sigsfr) & \logten(O32) & \logten(EW(H$\beta$)) & E(B-V)$_{\rm neb}$\\
\hline
\multicolumn{7}{c}{Top Variables by MC Rank Order} \\
\hline
 & EW(\hi,abs) [3.08]  & $\beta_{\rm 1550}$ [5.55] & EW(\hi,abs) [2.09] & EW(\hi,abs) [3.86] & \logten(O32) [1.50] &  \logten(O32) [1.88]  \\
 & $\beta_{\rm 1550}$ [4.70]  & EW(\hi,abs) [6.32]   & E(B-V)$_{\rm UV}$ [3.07] & E(B-V)$_{\rm UV}$ [3.91]  & $\beta_{\rm 1550}$ [3.02] & $\beta_{\rm 1550}$ [2.86] \\
 & \logten(O32) [7.50] & \logten(O32) [6.39] & \fesclya\ [4.18] & \logten(O32) [4.15] & $M_{\rm 1500}$ [4.38] & $M_{\rm 1500}$ [3.52]  \\
 & $R_l$(\hi,abs) [7.89] & \fesclya\ [7.00]  & \logten(O32) [4.89] & \fesclya\ [4.52] & \logten(\sigsfr) [4.85] & E(B-V)$_{\rm neb}$ [5.06] \\
\enddata
\tablecomments{Numbers in brackets indicate the mean of the variable's ranks in the 100 MC runs.}
\end{deluxetable*} %
\end{rotatetable*}

To explore variations related to feedback, we also split the sample based on the median values of \logten(\sigsfr)=0.77 and O32=3.4. While splitting the LzLCS+ sample on mass resulted in slightly different variable selections, the high O32 and high \sigsfr\ subsample fiducial fits more closely resemble the full model, likely because they contain nearly the same population of strong LCEs; the high \sigsfr\ and high O32 subsets each include all but two of the strong LCEs with \fesc\ $>0.1$. The two most important predictors of \fesc\ remain \hi\ absorption (either $R_l$(\hi,abs) or EW(\hi,abs)) and dust attenuation. Interestingly, \sigsfr\ is no longer a statistically significant variable in the fiducial and JWST models for the high \sigsfr\ and high O32 subsamples. \sigsfr\ may be more useful for identifying non-leakers than in actually quantifying \fesc. Indeed, LCEs in the LzLCS+ sample nearly all lie above a threshold value of \sigsfr\ $=10$ \Msol\ yr$^{-1}$ kpc$^{-2}$, but above this value, no discernible relationship exists between \fesc\ and \sigsfr\ \citep{flury22b}. 

For the low \sigsfr\ subsample, only \fesclya\ and E(B-V)$_{\rm UV}$ are statistically significant in the fiducial model, while these same variables plus E(B-V)$_{\rm neb}$ are significant for the low O32 subset. As usual, in the variable rankings, EW(\hi,abs) and dust measurements are typically important variables. However, \lya\ measurements also appear to be relevant predictors. For the Full Model variable set, $L$(\lya) is the top-ranked variable for the low O32 subsample. Similarly, for the low \sigsfr\ subsample, \fesclya\ is the top-ranked variable for the Fiducial+HI model and the top-ranked variable in forward modeling for the Full Model. Because these subsamples contain few strong LCEs, this result emphasizes that \lya\ emission may be useful in distinguishing weak LCEs from non-leakers.

\subsection{Outliers in \fesc\ Trends}
\label{sec:outliers}
Many studies have identified relationships between \fesc\ and physical or observable parameters \citep[\eg,][]{verhamme17,izotov18b,flury22b,saldana22,chisholm22}, yet even the best trends show significant scatter, with individual galaxies showing \fesc\ significantly above or below the expected value. \citet{schaerer22} highlight one outlier galaxy in particular, J1248+4259. J1248+4259's high nebular ionization and strong \lya\ emission resemble those of other strong low-redshift LCEs, yet its \fesc\ is $\leq$1.3\%. Here, we investigate whether the Cox model improves \fesc\ predictions for J1248+4259 and other outlier galaxies.

We select outliers based on the variables that show the strongest trends with \fesc\ in the \citet{flury22b} analysis of the LzLCS+ sample: $r_{\rm 50,NUV}$, O32, \sigsfr/$M_*$, EW(\lya), \sigsfr, and \fesclya. From the strong LCEs (\fesc\ $\geq 0.05$) in the LzLCS+ sample, we choose the three  with the highest $r_{\rm 50,NUV}$, the three lowest in O32, the three lowest in \sigsfr/$M_*$, the three lowest in EW(\lya), the three lowest in \sigsfr, and the three lowest in \fesclya. Conversely, we choose three weak or non-LCEs (\fesc\ $< 0.05$) with the lowest $r_{\rm 50,NUV}$, highest O32, highest \sigsfr/$M_*$, highest EW(\lya), highest \sigsfr, or highest \fesclya. This selection gives us a list of 10 strong LCEs and 13 weak or non-LCEs that are outliers by one or more selection. In other words, applying a simple, single-variable trend as in \citet{flury22b} would incorrectly identify these galaxies as strong LCEs or as non-leakers. 

\begin{figure*}
\gridline{\fig{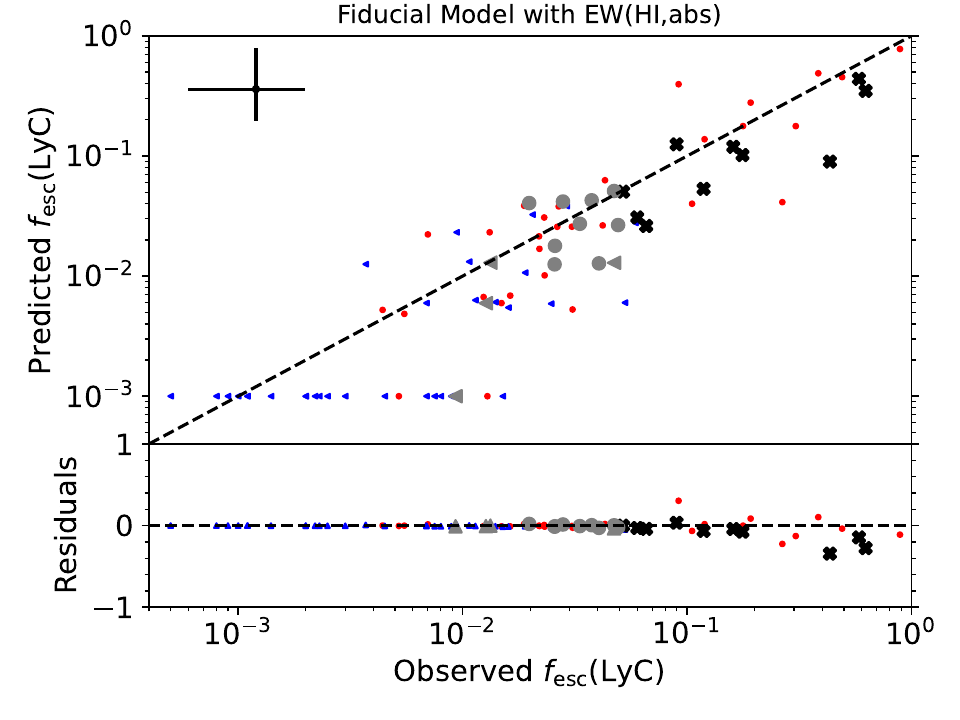}{0.5\textwidth}{(a)}
	\fig{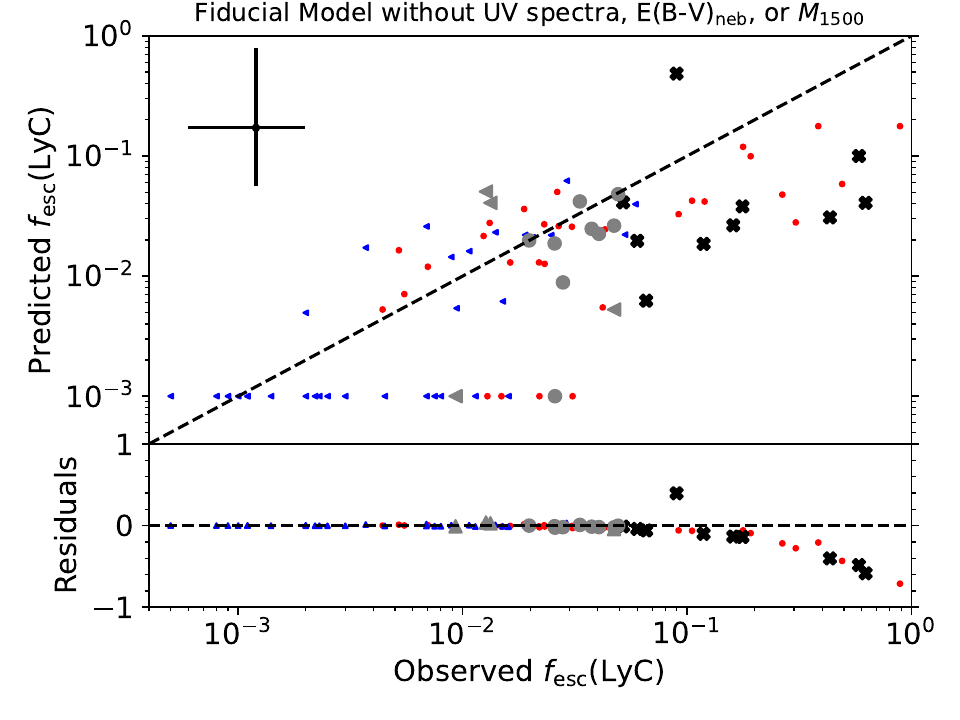}{0.5\textwidth}{(b)}}
\caption{The \fesc\ predictions from the fiducial model (a) modified to use EW(\hi,abs) and (b) modified to exclude information about the line of sight from \lya, absorption lines, dust attenuation, or the observed UV luminosity. Black crosses show strong LCEs that are outliers by one or more selection criteria; gray circles and triangles show weak and non-LCEs that are outliers by one or more selection criteria. Small red circles and blue triangles represent the other detections and non-detections in the LzLCS+ sample. The fiducial+EW(\hi,abs) model reproduces the \fesc\ of the outliers, whereas the model without line-of-sight information does not.
\label{fig:outliers}}
\end{figure*}

Figure~\ref{fig:outliers}a shows that the fiducial model modified to include EW(\hi,abs) predicts \fesc\ for these ``outliers" quite well. In fact, the outlier predictions are generally no more or less accurate than the predictions for other galaxies in the LzLCS+ sample. For example, the outlier galaxy J1248+4259, which ``should" be a strong LCE based on its high O32 ratio and high EW(\lya), has a predicted \fesc\ of 0.006 according to the model, consistent with its observed limit of \fesc$\leq$0.013. 
 
Crucially, the fiducial+EW(\hi,abs) model includes information about the optical depth along the line of sight, which is necessary to predict the \fesc\ observed along the same line of sight. As discussed in Sections~\ref{sec:modified_fiducial} and \ref{sec:results:variablerank}, EW(\hi,abs), \fesclya, and E(B-V)$_{\rm UV}$, all of which are sightline dependent, have some of the largest effects on the accuracy of the model predictions. If we remove these and other sightline-dependent parameters (E(B-V)$_{\rm neb}$, $M_{\rm 1500}$) from the fit (Figure~\ref{fig:outliers}b), we can no longer reproduce the line-of-sight \fesc\ of many galaxies. Figure~\ref{fig:outliers}b indicates that we would expect a different \fesc\ for these galaxies based on their global properties. Hence, if the only cause of the scatter in Figure~\ref{fig:outliers}b is chance galaxy orientation, the predicted \fesc\ should represent the global average across all sightlines. By this model, J1248+4259 should indeed be an LCE, although not an exceptionally strong one, with \fesc\ $= 0.051$. 

At high \fesc, the model without line-of-sight information seems to systematically under-predict the observed \fesc, which may suggest that the global properties of the strongest LCEs are not distinct from those of weaker LCEs. The observed properties of the strongest LCEs support this interpretation. For example, the strongest LCEs all tend to have high O32 \citep{flury22b}, yet the median observed \fesc\ for the 11 galaxies with O32 $\geq$10 is only 0.05. For these same 11 galaxies, the model without line-of-sight information finds the median of their predicted \fesc\ values to be 0.04, comparable to the observations. The only model that successfully reproduces the \fesc\ of the strongest LCEs is the fiducial+EW(\hi,abs) model. Hence, of the variables we consider, only EW(\hi,abs) distinguishes the strongest LCEs from other galaxies. Either the strongest LCEs differ in some other unknown global property or their high \fesc\ results from a favorable orientation compared to weaker LCEs. The high O32 galaxies may have \fesc$\sim0.04-0.05$ along most sightlines but may also contain several extremely optically thin channels, which lead to much higher observed \fesc, $\gtrsim0.5$ in some cases.

Although orientation bias likely causes some of the scatter in \fesc\ predictions, the Cox modeling results also suggest that some of the ``outlier" galaxies are not outliers at all, at least not when multiple global variables are considered. For example, J0919+4906 and J0232-0426 have high O32 ratios of 12.7 and 9.9 but \fesc\ of 0.049 and $<0.047$, respectively. However, both galaxies also have low stellar masses (\logten($M_*$)$=7.5$ and 7.4), such that their O32 values are not extreme compared to other galaxies of similar mass (see Figure 23 in \citealt{flury22b}). Furthermore, other criteria, such as their moderate or low \sigsfr\ values (\logten(\sigsfr)$=0.66$ and -0.39) would argue against a high \fesc. The combination of multiple properties leads to a moderate predicted \fesc\ for J0919+4906 (\fesc=0.048) and little escaping LyC (\fesc=0.005) for J0232-0426, even when the model includes only global, and not sightline-dependent, properties. Similarly, several galaxies have concentrated star formation (J0910+6105, J1209+3053, J1248+1234, J1349+5631, J1503+3644, J1517+3705) or strong \lya\ (J0122+0520, J0811+4141, J0901+2119, J1648+4957) but \fesc $< 0.05$; these galaxies' other properties yield consistent predictions of \fesc$ < 0.05$, even in models that neglect line-of-sight information.

The Cox model's success at reproducing \fesc\ even for these outliers demonstrates that accurately predicting \fesc\ requires multiple input variables. Moreover, because the observed \fesc\ depends on our line of sight, the best models also require sightline-specific information about optical depth from absorption lines, UV dust attenuation, and \lya. Although \lya\ and \hi\ absorption lines are inaccessible for most $z>6$ galaxies, high-redshift \fesc\ predictions can include information about UV dust attenuation as in the JWST models discussed in Section~\ref{sec:jwstmodels}. Information about dust attenuation is critical to predict \fesc, and large sample sizes can probe multiple random orientations to estimate the average \fesc\ for a galaxy population \citep[\eg,][]{cen15, saldana23}.

\section{Predicting Alternative Quantities}
\label{sec:results:other}
\subsection{Alternative Measures of LyC: Predicting \fratio\ and $L$(LyC)}
\label{sec:results:altlyc}
As shown above, multivariate Cox models can successfully predict \fesc; we now consider whether they can predict alternative measurements of the escaping LyC. To model how galaxies reionized the universe, we need to know the input rate of LyC photons into the IGM from different galaxy populations. Deriving this quantity from \fesc\ requires knowing the production rate of LyC photons in each galaxy, which is typically derived from SED fitting or Balmer line observations. To avoid this process and its associated systematics, we could instead predict a more directly observable quantity, such as the \fratio\ ratio or the total escaping LyC luminosity \logten($L$(LyC)). The \fratio\ flux ratio measures the ratio of observed LyC flux to the observed flux at 1100 \AA. As such, it depends on observed quantities and does not require an assumed stellar population model or dust attenuation law. To predict \fratio\ using the Cox proportional hazards model, we follow the same procedure as for \fesc\ and convert the \fratio\ measurements to right-censored data (i.e., detections and lower limits) by predicting 1-\fratio. Unlike \fesc\ and \fratio, which cannot be higher than 1, $L$(LyC) has no obvious maximum allowed value. To convert it to a right-censored format, we therefore adopt a maximum of \logten($L$(LyC))$ =  41$, greater than the maximum luminosity in the sample of \logten($L$(LyC))=40.6, and predict 41-\logten($L$(LyC). We present the goodness-of-fit metrics for the fiducial and JWST model predictions of \fratio, $L$(LyC), and \lya\ measurements in Table~\ref{table:metrics:altlyc}.

\begin{figure*}
\gridline{\fig{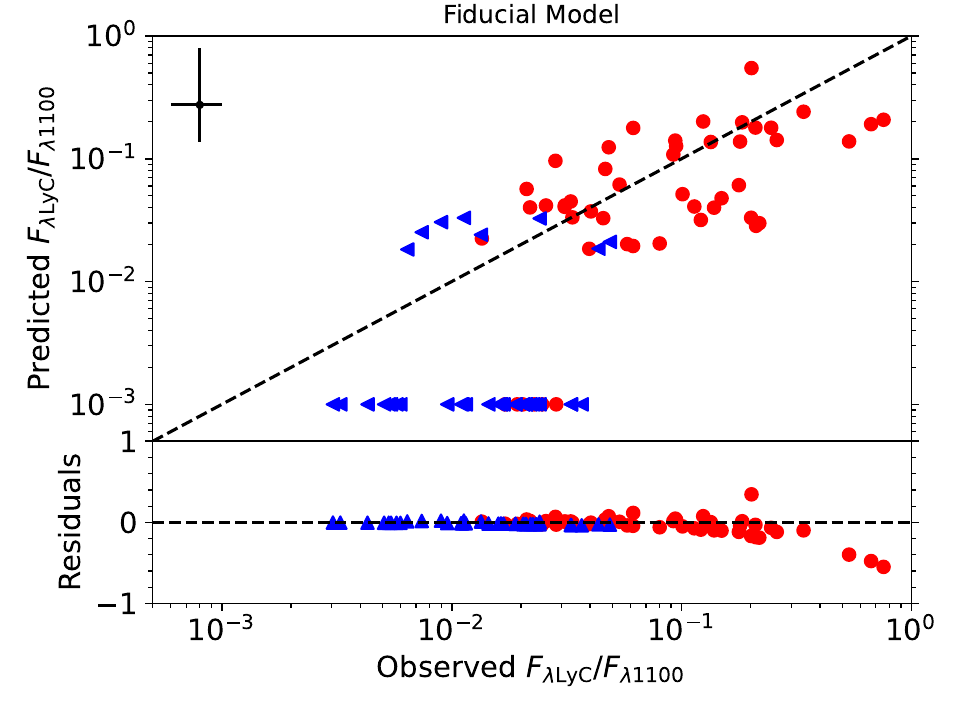}{0.5\textwidth}{(a)}
	\fig{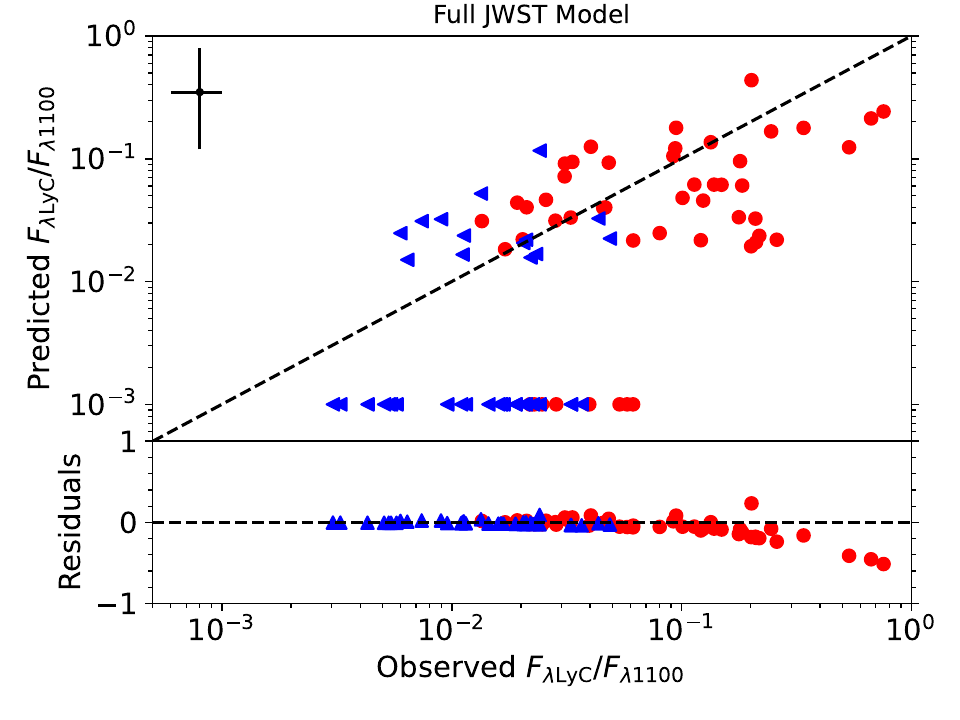}{0.5\textwidth}{(b)}}
\caption{Examples of \fratio\ model predictions using the variable sets for the fiducial model (a) and the JWST model (b). Symbols are the same as in Figure~\ref{fig:fiducialfit}. Predictions of \fratio\ are not as accurate as the model predictions for \fesc.
\label{fig:fratio}}
\end{figure*}

The \fratio\ models perform worse than the \fesc\ models for every variable combination we consider (see Figure~\ref{fig:fratio}) and by all three metrics ($R^2$, RMS, $C$). Although \fratio\ is simpler to measure, \fesc\ appears simpler to predict. It may be more directly linked to galaxy physical properties such as optical depth, porosity, and feedback. Furthermore, while measuring \fratio\ does not depend on assumptions about the galaxy SED, predicting it does. The galaxy SED shape changes with age and metallicity, which affects the ratio of \fratio\ even for \fesc $= 1$ \citep[see \eg,][]{chisholm19}. 

Although we derive both E(B-V)$_{\rm UV}$ and \fesc\ from the COS UV spectra, the greater accuracy of the \fesc\ Cox models does not appear to result from this interdependency. First, we see the same improvement in accuracy for \fesc\ vs.\ \fratio\ when we use $\beta_{\rm 1550}$ instead of E(B-V)$_{\rm UV}$; $\beta_{\rm 1550}$ is directly related to the observed spectral slope and is not sensitive to the choice of stellar population model or star formation history \citep{chisholm22}. Secondly, the \fesc\ models have higher $R^2$ and $C$ than the \fratio\ predictions, even for the models that contain no UV spectral information at all. We conclude that although \fratio\ is a useful quantity, accurately predicting it is difficult given its dependence on the intrinsic SED shape. We list and discuss the top-ranked variables for the \fratio\ predictions in the Appendix.

\begin{deluxetable*}{llllllll}
\tablecaption{Goodness-of-Fit Metrics for Different Dependent Variables}
\label{table:metrics:altlyc}
\tablehead{
\colhead{Dependent Variable} & \colhead{$R^2$} & \colhead{$R^2_{\rm adj}$} & \colhead{RMS} & \colhead{$C$} & \colhead{$R^2_{\rm CV}$} & \colhead{RMS$_{\rm CV}$} & \colhead{$C_{\rm CV}$}}
\startdata
\multicolumn{8}{c}{Fiducial Model \tablenotemark{a}} \\
\hline
\fesc\ & 0.60 & 0.49 & 0.36 & 0.89 & 0.55 & 0.36 & 0.85 \\ %U
\fratio\ & 0.07 & -0.22 & 0.40 & 0.85 & -0.04 & 0.40 & 0.81 \\
$L$(LyC) & -0.20 & -0.55 & 0.38 & 0.81 & -0.33 & 0.37 & 0.76\\
\fesclya\ & 0.33 & 0.24 & 0.32 & 0.75 & 0.29 & 0.32  & 0.71 \\
$L$(\lya)\tablenotemark{b} & 0.58 & 0.53 & 0.23 & 0.76 & 0.55 & 0.23 & 0.71 \\
\hline
\multicolumn{8}{c}{JWST Model}\\
\hline
\fesc\ & 0.29 & 0.14 & 0.47 & 0.83 & 0.21 & 0.47 & 0.79 \\%JWSTbeta
\fratio\ & -0.18 & -0.48 & 0.49 & 0.77 & -0.38 & 0.48 & 0.73 \\
$L$(LyC) & -0.25 & -0.53 & 0.39 & 0.78 & -0.34 & 0.38 & 0.74 \\
\fesclya\ & 0.07 & -0.04 & 0.38 & 0.75 & -0.01 & 0.37 & 0.72 \\
$L$(\lya)\tablenotemark{b} & 0.53 & 0.48 & 0.25 & 0.77 & 0.50 & 0.24 & 0.73 \\
\enddata
\tablenotetext{a}{The \fesclya\ variable is not included in the fiducial model for the \fesclya\ and $L$(\lya) predictions.}
\tablenotetext{b}{The $L$(\lya) model sample excludes six galaxies with \lya\ in absorption.}
\end{deluxetable*}

The \fratio\ predictions rely on much the same variables as the \fesc\ predictions, except for a weaker dependence on the UV dust attenuation, as expected. For the fiducial model, the same variables are statistically significant except for E(B-V)$_{\rm UV}$, and excluding E(B-V)$_{\rm UV}$ from the fiducial model changes $R^2$, RMS, and $C$ by only $\leq0.03$. However, when we substitute EW(\hi,abs) in the fiducial model, E(B-V)$_{\rm UV}$ does appear statistically significant, indicating that its influence is not always negligible. Table~\ref{table:rankorder_fratio} in the Appendix lists the top-ranked variables and likewise indicates a reduced role for dust attenuation in predicting \fratio\ compared to its role in predicting \fesc.

\begin{figure*}
\gridline{\fig{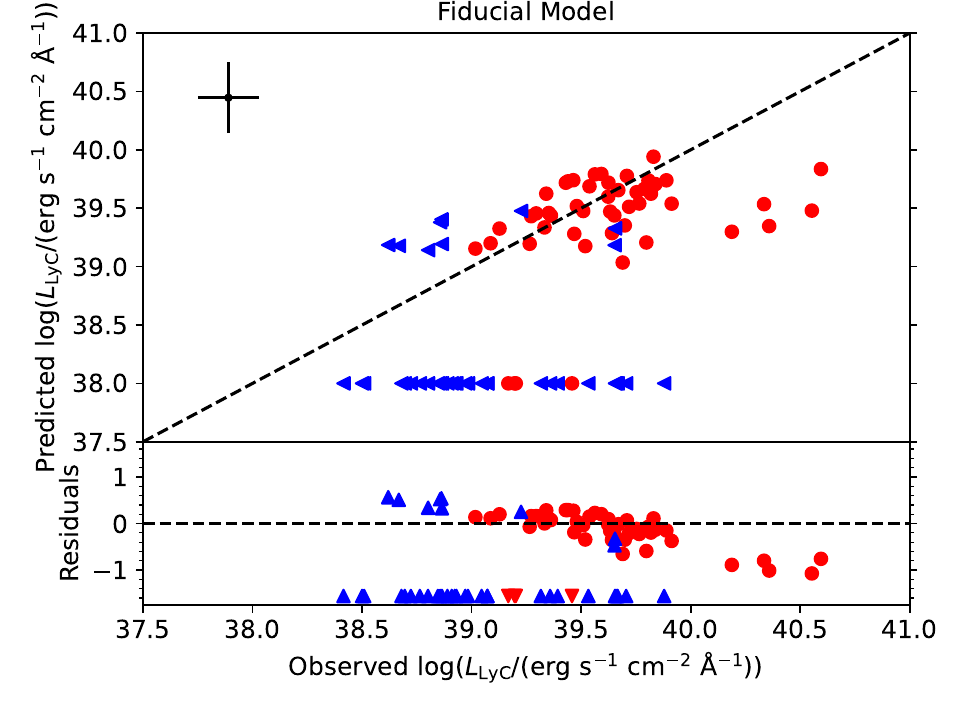}{0.5\textwidth}{(a)}
	\fig{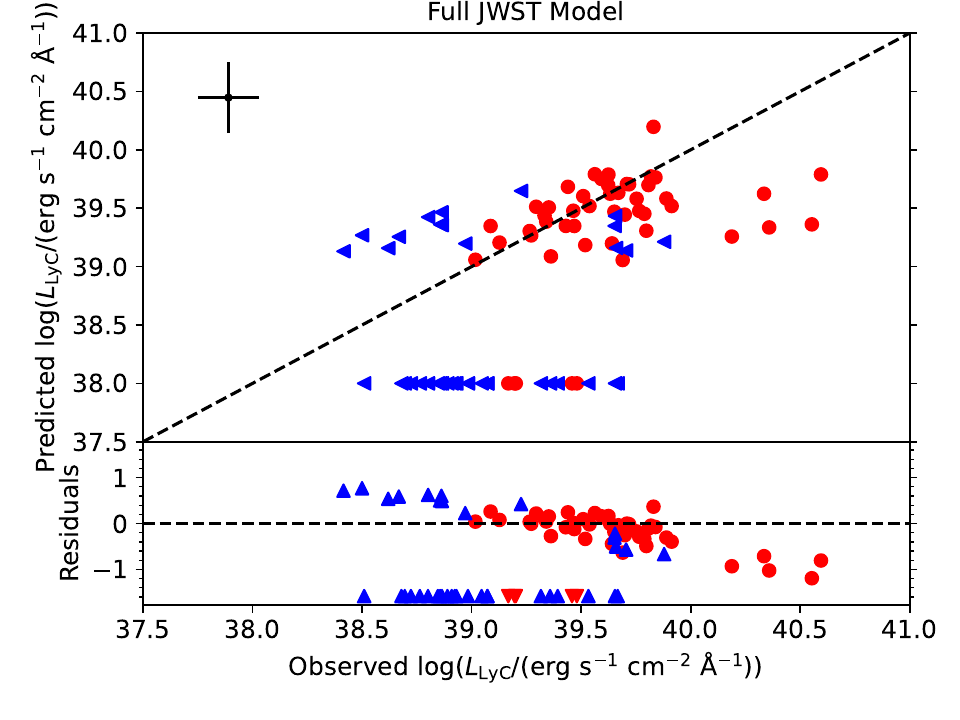}{0.5\textwidth}{(b)}}
\caption{Examples of $L$(LyC) model predictions using the variable sets for the fiducial model (a) and the JWST model (b). We plot galaxies with predicted $L$(LyC)$\sim0$ at the bottom of each panel. Symbols are the same as in Figure~\ref{fig:fiducialfit}. The models fail to accurately predict $L$(LyC).
\label{fig:llyc}}
\end{figure*}

Like \fratio, the predictions for $L$(LyC) (Figure~\ref{fig:llyc}, Table~\ref{table:metrics:altlyc}) are also less accurate than for the \fesc\ predictions. The concordance drops for every model, and $R^2$ is negative in all cases, meaning that simply adopting the mean $L$(LyC) for the detections would result in a better fit than using the model predictions. The negative $R^2$ arises from the fact that the $R^2$ calculation (Equation~\ref{eqn:r2}) uses only the detections and their mean, whereas the Cox model fits to the full set of galaxies; $R^2$ becomes positive if we use the median $L$(LyC) from the full sample of limits and detections, instead of the mean $L$(LyC) of the detections alone. Not surprisingly, other measures of luminosity show up as important variables for the $L$(LyC) models. $M_{\rm 1500}$ is statistically significant in the fiducial model and is one of top three ranked parameters by forward and backward selection, while $L$(\lya) holds the top place in forward selection for the full variable list (see Table~\ref{table:rankorder_llyc} in the Appendix for a list of top-ranked variables). However, the $L$(LyC) models' poor performance indicates that key information is missing. Specifically, a lack of detailed information about the SED likely prevents an estimate of the intrinsic $L$(LyC) from $M_{\rm 1500}$. The escaping $L$(\lya) also imperfectly traces the escaping $L$(LyC) because of scattering. In addition, the production of the \lya\ recombination line requires LyC absorption, which could cause $L$(\lya) to drop for the few extreme LCEs in the LzLCS+ sample \citep[\eg,][]{nakajima14}. In conclusion, neither the $L$(LyC) nor the \fratio\ models provide satisfactory alternatives to the \fesc\ predictions.

\subsection{Predicting \fesclya}
\label{sec:results:fesclya}
Although closely related to the \hi\ optical depth, the escape of \lya\ may depend on different parameters than the LyC escape fraction does. Here, we apply the Cox model to the \fesclya\ data for the LzLCS+ sample and assess the differences between predictions for \fesclya\ vs.\ \fesc. We use the same input variables, with one exception: we exclude any variables containing \lya\ measurements (\fesclya, EW(\lya), $L$(\lya)) from the models. Also, unlike our \fesc\ data, the \fesclya\ measurements include no upper limits, since \lya\ can appear in absorption or emission. For the negative values of \fesclya, representing net absorption, we cannot calculate \logten(\fesclya) and do not include these galaxies in the reported $R^2$ or RMS; they are included in the concordance.

\begin{figure*}
\gridline{\fig{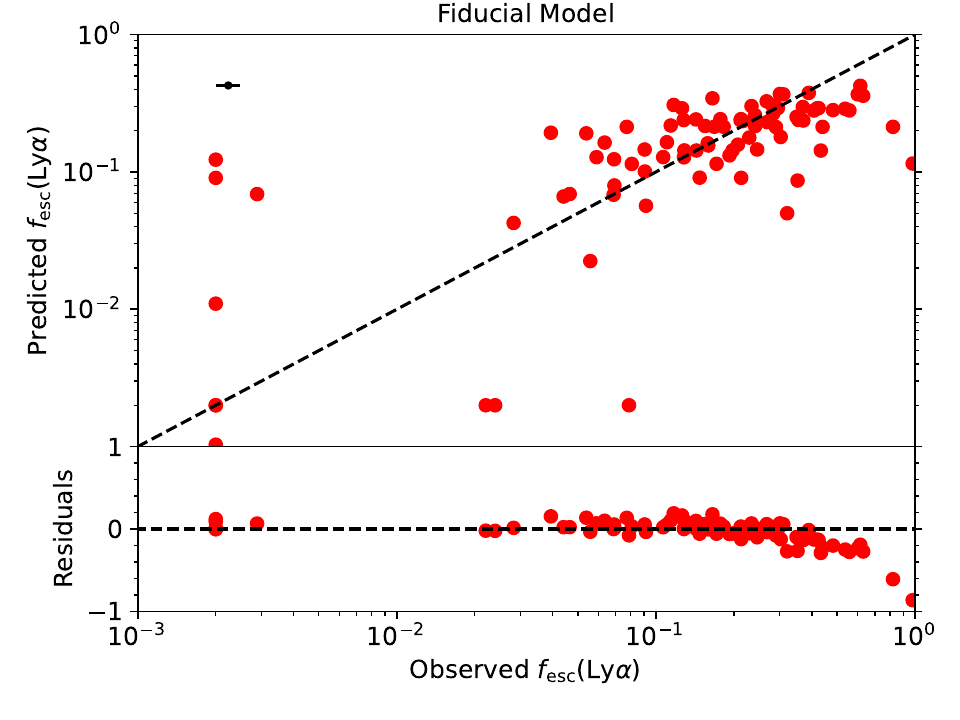}{0.3\textwidth}{(a)}
	\fig{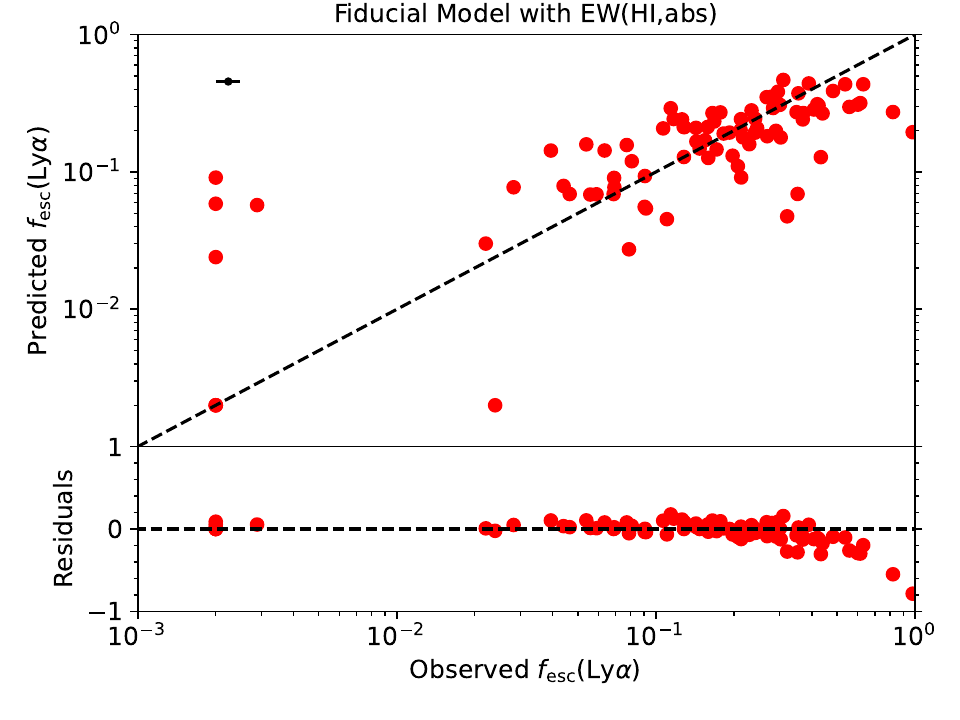}{0.3\textwidth}{(b)}
	\fig{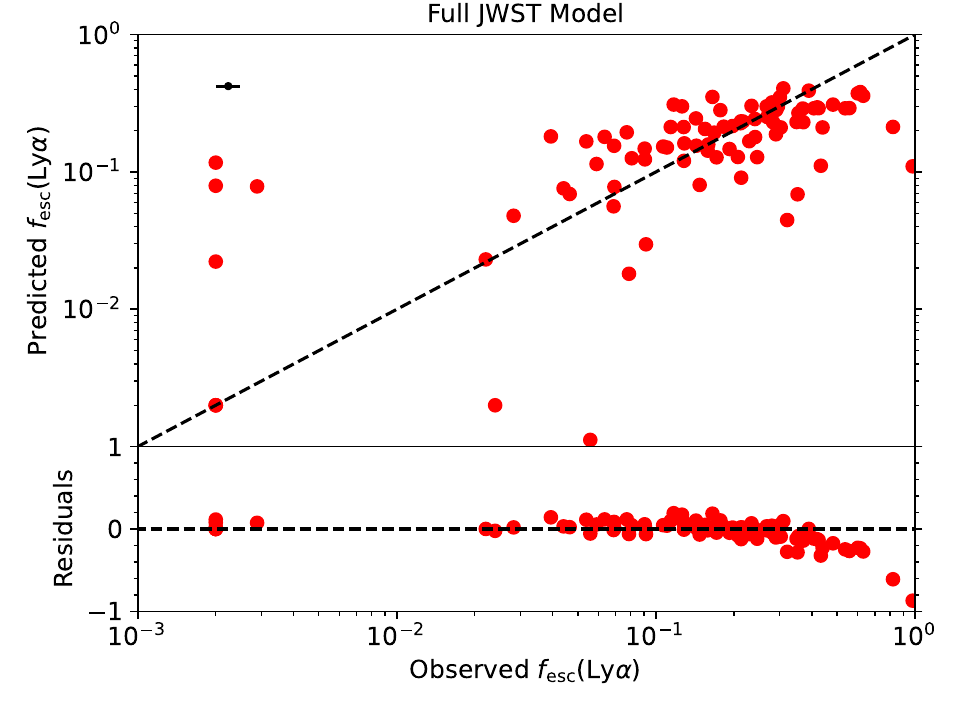}{0.3\textwidth}{(c)}}
\caption{Examples of \fesclya\ model predictions using the variable sets for the fiducial model (a), the fiducial model with EW(\hi,abs) (b) and the JWST model (c). Symbols are the same as in Figure~\ref{fig:fiducialfit}. We plot galaxies with negative \fesclya\ at \fesclya=2E-3. Of the three models, the fiducial model with EW(\hi,abs) predicts \fesclya\ with the lowest RMS scatter.
\label{fig:fesclya}}
\end{figure*}

The fiducial model is not very successful at predicting \fesclya\ (Figure~\ref{fig:fesclya}a, Table~\ref{table:metrics:altlyc}), with $R^2=0.33$, $R^2_{\rm adj}=0.24$, RMS=0.32, $C=0.75$. The lack of upper limits and need to precisely rank galaxies with even very low \fesclya\ may explain the reduced $C$ values for the \fesclya\ model. Including EW(\hi,abs) improves the \fesclya\ predictions  ($R^2=0.44$, RMS=0.30, $C=0.78$; Figure~\ref{fig:fesclya}b). As with the \fesc\ predictions, the JWST \fesclya\ model performs worse than the fiducial model (Figure~\ref{fig:fesclya}c), which demonstrates that the observed \fesclya\ does depend on line-of-sight \hi\ properties such as the neutral gas covering fraction or \hi\ column density. Unlike the \fesc\ models, for \fesclya, the fiducial model without E(B-V)$_{\rm UV}$ performs comparably to the fiducial model ($R^2=0.37$, RMS=0.29, $C=0.75$), because \lya\ is more sensitive to the nebular attenuation rather than the stellar attenuation.%aperture 

\begin{figure*}
\gridline{\fig{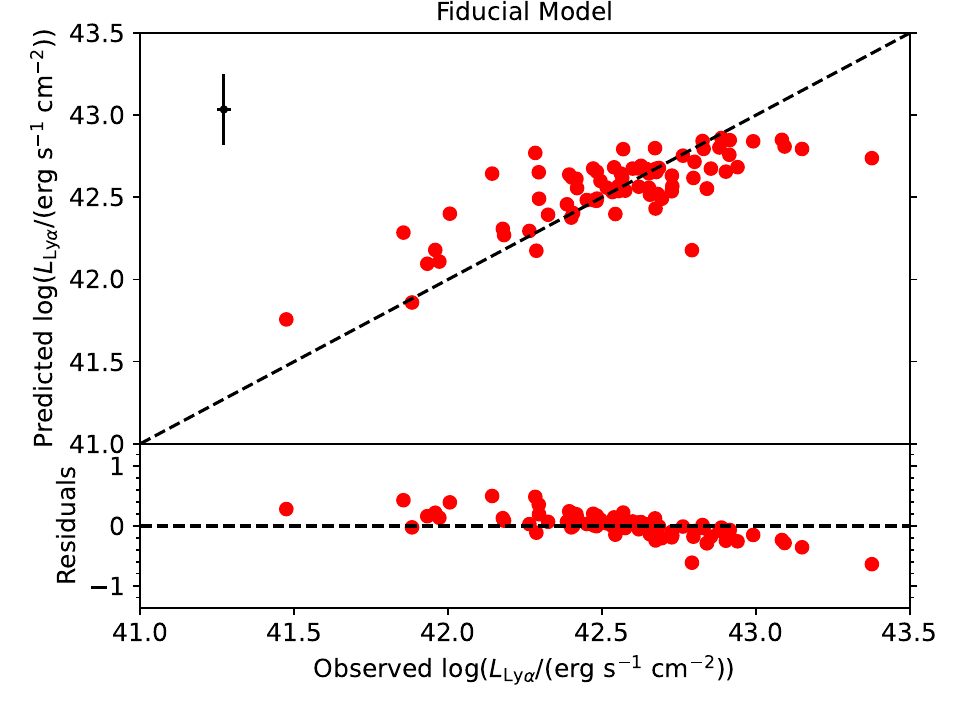}{0.5\textwidth}{(a)}
	\fig{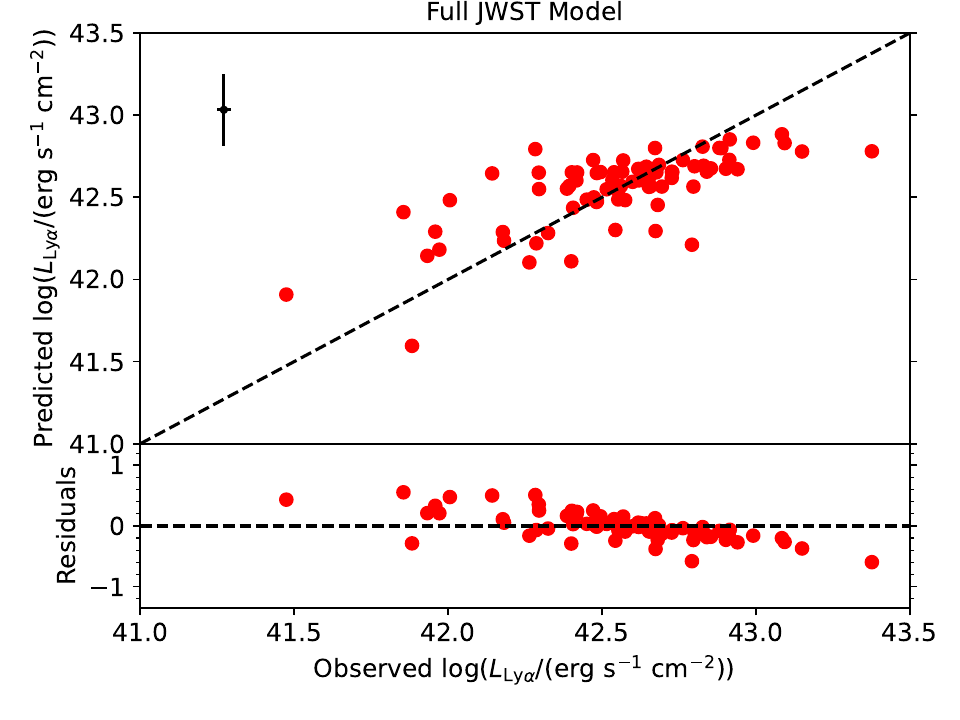}{0.5\textwidth}{(b)}}
\caption{Examples of $L$(\lya) model predictions using the variable sets for the fiducial model (a) and the JWST model (b). Symbols are the same as in Figure~\ref{fig:fiducialfit}. We exclude galaxies with negative $L$(\lya). The models do a better job at predicting $L$(\lya) compared to \fesclya.
\label{fig:llya}}
\end{figure*}

We also investigate $L$(\lya), using a similar method as for $L$(LyC) and predicting 43.5-\logten($L$(\lya)). We exclude all galaxies with net \lya\ absorption and predict only the luminosities of \lya\ Emitters. Whereas we found that $L$(LyC) was harder to predict than \fesc, for $L$(\lya), the situation is reversed: the $L$(\lya) models are actually slightly better than the \fesclya\ models (Table~\ref{table:metrics:altlyc} and Figure~\ref{fig:llya}). The $L$(LyC) predictions suffered from the large uncertainty in estimating the intrinsic LyC production from the FUV continuum. The available Balmer line measurements make estimating the intrinsic \lya\ luminosity less of a concern. Another explanation for the better performance of the $L$(\lya) predictions may be mundane; since it depends on predicting the intrinsic \lya, \fesclya\ is a more model-dependent measurement than $L$(\lya) and consequently has a greater uncertainty (26\% on average for \fesclya\ vs 21\% for the \lya\ flux alone). Again the $L$(\lya) JWST model (Table~\ref{table:metrics:altlyc} and Figure~\ref{fig:llya}b) has slightly worse $R^2$ and RMS metrics than the fiducial model because of the lack of information about line-of-sight gas.

Tables~\ref{table:rankorder_fesclya} and \ref{table:rankorder_llya} in the Appendix list the top-ranked variables for predicting \fesclya\ and $L$(\lya). For \fesclya, we find that \hi\ absorption strength and nebular dust attenuation are two of the most important predictive variables. In contrast, the UV dust attenuation is less important, always ranking after E(B-V)$_{\rm neb}$ in forward and backward selection. As a nebular emission line, \lya\ escape is more closely linked to the nebular attenuation rather than the attenuation experienced by the stellar light. 

For $L$(\lya), the most important variables include \sigsfr, $M_{\rm 1500}$, E(B-V)$_{\rm neb}$, EW(H$\beta$), 
and EW(\hi,abs). These five variables are also statistically significant in either the fiducial model (\sigsfr), the fiducial+EW(\hi,abs) model (EW(H$\beta$), EW(\hi,abs)), or both (E(B-V)$_{\rm neb}$ and $M_{\rm 1500}$). The dependence on UV luminosity reflects the fact that higher SFRs enhance \lya\ production, while the dependence on E(B-V)$_{\rm neb}$ and EW(\hi,abs) arises because lower dust attenuation and lower \hi\ column densities enhance \lya\ escape. As a recombination line, the H$\beta$ EW is also directly linked to nebular \lya\ production. While the dependence on \sigsfr\ may reflect the role of feedback in \lya\ escape \citep[\eg][]{kim21}, it may also serve to quantify aperture losses, as more extended galaxies may have more of their \lya\ emission scatter outside the COS aperture. 

\citet{runnholm20} also assess the most important variables in predicting $L$(\lya) using the Lyman-$\alpha$ Reference Sample (LARS). They find that the top five ranked variables by both forward and backward selection are SFR, E(B-V)$_{\rm neb}$, $M_*$, $r_{\rm 50}$, and the gas covering fraction. If using observables, the top variables are the FUV luminosity and the UV size. Despite the different methodologies between the LARS \lya\ measurements and the LzLCS+ \lya\ measurements, our results generally agree with their findings, as \sigsfr, E(B-V)$_{\rm neb}$, and the galaxy FUV luminosity also appear among our most important predictors. 

\section{Discussion}
\label{sec:discuss}
Both theoretical and observational studies have long recognized that LyC escape is likely anisotropic due to the inherently inhomogeneous structure of the ISM \citep[\eg,][]{gnedin08, wise09, kim13, zastrow13, paardekooper15, cen15, trebitsch17, keenan17, riverathorsen17a, riverathorsen19, kim23}. LyC may escape in a ``picket-fence" geometry, emerging only through holes in the ISM \citep[\eg,][]{heckman01}, and simulations find that supernova feedback may carve these low-density paths in the ISM \citep[\eg,][]{wise09, cen15, paardekooper15, trebitsch17}. Observationally, the detection of residual flux from saturated low-ionization absorption lines \citep[\eg,][]{heckman01, heckman11, gazagnes18} and the co-existence of narrow \lya\ emission with underlying broad absorption \citep[\eg,][]{mckinney19} provide evidence for optical depth variations within galaxies. Highly ionized channels observed in some low-redshift galaxies \citep[\eg,][]{zastrow13, bik18} and spatially resolved LyC observations in the lensed Sunburst Arc at $z=2.37$ \citep{riverathorsen19, kim23} likewise imply that LyC may escape through narrow opening angles, perhaps driven by outflows \citep[\eg,][]{amorin24}. Even when \fesc\ is large, with escape occurring in most or all directions, \hi\ column density variations will still lead to variations in the observed \fesc\ along different lines of sight \citep[\eg,][]{gazagnes20}.

Because of the anisotropy of LyC escape, estimating global \fesc\ values from line-of-sight LyC detections is difficult, particularly since the observed \fesc\ depends both on galaxy properties and on orientation. The multivariate Cox models present new evidence for the anisotropy of LyC escape by showing that the global properties of the strongest LCEs (\fesc $\gtrsim0.4$) do not distinguish them from weaker LCEs (\fesc $\sim0.04$; see \S\ref{sec:outliers}). At $z\sim3$, \citet{nakajima20} reach a similar conclusion, finding that high O32 galaxies can have wide variations in \fesc, despite sharing comparable spectral properties. We find that the only variable that successfully predicts the high observed \fesc\ of the strongest LCEs is EW(\hi,abs), which traces the line-of-sight \hi\ absorption. This result supports the picture described by \citet{gazagnes20}, where galaxies with high O32 may have sufficient ionizing radiation to ionize most lines of sight, but where the highest \fesc\ occurs in only the most diffuse channels. 

Multivariate survival analysis models such as those presented here may offer a tool for quantifying the line-of-sight variation in \fesc\ among galaxy populations. Given a large reference sample of galaxies with measurements of relevant global properties, the Cox models generate the {\it probability distribution} of \fesc\ for a given set of variables. If the variation in the observed \fesc\ for similar galaxies results only from orientation effects, the derived probability distributions should reflect the distribution of \fesc\ across all sightlines. 

Although the outsized importance of EW(\hi,abs) in accurately predicting high observed \fesc\ suggests that favorable orientation accounts for the strongest LCEs in the LzLCS+, orientation alone may not fully explain the scatter between predicted and observed \fesc\ across the entire LzLCS+ sample. LyC escape will depend on local properties, which global variables may not accurately capture \citep[\eg,][]{riverathorsen19}. Observational studies at higher spatial resolution and simulations that track the evolution of \fesc\ with high spatial resolution will help reveal the origin of \fesc\ variations in otherwise similar galaxies.

The variable rankings in \S\ref{sec:results:variablerank} show that the two most essential variables in predicting the line-of-sight \fesc\ are variables sensitive to the line-of-sight \hi\ gas (EW(\hi,abs)) and the line-of-sight dust (E(B-V)$_{\rm UV}$ or $\beta_{\rm 1550}$). Although $z>6$ observations will not be able to measure the former, $\beta_{\rm 1550}$ is easily accessible. {\it JWST} observations of the epoch of reionization are rapidly expanding the available measurements of $\beta_{\rm 1550}$ at high redshift \citep[\eg,][]{schaerer22b, mascia23, saxena24, morishita24}, and this variable alone can provide useful constraints on \fesc\ \citep{chisholm22}. By examining the spread of $\beta_{\rm 1550}$ values for a fixed intrinsic $M_{\rm 1500}$ or set of galaxy properties, high-redshift studies could gain insight into the possible variation in the line-of-sight \fesc\ among a population, at least the variation due to changing levels of dust attenuation. To understand the \hi\ component of the LyC absorption and its variation, high-redshift studies will need to rely on simulations and lower-redshift observations.

Our models provide some clues as to which physical properties most affect a galaxy's \hi\ distribution and resulting \fesc. In the absence of EW(\hi,abs) and \lya\ measurements, which more directly trace \hi, the models find that \sigsfr\ and O32 are the most important variables in predicting \fesc, given constraints on the dust attenuation from $\beta_{\rm 1550}$ (\S\ref{sec:jwstmodels}). These variables suggest that feedback is key in determining a galaxy's average \hi\ column density. Interestingly, we find in \S\ref{sec:subsamples} that O32 may be more important in lower mass galaxies (\logten($M_*$/\Msol) $<8.8$), whereas \sigsfr\ takes on greater importance in the higher mass LzLCS+ galaxies. Previous studies have suggested that the mechanism responsible for LyC escape may be different in different galaxy populations \citep[\eg,][]{flury22b, katz23}. Here, we find that differences in the multivariate \fesc\ models are specifically associated with galaxy mass, rather than luminosity. We hypothesize that the typical star formation histories of the low- and high-mass LzLCS+ galaxies may differ, with the lower-mass starbursts exhibiting burstier star formation \citep[\eg,][]{lee07}. Radiative feedback in extremely young, strong starbursts may drive LyC escape over wide opening angles in these low-mass galaxies, whereas the cumulative effect of supernovae in longer-lived concentrated starbursts could regulate LyC escape at higher masses \citep[\eg,][]{bremer23}.

While the same physical processes, such as bursty star formation with radiative feedback or an extended star formation episode with supernova feedback, may occur at high redshift, they may not occur in exactly the same type of galaxies as the LzLCS+. In particular, at a given stellar mass, a high-redshift galaxy could have a different gas mass, total mass, or star formation history compared to the same stellar mass galaxy at low redshift. Nevertheless, the distinct models for the low- vs.\ high-mass subsets in the LzLCS+ demonstrate that different parameters and physical mechanisms may regulate LyC escape in different types of galaxies. These differences could explain why some studies disagree regarding the properties of LCEs \citep[\eg,][]{alexandroff15, rutkowski17, izotov18b, naidu18}. Depending on the sample probed, high O32 or high \sigsfr\ may or may not be a relevant predictor of \fesc. Understanding the origin of the relationships between O32, \sigsfr, and \fesc\ in low- vs.\ high-mass LzLCS+ galaxies will require more observational and theoretical studies examining the interplay between star formation history, geometry, and feedback in these types of galaxies.

\section{Summary}
\label{sec:conclusions}
LyC escape depends on a range of galaxy properties, from gas and dust distributions in the ISM to galaxy star formation histories and the resulting stellar feedback. Because of this complexity, \fesc\ correlates observationally with a variety of parameters, and correlations with individual variables show high scatter. Using the results of the LzLCS and other $z\sim0.3$ LyC observations from the literature, we have generated new multivariate models to predict \fesc. We use the statistical technique of survival analysis, specifically the Cox proportional hazards model \citep{cox72}, which accounts for data from both LyC detections and \fesc\ upper limits from non-detections. With these models, we explore which combinations of variables best predict \fesc\ and discuss the possible physical basis for these trends.

We summarize our main findings below.
\begin{enumerate}
\item The \fesc\ predictions from our fiducial model match the observed \fesc\ of $z\sim0.3$ LCEs with an RMS of 0.36 dex, and our best-performing model, which includes \hi\ absorption lines, reaches an RMS of 0.31 dex. These multivariate models offer a substantial improvement over correlations with single variables, which can span $\gtrsim$ 2 dex in \fesc\ for a given galaxy property \citep[\eg,][]{wang21, flury22b, saldana22, chisholm22}. (\S\S\ref{sec:fiducial}-\ref{sec:modified_fiducial}).

\item Our best-performing model accurately predicts \fesc\ even for galaxies that are outliers in a single-variable selection. This model, which includes \hi\ absorption line measurements, is the only model that reproduces the high observed \fesc\ values of the strongest LCEs in the LzLCS+. Among the variables we consider, weak \hi\ absorption is the only distinctive property of these strong LCEs, which suggests that these high \fesc\ values may occur only along fortuitous sightlines. (\S\ref{sec:outliers}).

\item The most important variables in the models, as determined by forward and backward selection, are the EW of the \hi\ Lyman series absorption lines and either E(B-V)$_{\rm UV}$ or the UV slope $\beta_{\rm 1550}$. These properties trace the line-of-sight \hi\ absorption and UV dust attenuation, which determine the line-of-sight \fesc\ that we measure \citep[\eg,][]{saldana22}. In our fiducial model, E(B-V)$_{\rm UV}$ is a statistically significant variable, along with \fesclya, \sigsfr, E(B-V)$_{\rm neb}$, and O32. Excluding E(B-V)$_{\rm UV}$ or \fesclya\ has the greatest adverse effects on model performance, and including \fesclya\ appears to be particularly important for identifying non-LyC emitters. These analyses show which variables are the most important to include for accurate predictions; models using only the statistically significant variables or the 3-5 top-ranked variables achieve comparable accuracy to models with a longer list of inputs (\S\S\ref{sec:modified_fiducial}-\ref{sec:results:variablerank}).

\item We generate a ``JWST model" for predicting \fesc, which can apply to $z>6$ observations. The model excludes \lya\ and absorption line measurements, which will be inaccessible for most $z>6$ galaxies. When compared with the $z\sim0.3$ LCEs, this model has a higher scatter (RMS$=0.47$ dex) than the fiducial model, but it can distinguish LCEs from non-emitters. Three variables in this model are statistically significant: $\beta_{\rm 1550}$, \sigsfr, and O32. These three variables alone can predict \fesc\ with an RMS of 0.46 dex in the $z\sim0.3$ sample. (\S\ref{sec:jwstmodels}). 

\item We find that \fesc\ may depend on different variables for high- vs.\ low-mass galaxies. \sigsfr\ is statistically significant for the high-mass subsample but not the low-mass subsample and is the top-ranked variable for the high-mass subsample by forward and backward selection. In contrast, $M_{\rm 1500}$ and O32 are statistically significant for the low-mass subsample but not for the high-mass subsample. These differences suggest that the physical causes of LyC escape and the feedback mechanisms involved could differ across the galaxy population, with ionization and radiative feedback potentially playing a greater role in low-mass galaxies. (\S\ref{sec:subsamples}). 

\item The multivariate Cox models are less successful at accurately predicting alternative measures of LyC escape, such as \fratio\ and the LyC luminosity. We hypothesize that accurately predicting these properties requires information about the unattenuated SED shape, which is not included in our set of variables. (\S\ref{sec:results:altlyc}).

\item We also create Cox models to predict \lya, specifically the \fesclya\ and \lya\ luminosity. Unlike the LyC \fesc\ predictions, \hi\ absorption EW and E(B-V)$_{\rm neb}$, rather than E(B-V)$_{\rm UV}$, are two of the most important variables for predicting \fesclya. For $L$(\lya), top-ranked and statistically significant variables are the \hi\ absorption EW, E(B-V)$_{\rm neb}$, $M_{\rm 1500}$, EW(H$\beta$), and \sigsfr, which are similar to results found for the LARS sample \citep{runnholm20}. (\S\ref{sec:results:fesclya}). 

\end{enumerate}

Our models highlight the complexity of LyC escape and the dependence of \fesc\ on multiple variables. Multivariate Cox models represent a new tool to investigate the relationships between \fesc\ and galaxy properties and to predict \fesc\ in the epoch of reionization. These models offer a substantial improvement over \fesc\ predictions based on a single variable, and the models presented here can be refined and expanded to incorporate future observations. Our results show which variables have the greatest effect on \fesc\ predictions for the LzLCS+ sample and for galaxies in different mass ranges. By exploring these same variables, future simulations could try to reproducing these scalings and determine their physical origin. These flexible multivariate models also offer an empirical approach to predicting \fesc\ at high redshift. In a follow-up paper (Paper II, Jaskot et al., in prep.), we apply Cox models derived from the LzLCS+ to $z\sim3$ and $z>6$ samples and provide the best-fit models to the community for application to future observations.\footnote{We have also created a python script, available at \url{https://github.com/sflury/LyCsurv}, that generates Cox model \fesc\ predictions using any desired combination of variables from the LzLCS+. The LzLCS+ dataset used in this paper is available at the same page, and version 0.1.0 of the code is archived in Zenodo \citep{lycsurv}.} 

\begin{acknowledgments}
\nolinenumbers
We thank Kate Follette for interesting discussions that led to this project, and we thank the anonymous referee for comments that improved the clarity of this work. AEJ and SRF acknowledge support from NASA through grant HST-GO-15626. STScI is operated by AURA under NASA contract NAS-5-26555. ASL acknowledges support from Knut and Alice Wallenberg Foundation. MT acknowledges support from the NWO grant 0.16.VIDI.189.162 (``ODIN”).
\end{acknowledgments}

\appendix
Tables~\ref{table:rankorder_fratio}-\ref{table:rankorder_llya} list the four top-ranked variables by forward selection, backward selection, and an MC rank ordering for the \fratio, $L$(LyC), \fesclya, and $L$(\lya) predictions. We describe these ranking methods in \S\ref{sec:selection} and discuss the results in \S\S\ref{sec:results:altlyc} and \ref{sec:results:fesclya}. The ``Full Model" uses $\beta_{\rm 1550}$ as a measure of UV dust attenuation and includes multiple variables that trace \lya\ or UV absorption line strength. The ``Fiducial+HI Model" uses E(B-V)$_{\rm UV}$ to quantify dust attenuation, \fesclya\ as a measure of \lya, and EW(\hi,abs) as the only absorption line variable. The ``JWST Model" excludes absorption line and \lya\ variables.

\movetabledown=2.25in
\begin{rotatetable*}
\begin{deluxetable*}{lllllll}
\tablecaption{Most Important Variables for \fratio\ Predictions}
\label{table:rankorder_fratio}
\tablehead{
\colhead{Rank} & \multicolumn{2}{c}{Full Model} & \multicolumn{2}{c}{Fiducial+HI Model} & \multicolumn{2}{c}{JWST Model} \\
\colhead{} & \colhead{Forward} & \colhead{Backward} & \colhead{Forward} & \colhead{Backward} & \colhead{Forward} & \colhead{Backward}}
\startdata
\multicolumn{7}{c}{Top-Ranked Variables} \\
\hline
1 & EW(\hi,abs) & EW(\hi,abs) & EW(\hi,abs)  & EW(\hi,abs) & \logten(\sigsfr) & \logten(\sigsfr) \\
2 & $R_l$(\hi,abs)  & $M_{\rm 1500}$ & $M_{\rm 1500}$ & $M_{\rm 1500}$ & $\beta_{\rm 1550}$ & \logten(O32) \\
3 & $M_{\rm 1500}$ & EW(\lya) & \fesclya\  & \fesclya\ & 12+\logten(O/H) & \logten(EW(H$\beta$))  \\
4 & $\beta_{\rm 1550}$ & $\beta_{\rm 1550}$ & \logten(\sigsfr) & E(B-V)$_{\rm neb}$ & \logten(EW(H$\beta$)) & $\beta_{\rm 1550}$\\
\hline
\multicolumn{7}{c}{Top Variables by MC Rank Order} \\
\hline
 & EW(\hi,abs) [1.01] & EW(\hi,abs) [2.87] & EW(\hi,abs) [1.00] & EW(\hi,abs) [1.94] & \logten(\sigsfr) [1.13] &  \logten(\sigsfr) [1.09]  \\
 &$R_l$(\hi,abs) [2.86]  & $R_l$(\hi,abs) [5.83]  & \fesclya\ [3.57] & \fesclya\ [3.64]  & \logten(O32) [3.22] & \logten(O32) [2.14] \\
 & \fesclya\ [7.20]  &  \logten(EW(H$\beta$)) [5.95] & $M_{\rm 1500}$ [3.79] & \logten(\sigsfr) [4.58] & \logten(EW(H$\beta$)) [5.45] & \logten(EW(H$\beta$)) [3.11]  \\
 & EW(\lya) or \logten(\sigsfr) [7.50] & EW(\lya) [6.68]  & \logten(\sigsfr) [4.21] &  \logten(O32) [4.85] & $\beta_{\rm 1550}$ [3.97] & $\beta_{\rm 1550}$ [4.73] \\
\enddata
\tablecomments{Numbers in brackets indicate the mean of the variable's ranks in the 100 MC runs.}
\end{deluxetable*} %
\end{rotatetable*}

\movetabledown=2.25in
\begin{rotatetable*}
\begin{deluxetable*}{lllllll}
\tablecaption{Most Important Variables for $L$(LyC) Predictions}
\label{table:rankorder_llyc}
\tablehead{
\colhead{Rank} & \multicolumn{2}{c}{Full Model} & \multicolumn{2}{c}{Fiducial+HI Model} & \multicolumn{2}{c}{JWST Model} \\
\colhead{} & \colhead{Forward} & \colhead{Backward} & \colhead{Forward} & \colhead{Backward} & \colhead{Forward} & \colhead{Backward}}
\startdata
\multicolumn{7}{c}{Top-Ranked Variables} \\
\hline
1 & $L$(\lya) & EW(\lya) & \logten(\sigsfr)  & EW(\hi,abs) & \logten(\sigsfr) & \logten(O32) \\
2 & E(B-V)$_{\rm neb}$ & $M_{\rm 1500}$ & \fesclya\ & $M_{\rm 1500}$ & $\beta_{\rm 1550}$ & $M_{\rm 1500}$\\
3 & $M_{\rm 1500}$ & $\beta_{\rm 1550}$ & $M_{\rm 1500}$  & \logten(O32) & $M_{\rm 1500}$ & \logten(\sigsfr)  \\
4 & EW(\lya) & EW(\hi,abs) & \logten(O32) & \fesclya\ & \logten(O32) & \logten(EW(H$\beta$))\\
\hline
\multicolumn{7}{c}{Top Variables by MC Rank Order} \\
\hline
 & $L$(\lya) [1.00] & $M_{\rm 1500}$ [3.05] & $M_{\rm 1500}$ [2.61] & $M_{\rm 1500}$ [2.95] & \logten(\sigsfr) [1.17] & \logten(O32) [1.71]  \\
 &$M_{\rm 1500}$ [4.43]  & EW(\lya) [4.78]  & EW(\hi,abs) [3.28]  & EW(\hi,abs) [3.39]  & $\beta_{\rm 1550}$ [3.40] & $M_{\rm 1500}$ [2.55] \\
 & E(B-V)$_{\rm neb}$ [6.21] & EW(\hi,abs) [6.08] & \logten(\sigsfr) [3.30] & \fesclya\ [4.27] & $M_{\rm 1500}$ [3.53] & \logten(\sigsfr) [2.87]  \\
 & EW(\lya) [6.96] & \logten(O32) [6.52]  & \fesclya\ [3.32] &  \logten(O32) [4.54] & \logten(O32) [4.38] & \logten(EW(H$\beta$)) [4.58] \\
\enddata
\tablecomments{Numbers in brackets indicate the mean of the variable's ranks in the 100 MC runs.}
\end{deluxetable*} %
\end{rotatetable*}

\movetabledown=2.25in
\begin{rotatetable*}
\begin{deluxetable*}{lllllll}
\tablecaption{Most Important Variables for \fesclya\ Predictions\tablenotemark{a}}
\label{table:rankorder_fesclya}
\tablehead{
\colhead{Rank} & \multicolumn{2}{c}{Full Model} & \multicolumn{2}{c}{Fiducial+HI Model} & \multicolumn{2}{c}{JWST Model} \\
\colhead{} & \colhead{Forward} & \colhead{Backward} & \colhead{Forward} & \colhead{Backward} & \colhead{Forward} & \colhead{Backward}}
\startdata
\multicolumn{7}{c}{Top-Ranked Variables} \\
\hline
1 & \logten($M_*$) & E(B-V)$_{\rm neb}$ & \logten($M_*$)  & EW(\hi,abs) & \logten($M_*$) & $\beta_{\rm 1550}$ \\
2 & E(B-V)$_{\rm neb}$  & EW(LIS) & EW(\hi,abs) & E(B-V)$_{\rm neb}$ & E(B-V)$_{\rm neb}$ & E(B-V)$_{\rm neb}$ \\
3 & EW(\hi,abs) & EW(\hi,abs) & E(B-V)$_{\rm neb}$ & E(B-V)$_{\rm UV}$ & 12+\logten(O/H) & 12+\logten(O/H)  \\
4 & EW(LIS) & \logten(\sigsfr) & E(B-V)$_{\rm UV}$ & 12+\logten(O/H) & $\beta_{\rm 1550}$ & \logten($M_*$)\\
\hline
\multicolumn{7}{c}{Top Variables by MC Rank Order} \\
\hline
 & E(B-V)$_{\rm neb}$ [1.61] & E(B-V)$_{\rm neb}$ [2.01] & E(B-V)$_{\rm neb}$ [2.36] & E(B-V)$_{\rm neb}$ [2.43] & E(B-V)$_{\rm neb}$ [2.18] &  E(B-V)$_{\rm neb}$ [1.93]  \\
 &EW(\hi,abs) [3.19]  & EW(\hi,abs) [2.85]  & EW(\hi,abs) [2.89] & EW(\hi,abs) [2.51]  & 12+\logten(O/H) [3.34] & 12+\logten(O/H) [2.91] \\
 & EW(LIS) [5.65] & 12+\logten(O/H) [5.87] & E(B-V)$_{\rm UV}$ [3.92] & 12+\logten(O/H) [4.21 & \logten(O32) [3.40] & $\beta_{\rm 1550}$ [3.51]  \\
 & 12+\logten(O/H) [6.24] & EW(LIS) [6.39]  & \logten(O32) [3.97] & E(B-V)$_{\rm UV}$ [4.29] & $\beta_{\rm 1550}$ [3.45] & \logten($M_*$) [4.61] \\
\enddata
\tablecomments{Numbers in brackets indicate the mean of the variable's ranks in the 100 MC runs.}
\tablenotetext{a}{Variable lists exclude any measures of \lya.}
\end{deluxetable*} %
\end{rotatetable*}

\movetabledown=2.25in
\begin{rotatetable*}
\begin{deluxetable*}{lllllll}
\tablecaption{Most Important Variables for $L$(\lya) Predictions\tablenotemark{a}}
\label{table:rankorder_llya}
\tablehead{
\colhead{Rank} & \multicolumn{2}{c}{Full Model} & \multicolumn{2}{c}{Fiducial+HI Model} & \multicolumn{2}{c}{JWST Model} \\
\colhead{} & \colhead{Forward} & \colhead{Backward} & \colhead{Forward} & \colhead{Backward} & \colhead{Forward} & \colhead{Backward}}
\startdata
\multicolumn{7}{c}{Top-Ranked Variables} \\
\hline
1 & \logten(\sigsfr) & $M_{\rm 1500}$ & \logten(\sigsfr)  & \logten(EW(H$\beta$)) & \logten(\sigsfr) & \logten(EW(H$\beta$)) \\
2 & E(B-V)$_{\rm neb}$  & \logten(EW(H$\beta$)) & E(B-V)$_{\rm neb}$ & $M_{\rm 1500}$ & E(B-V)$_{\rm neb}$ & $M_{\rm 1500}$ \\
3 & $M_{\rm 1500}$ & EW(\hi,abs) & $M_{\rm 1500}$ & EW(\hi,abs) & $M_{\rm 1500}$ & 12+\logten(O/H)  \\
4 & 12+\logten(O/H) & E(B-V)$_{\rm neb}$ & 12+\logten(O/H) & E(B-V)$_{\rm UV}$ & 12+\logten(O/H) & \logten(\sigsfr) \\
\hline
\multicolumn{7}{c}{Top Variables by MC Rank Order} \\
\hline
 & \logten(\sigsfr) [1.73] & $M_{\rm 1500}$ [1.93] & \logten(\sigsfr) [2.10] & $M_{\rm 1500}$ [1.98] & \logten(\sigsfr) [1.00] &  $M_{\rm 1500}$ [1.94]  \\
 &E(B-V)$_{\rm neb}$ [3.24]  & \logten(EW(H$\beta$)) [4.25]  & E(B-V)$_{\rm neb}$ [2.71] & EW(\hi,abs) [4.15]  & E(B-V)$_{\rm neb}$ [2.07] & \logten(EW(H$\beta$)) [3.71] \\
 & $M_{\rm 1500}$ [3.50] & EW(\hi,abs) [4.88]  & $M_{\rm 1500}$ [3.24] &  \logten(EW(H$\beta$)) [4.21] & $M_{\rm 1500}$ [3.61] &  \logten(\sigsfr) [4.36]  \\
 & 12+\logten(O/H) [5.08] & 12+\logten(O/H) [5.63]  & EW(\hi,abs) [4.78] &  12+\logten(O/H) [5.18] & 12+\logten(O/H) [3.97] & 12+\logten(O/H) [4.58] \\
\enddata
\tablecomments{Numbers in brackets indicate the mean of the variable's ranks in the 100 MC runs.}
\tablenotetext{a}{Variable lists exclude any measures of \lya.}
\end{deluxetable*} %
\end{rotatetable*}

\bibliography{latestrefs}
\bibliographystyle{aasjournal}

\end{document}